\documentclass[a4paper,showpacs,preprintnumbers,superscriptaddress,aps,prd,nofootinbib,floatfix,10pt]{revtex4-1}
\usepackage{graphicx}
\usepackage{dcolumn}
\usepackage{bm}
\usepackage{amsmath,amssymb,amsfonts}
\usepackage{latexsym}
\usepackage{color}
\usepackage{subcaption}
\usepackage{ulem}
\usepackage{mathrsfs}
\usepackage{braket}
\usepackage{yhmath}
\usepackage{cancel}
\usepackage{tensor}

\usepackage[usestackEOL]{stackengine} 
 
\usepackage{multirow}

\usepackage{physics}
\usepackage{titlesec}

\renewcommand{\thesection}{\arabic{section}}
\renewcommand{\thesubsection}{\arabic{subsection}}

\usepackage{hyperref}
\hypersetup{%
setpagesize=false,
 bookmarksnumbered=true,%
 bookmarksopen=true,%
 colorlinks=true,%
 linkcolor=blue,
 citecolor=magenta,
}

\usepackage{graphicx}% Include figure files
\usepackage{dcolumn}% Align table columns on decimal point
\usepackage{bm}% bold math
\usepackage{physics,amsmath,amssymb,amsfonts}
\usepackage{latexsym}
\usepackage{color}
\usepackage{slashed}
\def\nn    {\nonumber}

\usepackage{titlesec}

\titleformat{\section}
  {\normalfont\large\bfseries}{\thesection}{1em}{}
\titlespacing*{\section}
  {0pt}{1.ex plus .1ex minus .1ex}{0.ex plus .1ex minus .1ex}

\titleformat{\subsection}
  {\normalfont\bfseries}{\thesection.\thesubsection}{1em}{}
\titlespacing*{\subsection}
  {0pt}{1.ex plus .1ex minus .1ex}{0.ex plus .1ex minus .1ex}

\makeatletter
\renewcommand{\p@subsection}{\thesection.}
\renewcommand{\p@subsubsection}{\thesection.\thesubsection.}
\makeatother

\titleformat{\subsubsection}
  {\normalfont\itshape}{\thesection.\thesubsection.\thesubsubsection}{1em}{}
\titlespacing*{\subsubsection}
  {0pt}{1.ex plus .1ex minus .1ex}{0.ex plus .1ex minus .1ex}

\begin{document}
\allowdisplaybreaks
\flushbottom
%%%%%%%%%%%%%%%%%%%%%%%%%
\title{\boldmath %$SU(2) \times U(1)$ 
% Electroweak preheating and geometric baryogenesis in $R^2$-Higgs inflation
 \vspace*{2cm} Implication of preheating on gravity assisted baryogenesis in $R^2$-Higgs inflation
\vspace{5mm}
}

%%%%%%%%%%%%%%%%%%%%%%%%%
\author{Yann Cado}
\affiliation{Laboratoire de Physique Th\'eorique et Hautes Energies (LPTHE), Sorbonne Universit\'e et CNRS UMR 7589, 4 place Jussieu, 75252 Paris CEDEX 05, France\\[0.1cm]}
\author{Christoph Englert}
\affiliation{School of Physics \& Astronomy, University of Glasgow, Glasgow G12 8QQ, UK\\[0.1cm]}
\author{Tanmoy Modak}
\affiliation{Department of Physical Sciences, Indian Institute of Science Education and Research Berhampur, Berhampur 760010, Odisha, India\\[0.1cm]}
\author{Mariano Quir\'{o}s}
\affiliation{Institut de F\'{i}sica d'Altes Energies (IFAE) and The Barcelona Institute of Science and Technology (BIST),
Campus UAB, 08193 Bellaterra, Barcelona, Spain\\[0.1cm]}

\begin{abstract}
% \vspace{8.5cm}
% \begin{center}
% % \textbf{Abstract}  \\
% \end{center}
We investigate the impact of preheating on baryogenesis in $R^2$-Higgs inflation. In this scenario, the inclusion of a dimension-six operator ${(R/ \Lambda^2)} B_{\mu\nu} \widetilde{B}^{\mu\nu} $  abundantly generates helical hypermagnetic fields during inflation, leading to a baryon asymmetric Universe at the electroweak crossover. Focusing on the $R^2$-like regime, we first derive the relevant dynamics of preheating using a doubly-covariant formalism. We find that preheating can happen for the Higgs, transverse gauge and Goldstone bosons, however, it is dependent
on the value of the non-minimal coupling $\xi_H$ between the Standard Model Higgs field and the Ricci scalar. We identify the preheating temperature to determine the appropriate scale $\Lambda$ for driving baryogenesis, which is around $\Lambda \sim 2.2 \, (2.6) \times 10^{-5}\, M_{\rm P}$ for $\xi_H \sim 1 \, (10)$.  Our results represent the most accurate estimation of the scale of gravity induced baryogenesis in $R^2$-Higgs inflation to date. Areas for further improvement are identified.
\end{abstract}

\vspace{2ex}
\maketitle
\hrule
\vspace{2ex}
\tableofcontents
\vspace{2ex}
\numberwithin{equation}{section}
%\newpage
\hrule
\setlength{\parskip}{1\baselineskip}
\setlength{\parindent}{0pt}
%%%%%%%%%%%%%%%%%%%%%%%%%%%%%%%%%%%%%%%%%%%%%
\vspace{1ex}
%%%%%%%%%%%%%%%%%%%%%%%%%%%%%%%%%%%%%%%%%%%%%%
\section{Introduction}\label{sec:intro}
%%%%%%%%%%%%%%%%%%%%%%%%%%%%%%%%%%%%%%%%%%%%%%

The existence of the baryon asymmetry of the Universe (BAU) is firmly established by various cosmological observations such as the cosmic microwave background and big-bang nucleosynthesis. However, its origin still remains unclear. If the fundamental scale of the mechanism behind the BAU is tied to a higher scale than the electroweak one, it might be possible that telltale effects at present, or even future, colliders could remain absent. One such high-scale mechanism provides, in the Jordan frame~\cite{Cado:2023zbm}, an additional source of CP violation via a (CP-odd) dimension-six operator $ {(R/ \Lambda^2)}B_{\mu\nu} \widetilde{B}^{\mu\nu} $, where $R$ is the Ricci scalar and $B_{\mu\nu}$ is the field stress tensor of the hypercharge $U(1)_Y$. We will refer to this mechanism as gravity assisted baryogenesis from here onwards. In this case, even at the electroweak crossover of the Standard Model (SM), the out-of-equilibrium condition can be met if the source and washout decay rates are different and shut off at different epochs~\cite{Shaposhnikov:1987tw,Kamada:2016eeb}. The CP-violating dimension-six Chern-Simons density can abundantly produce helical hypermagnetic fields at the end of inflation~\footnote{This term, not directly connected to BAU but in the context of hypermagnetic field production, has been discussed in, e.g., Refs.~\cite{Durrer:2022emo,Durrer:2023rhc,Savchenko:2018pdr,Subramanian:2015lua,Durrer:2013pga}.}. These helical hypermagnetic fields~\footnote{See also Refs.~\cite{Anber:2006xt,Bamba:2006km,Bamba:2007hf,Anber:2009ua,Anber:2015yca,Cado:2016kdp} for the production of helical hypermagnetic fields due to inflaton dynamics.} may then create the observed baryon asymmetry at the electroweak crossover~\cite{Kamada:2016eeb,Kamada:2016cnb,Jimenez:2017cdr,Domcke:2019mnd,Cado:2021bia,Cado:2022evn,Cado:2022pxk}.

This mechanism can seamlessly be integrated into inflationary scenarios such as $R^2$-Higgs inflation~\cite{Salvio:2015kka,Ema:2017rqn,Pi:2017gih,Gorbunov:2018llf,Gundhi:2018wyz,He:2018mgb,Cheong:2019vzl,He:2020ivk,He:2020qcb} since the ${(R/ \Lambda^2)} B_{\mu\nu} \widetilde{B}^{\mu\nu} $ term can also be considered within the context of $f(R)$ gravity (or, rather here, $f(R,\Phi, B_\mu)$ gravity). In the dual scalar-tensor theory, the $R^2$ term is manifest as a dynamical scalar degree of freedom, which, along with the Higgs field $\Phi$, couples to the Chern-Simons density, resulting in the production of hypermagnetic fields at the end of inflation. It should also be noted that $R^2$-Higgs inflation, like $R^2$~\cite{Starobinsky:1980te, Starobinsky:1983zz,Vilenkin:1985md,Mijic:1986iv, Maeda:1987xf} inflation and Higgs inflation~\cite{Bezrukov:2007ep,Barvinsky:2008ia,Bezrukov:2010jz,Bezrukov:2013fka,DeSimone:2008ei,Bezrukov:2008ej,Barvinsky:2009ii} (for similar mechanisms, see e.g.~\cite{Spokoiny:1984bd,Futamase:1987ua,Salopek:1988qh,Fakir:1990eg,Amendola:1990nn,Kaiser:1994vs,Cervantes-Cota:1995ehs,Komatsu:1999mt}), is one of the best-fit models of the Planck data~\cite{Planck:2018jri}. Further, unlike Higgs inflation, where longitudinal gauge bosons are violently produced well beyond the unitarity cut-off scale~\cite{DeCross:2015uza,Ema:2016dny,Sfakianakis:2018lzf}, the unitarity scale is restored up to the Planck scale in $R^2$-Higgs inflation~\cite{Ema:2017rqn}. Thus, high-scale baryogenesis via ${(R/ \Lambda^2)} B_{\mu\nu} \widetilde{B}^{\mu\nu} $ can be elegantly connected to the $R^2$-Higgs model without requiring additional degrees of freedom beyond the SM ones.

It is not surprising that, in the gravity assisted baryogenesis, the production of hypermagnetic fields highly depends on the inflationary dynamics. The baryon-to-entropy ratio is also highly sensitive to the value of the magnetic Reynolds number which in turn highly depends on the exact (p)reheating temperature~\cite{Kamada:2016cnb,Cado:2021bia,Cado:2022evn,Cado:2022pxk,Cado:2023zbm}. In addition, the latter is also highly dependent on the value of quantum hypermagnetic energy density when preheating is complete. Therefore, the exact amount of baryon asymmetry will be model-dependent as it makes use of specific (p)reheating results~\footnote{In the absence of efficient preheating, thermalization proceeds through perturbative reheating.}. An improved understanding of these dynamics brings about major improvements in the precision of predictions since we do not need to rely on estimates as in previous studies but directly determine the thermal plasma temperature, the relevant energy densities and the scale factor ratio at reheating and the end of inflation. Previous studies performed by the current authors treated such effects as effective parameters~\cite{Cado:2023zbm}, while a detailed analysis of how (p)reheating impacts gravity assisted baryogenesis was missing. This will entail us to find a more precise value for the $\Lambda$ required for baryogenesis.

In this work, taking $R^2$-Higgs inflation as a benchmark model for inflation, and focusing primarily on the $R^2$-like regime, we study the impact of preheating and particle production in the mechanism of gravity assisted baryogenesis. Adopting the doubly-covariant formalism and including all perturbations at the linear order, we study the preheating dynamics of the scalar and gauge sectors. In Ref.~\cite{Sfakianakis:2018lzf}, Sfakianakis and van de Vis did provide a similar detailed analysis of linear fluctuations during preheating in Higgs inflation in the Einstein frame. In our case, the field space of the inflationary dynamics consists of five fields: one dynamical scalar degree of freedom arising due to the $R^2$ term and four from the Higgs $SU(2)_L$ doublet, corresponding to the physical Higgs and the three Goldstone bosons. The background dynamics are governed by the $R^2$ scalar alongside the physical Higgs boson; the three would-be Goldstone bosons are treated as perturbations. Our results include the full $SU(2)_L\times U(1)_Y$ gauge dynamics in a complete analysis.

The inflaton's self-resonance turns out to not be efficient enough for preheating. However, Higgs fluctuations can lead to preheating if the non-minimal coupling $\xi_H$ between the Ricci scalar $R$ and the Higgs is $\gtrsim 10$. We also show that for $\xi_H \sim 10$, the transverse modes of the $Z$ and $W$ boson can lead to gauge preheating. For tiny $\xi_H$ values, gauge preheating might be induced if $\Lambda$ is sufficiently small. We shall see that such a small value of $\Lambda$ would, however, lead to an overproduction of baryon asymmetry. Further, Goldstone bosons may also preheat the Universe even for $\xi_H\sim 1$ (see also Refs.~\cite{Ema:2016dny,Bezrukov:2019ylq,Garcia-Bellido:2008ycs,He:2020ivk,Bezrukov:2020txg,Figueroa:2021iwm,Jeong:2023zrv} for Goldstone/longitudinal gauge boson preheating in $R^2$-Higgs inflation). Our results show that the Goldstone bosons can preheat the Universe mildly faster than any other fields and, hence, they determine the reheating temperature needed for the baryogenesis computation.

We organize this paper as follows. First, we start with outlining the action and derive the relevant equations of motion (EoM) for the different fields in Sec.~\ref{sec:action}. We discuss the inflationary dynamics in the covariant formalism in Sec.~\ref{sec:infdynamics}. The production of inflaton and Higgs fluctuations are studied in Sec.~\ref{sec:Inf+Higgs-fluctuations}, and the production of the $Z$, $W$ and Goldstone bosons is presented in Sec.~\ref{sec:GWZ-production}. The production of hypermagnetic fields and subsequent generation of the BAU are discussed, respectively, in Sec.~\ref{sec:EM-production} and Sec.~\ref{sec:baryogensis}. We discuss reheating in Sec.~\ref{sec:reheating}. Finally, we summarize and conclude in Sec.~\ref{sec:disc}. 
Some technical and computational details are relegated to App.~\ref{app:gaugegoldspectra}.

%%%%%%%%%%%%%%%%%%%%%%%%%%%%%%%%%%%%%%%%%%%%%%
\section{The action}\label{sec:action}
%%%%%%%%%%%%%%%%%%%%%%%%%%%%%%%%%%%%%%%%%%%%%%
We start with the action in the Jordan frame given by
\begin{equation} \begin{aligned}
  S_J  &=  \int d^4 x \sqrt{-g_J} \bigg[ \frac{M_{\rm P}^{2}}{2} f(R_J, \Phi, B_\mu, W^i_\mu)
  -g_J^{\mu\nu}(\nabla_\mu\Phi)^\dagger \nabla_\nu\Phi - V(\Phi, \Phi^\dagger)  -  \dfrac{1}{4} g_J^{\mu\rho} g_J^{\nu\sigma} B_{\mu\nu}B_{\rho\sigma}
  - \dfrac{1}{4} g_J^{\mu\rho} g_J^{\nu\sigma} W^i_{\mu\nu}W^i_{\rho\sigma}
 \bigg],\label{eq:actionJ1}
\end{aligned} \end{equation}
where $M_{\rm P}=\sqrt{1/\left(8\pi G\right)}\approx 2.4\times 10^{18}~\text{GeV}$ is the reduced Planck mass and $G$ is Newton's constant. Throughout this work, we follow the mostly-plus convention $(-1,+1,+1,+1)$ for the metric, $\sqrt{-g_J}$ is the determinant of the metric, and we choose the $\epsilon^{0123}=1$ convention for the Levi-Civita tensor. $R_J$ and $\Phi$ denote the space-time Ricci scalar and the Higgs doublet, $B_{\mu\nu}$ and $W^i_{\mu\nu}$ are the field stress tensors of the $U(1)_Y$ and $SU(2)_L$ gauge groups, respectively. The covariant derivative with the SM gauge groups is defined as
\begin{align}
\nabla_\mu = D_\mu + i g' \frac{1}{2} Q_{Y_f} B_\mu + i g \, \bm{T} \cdot \bm{W}_\mu,
\end{align}
where $g'$ and $g$ are $U(1)_Y$ and $SU(2)_L$ couplings. $Q_{Y_f}$ is $U(1)_Y$ hypercharge and $\bm{T}={\bm{\tau}}/{2}$ are the weak isospin generators derived from the three Pauli matrices $\bm{\tau}$. The field-stress tensors for the $U(1)_Y$ and $SU(2)_L$ gauge fields can be written as
\begin{align}
B_{\mu\nu} = D_\mu B_\nu - D_\nu B_\mu, \hspace{1.8cm} W^i_{\mu\nu} = D_\mu W^i_\nu - D_\nu W^i_\mu - g \sum_{j,k=1}^{3} \epsilon_{ijk} W^j_\mu W^k_\nu,
\end{align}
where $D_\mu$ is the covariant derivative of the space-time metric $g_{J\mu\nu}$. We have thus far ignored the fermions in Eq.~\eqref{eq:actionJ1} but shall return later part of the paper.
The Higgs potential $V(\Phi, \Phi^\dagger)$ and $f(R_J, \Phi, B_\mu, W^i_\mu)$ are given as
\begin{subequations} \begin{align}
V(\Phi, \Phi^\dagger) &= \lambda|\Phi|^4, \\ f(R_J, \Phi, B_\mu, W^i_\mu) &= R_J  + \frac{\xi_R}{2 M_{\rm P}^2} R_J^2 + \frac{2\xi_H}{M_{\rm P}^2} |\Phi|^2 R_J
- \frac{2}{\Lambda^2M_{\rm P}^2}\frac{\epsilon^{\mu\nu\rho\sigma}}{\sqrt{-g_J}} B_{\mu\nu} B_{\rho\sigma} R_J - \frac{2}{\Lambda^2 M_{\rm P}^2}\frac{\epsilon^{\mu\nu\rho\sigma}}{\sqrt{-g_J}} W^i_{\mu\nu} W^i_{\rho\sigma} R_J. \label{def:FR+CScouplings}
\end{align} \end{subequations}
The mass term in the Higgs potential is neglected as it plays no role for the inflationary dynamics.

To transition from a generic $f(R_J, \Phi, B_\mu, W^i_\mu)$ gravity to the respective scalar-tensor theory, we perform a Legendre transformation by first introducing an auxiliary field $\Psi$ and rewrite Eq.~\eqref{eq:actionJ1} as
\begin{equation} \begin{aligned}
  S_J  =  \int d^4 x \sqrt{-g_J}  \bigg[&\frac{M_{P}^{2}}{2} \left(f(\Psi,  \Phi, B_\mu,W^i_\mu)
  + \frac{\partial f(\Psi,  \Phi, B_\mu,W^i_\mu)}{\partial \Psi} (R_J-\Psi)\right)
 \\
  &   -g_J^{\mu\nu}(\nabla_\mu\Phi)^\dagger \nabla_\nu\Phi - V(\Phi, \Phi^\dagger) -  \dfrac{1}{4} g_J^{\mu\rho} g_J^{\nu\sigma} B_{\mu\nu}B_{\rho\sigma}
  - \dfrac{1}{4} g_J^{\mu\rho} g_J^{\nu\sigma} W^i_{\mu\nu}W^i_{\rho\sigma}\bigg]\label{eq:actionJ3}.
\end{aligned} \end{equation}
We can introduce a physical degree of freedom
\begin{align}
\Theta = \frac{\partial f(\Psi,  \Phi, B_\mu,W^i_\mu)}{\partial \Psi}
\end{align}
and re-express Eq.~\eqref{eq:actionJ3} as
\begin{equation} \begin{aligned}
  S_J  =  \int d^4 x \sqrt{-g_J} \bigg[& \frac{M_{P}^{2}}{2} \Theta R_J - U(\Theta,   \Phi, B_\mu,W^i_\mu)
  -g_J^{\mu\nu}(\nabla_\mu\Phi)^\dagger \nabla_\nu\Phi - V(\Phi, \Phi^\dagger)  \\
  &-  \dfrac{1}{4} g_J^{\mu\rho} g_J^{\nu\sigma} B_{\mu\nu}B_{\rho\sigma}
  - \dfrac{1}{4} g_J^{\mu\rho} g_J^{\nu\sigma} W^i_{\mu\nu}W^i_{\rho\sigma}\bigg]\label{eq:actionJ4}
\end{aligned} \end{equation}
with
\begin{equation} \begin{aligned}
U(\Theta,   \Phi, B_\mu,W^i_\mu) &= \frac{M_{\rm P}^2}{2}\left[\Psi(\Theta) \Theta -  f(\Psi(\Theta),  \Phi, B_\mu)\right] \\
                          &= \frac{M_{\rm P}^4}{4 \xi_R}\bigg[\bigg(1- \Theta + \frac{\xi_H}{M_{\rm P}^2} |\Phi|^2
                          - \frac{2 }{\Lambda^2 M_{\rm P}^2}\frac{\epsilon^{\mu\nu\rho\sigma}}{\sqrt{-g_J}} B_{\mu\nu} B_{\rho\sigma}
                          - \frac{2 }{\Lambda^2 M_{\rm P}^2}\frac{\epsilon^{\mu\nu\rho\sigma}}{\sqrt{-g_J}} W^i_{\mu\nu} W^i_{\rho\sigma}\bigg)^2\bigg].
\end{aligned} \end{equation}
Next, we Weyl-rescale the metric
\begin{align}
g_{J\mu\nu} = \frac{1}{\Theta} \ g_{E\mu\nu}, \hspace{1.5cm} g^{\mu\nu}_J = \Theta \ g^{\mu\nu}_E,   \hspace{1.5cm}  \sqrt{-g_{J}}= \frac{1}{\Theta^2}\sqrt{-g_E},
\end{align}
to write the action Einstein frame as
\begin{equation} \begin{aligned}
S_E  = \int d^4 x \sqrt{-g_E}\bigg[ & \frac{M_{\rm P}^2}{2} R_E - \frac{3 M_{\rm P}^2}{4} g_E^{\mu \nu} D_\mu (\ln\Theta) D_\nu(\ln\Theta)
-\frac{1}{2 \Theta} g_E^{\mu\nu}(\nabla_\mu\Phi)^\dagger \nabla_\nu\Phi - V_E  \\
&  -  \dfrac{1}{4} g_E^{\mu\rho} g_E^{\nu\sigma} B_{\mu\nu}B_{\rho\sigma}
  - \dfrac{1}{4} g_E^{\mu\rho} g_E^{\nu\sigma} W^i_{\mu\nu}W^i_{\rho\sigma}\bigg]\label{eq:actionE1}
\end{aligned} \end{equation}
with 
\begin{align}
V_E &= \frac{1}{\Theta^2}\left[V(\Phi, \Phi^\dagger) +U(\Theta, \Phi, B_\mu,W^i_\mu)\right],\\
R_J &= \Theta \left[R_E +3 \Box_E  \Theta- \frac{3}{2} g_{E}^{\mu\nu} D_\mu (\ln\Theta) D_\nu(\ln\Theta) \right].
\end{align}
We have ignored the surface term $\Box_E = g_{E}^{\mu\nu} D_\mu D_\nu$ in the action $S_E$. With the field redefinition
\begin{align}
\phi = M_{\rm P} \sqrt{\frac{3}{2}} \ln\Theta,
\end{align}
Eq.~\eqref{eq:actionE1} finally becomes
\begin{equation} \begin{aligned}
S_E  = \int d^4 x \sqrt{-g_E}\bigg[&\frac{M_{\rm P}^2}{2} R_E - \frac{1}{2} g_E^{\mu \nu} D_\mu \phi D_\nu \phi
- e^{-\sqrt{\frac{2}{3}}\frac{\phi}{M_{\rm P}}} g_E^{\mu\nu}(\nabla_\mu\Phi)^\dagger \nabla_\nu\Phi - V_E \\
&  -  \dfrac{1}{4} g_E^{\mu\rho} g_J^{\nu\sigma} B_{\mu\nu}B_{\rho\sigma}
  - \dfrac{1}{4} g_E^{\mu\rho} g_E^{\nu\sigma} W^i_{\mu\nu}W^i_{\rho\sigma}\bigg]\label{eq:actionE2}.
\end{aligned} \end{equation}

We now turn our attention to the Higgs and gauge fields. The $Q_Y=+1$ Higgs field decomposes as
\begin{align}
\Phi =
\frac{1}{\sqrt{2}} \begin{pmatrix}
  \phi_3+ i \phi_4 \\
  h + i \phi_2 \\
\end{pmatrix} \label{eq:higgsuni}.
\end{align}
It is customary to perform a basis transformation for gauge bosons from the electroweak $W^i_\mu,B_\mu$ to the mass and QED charge basis $W_\mu,Z_\mu,A_\mu$ as
\begin{equation} \begin{aligned}
&W^1_\mu = \frac{W^+_{\mu} + W^-_{\mu}}{\sqrt{2}}, \hspace{2.4cm} W^2_\mu = \frac{i}{\sqrt{2}} \left(W^+_{\mu} - W^-_{\mu}\right),\\
&W^3_\mu = s_W A_\mu  + c_W Z_\mu, \hspace{2cm} B_\mu =  c_W A_\mu    -  s_W Z_\mu,\label{eq:gaugetomass}
\end{aligned} \end{equation}
where $e = s_W g = c_W g'$ with shorthand $s_W$ and $c_W$ for the sine and cosine of the electroweak Weinberg angle $\theta_W$.
In the following, we will consider equations of motion (EoM) at the linear order. We therefore expand the action Eq.~\eqref{eq:actionE2}
to quadratic order
\begin{equation} \begin{aligned}
S_E  &= \int d^4 x \sqrt{-g_E}\bigg[\frac{M_{\rm P}^2}{2} R_E - \frac{1}{2} G_{IJ} g_E^{\mu \nu} D_\mu \phi^I D_\nu \phi^J - 
V_E(\phi^I) -  \dfrac{1}{4} g_E^{\mu\rho} g_E^{\nu\sigma} F_{A\mu\nu}F_{A\rho\sigma}-  
\dfrac{1}{4} g_E^{\mu\rho} g_E^{\nu\sigma} F_{Z\mu\nu}F_{Z\rho\sigma} \\
&  - \dfrac{1}{2} g_E^{\mu\rho} g_E^{\nu\sigma} F^+_{W\mu\nu}F^-_{W\rho\sigma} - e^{-\sqrt{\frac{2}{3}}\frac{\phi}{M_{\rm P}}} g_E^{\mu\nu}\bigg(\frac{g_Z^2}{8}h^2 Z_\mu Z_\nu+
\frac{g_Z}{2}\left[ (D_\mu h) \ \phi_2- (D_\mu\phi_2) h\right] Z_\nu +\frac{e^2}{4 s_W^2} h^2  W^+_\mu W^-_\nu+\\
&\frac{i e}{2 \sqrt{2} s_W} D_\mu h \left[W^-_\nu  (\phi_3+i \phi_4)  - W^+_\nu   (\phi_3-i \phi_4) \right]-
\frac{i e}{2 \sqrt{2} s_W}  \left[ W^-_\nu D_\mu(\phi_3+i\phi_4)  -  W^+_\nu D_\mu(\phi_3-i\phi_4) \right] h\bigg)\label{eq:actionfinal},
\end{aligned} \end{equation}
introducing $\phi^I \in \{\phi, h, \phi_2, \phi_3, \phi_4 \}$, $g_Z= {e}/({s_W c_W})$, and the $5\times 5$ field space metric $G_{IJ}$ whose non-vanishing elements are
\begin{align}
G_{\phi\phi} = 1, \hspace{1.5cm} G_{ h h} = e^{-\sqrt{\frac{2}{3}}\frac{\phi}{M_{\rm P}}},\hspace{1.5cm} G_{\phi_i \phi_i} = e^{-\sqrt{\frac{2}{3}}\frac{\phi}{M_{\rm P}}}~~\mbox{with}~i=2,3,4.
\end{align}
We have treated Goldstone and gauge bosons as perturbations, i.e. they do not acquire any background values while deriving Eq~\eqref{eq:actionfinal}.
The potential $V_E(\phi^I)$ reads
\begin{equation} \begin{aligned}
V_E(\phi^I) &= e^{-2\sqrt{\frac{2}{3}}\frac{\phi}{M_{\rm P}}} \left[\frac{\lambda}{4} \left(h^2+\sum^4_{i=2} \phi_i^2\right)^2 
+ \frac{M_{\rm P}^4}{4 \xi_R}\Biggl\{1- e^{\sqrt{\frac{2}{3}}\frac{\phi}{M_{\rm P}}} + \frac{\xi_H}{M_{\rm P}^2} \left(h^2+\sum^4_{i=2} \phi_i^2\right) \right.\\
&\left.\hspace{2cm} - \frac{2}{\Lambda^2 M_{\rm P}^2}\frac{\epsilon^{\mu\nu\rho\sigma}}{\sqrt{-g_E}} e^{2 \sqrt{\frac{2}{3}}\frac{\phi}{M_{\rm P}}}  B_{\mu\nu} B_{\rho\sigma} - \frac{2}{\Lambda^2 M_{\rm P}^2}\frac{\epsilon^{\mu\nu\rho\sigma}}{\sqrt{-g_E}} e^{2 \sqrt{\frac{2}{3}}\frac{\phi}{M_{\rm P}}}  W^i_{\mu\nu} W^i_{\rho\sigma}\Biggr\}^2\right]\\
                          &\approx   V_0(\phi^I) + \frac{2 M_{\rm P}^2}{\xi_R\Lambda^2}
F(\phi^I)e^{\sqrt{\frac{2}{3}}\frac{\phi}{M_{\rm P}}}\, \bigg(F_{A\mu\nu} \widetilde{F}_A^{\mu\nu} + F_{Z\mu\nu} \widetilde{F}_Z^{\mu\nu} + 2 F^+_{W \mu\nu} \widetilde{F}_{W}^{-\mu\nu} \bigg),
\label{def:VE-final}
\end{aligned} \end{equation}
where we have further introduced
\begin{align}
&F(\phi^I)=1-e^{-\sqrt{\frac{2}{3}}\, \frac{\phi}{M_{\rm P}}}-\frac{\xi_H}{M_{\rm P}^2}\left(h^2+\sum^4_{i=2} \phi_i^2\right)e^{-\sqrt{\frac{2}{3}}\, \frac{\phi}{M_{\rm P}}},\\
&V_0(\phi^I)= \frac{\lambda}{4}\left(h^2+\sum^4_{i=2} \phi_i^2\right)^2e^{-2\sqrt{\frac{2}{3}}\,\frac{\phi}{M_{\rm P}}}+\frac{M_{\rm P}^4}{4\xi_R}F^2(\phi^I),\\
&\widetilde{F}_A^{\mu\nu}=\frac{1}{2 \sqrt{-g_E}}\epsilon^{\mu\nu\rho\sigma} F_{A\rho\sigma},~~\widetilde{F}_Z^{\mu\nu}=\frac{1}{2 \sqrt{-g_E}}\epsilon^{\mu\nu\rho\sigma}F_{Z\rho\sigma},~~
\widetilde{F}_W^{\pm\mu\nu}=\frac{1}{2 \sqrt{-g_E}}\epsilon^{\mu\nu\rho\sigma}F^\pm_{W\rho\sigma}.
\end{align}
The field stress tensors for the massive and massless gauge bosons are
\begin{subequations} \label{def:all-field-strength} \begin{eqnarray}
F_{A\mu\nu} &=& D_\mu A_\nu - D_\nu A_\mu = \partial_\mu A_\nu -\partial_\nu A_\mu,\\
F_{Z\mu\nu} &=& D_\mu Z_\nu - D_\nu Z_\mu = \partial_\mu Z_\nu -\partial_\nu Z_\mu,\\
F^{\pm}_{W\mu\nu} &=& D_\mu W^{\pm}_\nu - D_\nu W^{\pm}_\mu = \partial_\mu W^{\pm}_\nu -\partial_\nu W^{\pm}_\mu.
\end{eqnarray}  \end{subequations}
Since the field-stress tensors are torsionless, the covariant derivatives become partial derivatives. Furthermore,
in the linearized approximation, the field-stress tensors have reduced to the abelian case.

Varying the action, the equation of motion of the scalars fields $\phi^I$ can be written as
\begin{equation} \begin{aligned}
&\Box \phi^K + \Gamma^{K}_{\ IJ} \ g_E^{\alpha \nu} D_\alpha  \phi^I D_\nu \phi^J - G^{KM} V_{E,M} \\&+ G^{KM}
e^{-\sqrt{\frac{2}{3}}\frac{\phi}{M_{\rm P}}} \left(\sqrt{\frac{2}{3}}\frac{1}{M_{\rm P}}\right) g_E^{\alpha\nu} (\partial_\alpha \phi)
\bigg(\frac{g_Z}{2} \delta^3_M h Z_\nu +\frac{i e}{2 \sqrt{2} s_W}  \left[ W^-_\nu  (\delta^4_M+i\delta^5_M)  -  W^+_\nu  (\delta^4_M-i\delta^5_M) \right] h\bigg)\\
&-G^{KM}e^{-\sqrt{\frac{2}{3}}\frac{\phi}{M_{\rm P}}} g_E^{\alpha\nu}
\bigg(\frac{g_Z}{2}   \ D_\alpha\left(\delta^3_M h Z_\nu\right)
+\frac{i e}{2 \sqrt{2} s_W}  D_\alpha\left[\left( W^-_\nu  (\delta^4_M+i\delta^5_M)  -  W^+_\nu  (\delta^4_M-i\delta^5_M) \right] h\right)\bigg)  \\
&-G^{KM}\bigg[e^{-\sqrt{\frac{2}{3}}\frac{\phi}{M_{\rm P}}} g_E^{\mu\nu}\delta^3_M\bigg(\frac{g_Z}{2}(D_\mu h) Z_\nu\bigg)+e^{-\sqrt{\frac{2}{3}}\frac{\phi}{M_{\rm P}}} g_E^{\mu\nu}\delta^4_M\bigg(\frac{i e}{2 \sqrt{2} s_W} D_\mu h \left(\ W^-_\nu - W^+_\nu \right)\bigg)\\
&+e^{-\sqrt{\frac{2}{3}}\frac{\phi}{M_{\rm P}}} g_E^{\mu\nu}\delta^5_M\bigg(\frac{i e}{2 \sqrt{2} s_W} D_\mu h \left(\ i W^-_\nu + i W^+_\nu \right)\bigg)\bigg]
= 0\label{eom:scalar},
\end{aligned} \end{equation}
where $\Gamma^{K}_{\ IJ}$ are the Christoffel symbols associated with field-space metric.
The EoMs for the $Z$ boson, $W^\pm$ bosons and photon are
\begin{align}
g_E^{\mu \alpha} g_E^{\nu \beta} D_\alpha F_{Z\mu\nu}+&  \frac{8 M_{\rm P}^2}{\xi_R \Lambda^2 } \partial_\alpha \left( F(\phi^I)e^{\sqrt{\frac{2}{3}}\, \phi/M_{\rm P}} \right)
  \widetilde{F}_{Z}^{\alpha\beta}
-   e^{-\sqrt{\frac{2}{3}}\frac{\phi}{M_{\rm P}}} g_E^{\mu\beta} \bigg(  \frac{g_Z^2}{4} h^2 Z_\mu + \frac{g_Z}{2}
\left(\phi_2 D_\mu h - h D_\mu \phi_2\right)\bigg)= 0, \label{eq:Zeom} \\
g_E^{\mu \alpha} g_E^{\nu \beta} D_\alpha F^\pm_{W\mu\nu} +&  \frac{8 M_{\rm P}^2}{\xi_R \Lambda^2 } \partial_\alpha \left( F(\phi^I)e^{\sqrt{\frac{2}{3}}\, \phi/M_{\rm P}} \right)
  \widetilde{F}_{W}^{\pm\alpha\beta}
-  e^{-\sqrt{\frac{2}{3}}\frac{\phi}{M_{\rm P}}} g_E^{\mu\beta} \bigg(\frac{e^2}{4 s_W^2} h^2 W^\pm_\mu
\pm\frac{i e}{2 \sqrt{2} s_W}D_\mu h \left(\phi_3\pm i \phi_4\right)
\nn\\
&\mp\frac{i e}{2 \sqrt{2} s_W}D_\mu(\phi_3 \pm i \phi_4) h\bigg)
= 0, \label{eq:Wpmeom} \\
%%%%%%%%%%%%
 g_E^{\mu \alpha} g_E^{\nu \beta} D_\alpha F_{A\mu\nu}+& \frac{8 M_{\rm P}^2}{\xi_R \Lambda^2 } \partial_\alpha\left( F(\phi^I)e^{\sqrt{\frac{2}{3}}\, \phi/M_{\rm P}} \right)
   \widetilde{F}_{A}^{\alpha\beta}  = 0. 
   \label{eq:photon}
\end{align}

%%%%%%%%%%%%%%%%%%%%%%%%%%%%%%%%%%%%%%%%%%%%%%%%%%%%%%%%%%%%%%%%%
\section{Inflationary dynamics}
\label{sec:infdynamics}
The preheating after inflation depends on the background and perturbation dynamics. We closely follow covariant formalism as discussed in Ref.~\cite{Gong:2011uw,Sfakianakis:2018lzf,Kaiser:2012ak}, which is suited for multifield inflation with non-canonical kinetic terms as encountered in our scenario. We decompose the $\phi^I(x^\mu)$ fields into a homogeneous classical background part (${\varphi}^I $) and a perturbation ($\delta\phi^I$) part
\begin{align}
\phi^I(x^\mu) = \varphi^I(t) + \delta\phi^I(x^\mu)\label{fieldexpan}.
\end{align}
In the following, $t$ labels the cosmic time and $\varphi^I(t) = \{\varphi(t),h_0(t)\}$, i.e. only the Higgs and inflaton fields acquire background field values while the Goldstone modes $\phi_2$, $\phi_3$ and $\phi_4$ are perturbations. The perturbed spatially flat Friedmann-Robertson-Walker (FRW) metric can be expanded to linear order as~\cite{Kodama:1984ziu,Mukhanov:1990me,Malik:2008im}
\begin{align}
ds^2 &= -(1+2\mathcal{A}) dt^2 +2 a(t) (\partial_i \mathcal{B}) dx^i dt +
a(t)^2 \left[(1-2\psi) \delta_{ij}+ 2 \partial_i \partial_j \mathcal{E}\right] dx^i dx^j,\label{eq:frwmetric}
\end{align}
where $a(t)$ is scale factor. $\mathcal{A}, \mathcal{B}, \psi$ and $\mathcal{E}$ characterize the scalar metric perturbations. In this work, we adopt the longitudinal gauge where the scalar perturbations $ \mathcal{B}$ and $\mathcal{E}$ vanish.

Utilizing Eq.~\eqref{eom:scalar} and Eq.~\eqref{fieldexpan}, we find the EoMs for the background  fields as
\begin{subequations} \begin{eqnarray}
&\mathcal{D}_t \dot{\varphi} + 3 H\dot{\varphi} + G^{\phi J} V_{0,J}= 0\label{eq:bkg_inf},\\
&\mathcal{D}_t \dot{h}_0 + 3 H\dot{h}_0 + G^{hJ} V_{0,J}= 0,\label{eq:bkg_higgs}
\end{eqnarray} \label{eq:bkg_phi+h}  \end{subequations}
where
\begin{subequations} \begin{eqnarray}
\mathcal{D}_t A^I &\equiv& \dot{\varphi}^J \mathcal{D}_J A^I  = \dot{A}^I  + \Gamma^I_{\; JK} \dot{\varphi}^J A^K,\\
 \mathcal{D}_J A^I &\equiv& \partial_J A^I + \Gamma^I_{\; JK} A^K.
\end{eqnarray}  \end{subequations}
We draw the reader's attention to the fact that the covariant derivative $\mathcal{D}_I$ of field space $G_{IJ}$ shall not be confused with the covariant derivative $D_\mu$ of space-time metric $g_{J\mu\nu}$.
The Hubble function is defined as
\begin{align}
H^2 &= \left(\frac{\dot{a}}{a}\right)^2 = \frac{1}{3 M_{\rm P}^2} \bigg(\frac{1}{2} G_{IJ} \dot{\varphi}^I \dot{\varphi}^J + V_0(\varphi^I)\bigg),\label{hubble1}\\
\dot{H} &= -\frac{1}{2 M_{\rm P}^2} \bigg(G_{IJ} \dot{\varphi}^I \dot{\varphi}^J\bigg).\label{hubble2}
\end{align}
We solve the equations \eqref{eq:bkg_phi+h} and \eqref{hubble1} together while simultaneously performing consistency check such that $\dot{H}$ estimated from Eq.~\eqref{hubble1} matches Eq.~\eqref{hubble2} with adequate precision. We define the number of $e$-foldings relative to the end of inflation as
\begin{align}
\mathcal{N} \equiv \ln \frac{a(t)}{a(t_{\rm{end}})},
\end{align}
which we will use in the following interchangeably with the cosmic time $t$. The background energy density is
\begin{align}
&\rho_{\mathrm{inf}} = \frac{1}{2} G_{IJ} \dot{\varphi}^I \dot{\varphi}^J + V_0(\varphi^I),\label{eq:infenergy}
\end{align}
where $G_{IJ}$ is evaluated at the background field order.

The field fluctuations $\delta\phi^I(x^\mu)$ are gauge-dependent quantities. However, we can construct covariant field fluctuations $\mathcal{Q}^I(x^\mu)$ which connect the scalar fields $\phi^I(x^\mu)$ with their background fields $\varphi^I(t)$ along a unique geodesic of the field-space manifold such that the field fluctuations can be written as~\cite{Gong:2011uw,Elliston:2012ab}
\begin{align}
\delta\phi^I &= \mathcal{Q}^I -\frac{1}{2} \Gamma^I_{\ JK} \mathcal{Q}^K \mathcal{Q}^J+\frac{1}{3!} \big(\Gamma^I_{\ MN} \Gamma^N_{\ JK}-\Gamma^I_{\ JK,M}\big)  \mathcal{Q}^K \mathcal{Q}^J  \mathcal{Q}^M+\dots~.
\end{align}
This motivates one to consider gauge-independent Mukhanov-Sasaki variables for the field fluctuations expressed as~\cite{Sasaki:1986hm,Mukhanov:1988jd,Mukhanov:1990me}
\begin{align}
Q^I = \mathcal{Q}^I + \frac{\dot{\varphi}^I}{H}\psi,\label{mukh-sasaki}
\end{align}
which is doubly covariant with respect to field-space and space-time transformations. The quantities $\mathcal{Q}^I$, $\dot{\varphi}^I$ and $Q^I$ transform like vectors in the field-space manifold. In our five-field case, at the linear order, we thus have
\begin{subequations} \begin{eqnarray}
Q^\phi &=& \mathcal{Q}^\phi + \frac{\dot{\varphi}}{H}\psi=\delta\varphi+ \frac{\dot{\varphi}}{H}\psi,\\
Q^h &=& \mathcal{Q}^h + \frac{\dot{h}_0}{H}\psi= \delta h+ \frac{\dot{h}_0}{H}\psi,\\
Q^{\phi_i}& =& \mathcal{Q}^{\phi_i} = \delta\phi_i=\phi_i, \hspace{.5cm} (i=2,3,4)%Q^{\phi_3} &=& \mathcal{Q}^{\phi_3}= \delta\phi_3 ,\hspace{1.5cm} Q^{\phi_4} = \mathcal{Q}^{\phi_4}= \delta\phi_3.
\end{eqnarray} \end{subequations}
Inserting Eq.~\eqref{mukh-sasaki} and Eq.~\eqref{eq:frwmetric} in Eq.~\eqref{eom:scalar}, we find the EoMs for the gauge independent $Q^I$ at linear order as
\begin{align}
\mathcal{D}_t^2 Q^\phi &+ 3 H \mathcal{D}_t Q^\phi -\frac{\nabla^2}{a^2} Q^\phi + \mathcal{M}^\phi_{\ \ \phi} Q^\phi = 0,\label{eqpertur:phi}\\
\mathcal{D}_t^2 Q^h &+ 3 H \mathcal{D}_t Q^h -\frac{\nabla^2}{a^2} Q^h + \mathcal{M}^h_{\ \ h} Q^h = 0,\label{eqpertur:h}\\
\mathcal{D}_t^2 Q^{\phi_2} &+ 3 H \mathcal{D}_t Q^{\phi_2} -\frac{\nabla^2}{a^2} Q^{\phi_2} + \mathcal{M}^{\phi_2}_{\ \ \phi_2} Q^{\phi_2} 
+ \frac{g_Z}{2} \bigg[\left(\sqrt{\frac{2}{3}}\frac{1}{M_{\rm P}}\right)  \dot{\varphi} h_0 Z_0 -  2 \dot{h}_0 Z_0 + h_0 g_E^{\alpha \nu} (D_\alpha Z_\nu)\bigg]
= 0,\label{eqpertur:phi_2}\\
\mathcal{D}_t^2 Q^{\phi_3} &+ 3 H \mathcal{D}_t Q^{\phi_3} -\frac{\nabla^2}{a^2} Q^{\phi_3} + \mathcal{M}^{\phi_3}_{\ \ \phi_3} Q^{\phi_3}
+ \frac{i e}{2 \sqrt{2} s_W} \bigg[  \left(\sqrt{\frac{2}{3}}\frac{1}{M_{\rm P}}\right)  \dot{\varphi} h_0 \left(W^-_0-W^+_0\right)
- 2 \dot{h}_0  \left(W^-_0-W^+_0\right)\nn\\
&+ g_E^{\alpha \nu} h_0 D_\alpha\left(W^-_\nu-W^+_\nu\right) \bigg]=0,\label{eqpertur:phi_3}\\
\mathcal{D}_t^2 Q^{\phi_4} &+ 3 H \mathcal{D}_t Q^{\phi_4} -\frac{\nabla^2}{a^2} Q^{\phi_4} + \mathcal{M}^{\phi_4}_{\ \ \phi_4} Q^{\phi_4}
+ \frac{i e}{2 \sqrt{2} s_W} \bigg[  \left(\sqrt{\frac{2}{3}}\frac{1}{M_{\rm P}}\right)  \dot{\varphi} h_0 \left(i W^-_0+ i W^+_0\right)
-2 \dot{h}_0  \left(i W^-_0+ i W^+_0\right)\nn\\
&+ g_E^{\alpha \nu} h_0 D_\alpha\left(i W^-_\nu+i W^+_\nu\right) \bigg]=0,\label{eqpertur:phi_4}
\end{align}
with
\begin{align}
\mathcal{M}^{I}_{\ L} = G^{IJ} (\mathcal{D}_L\mathcal{D}_J V_E)- \mathcal{R}^I_{\ JKL} \dot{\varphi}^J \dot{\varphi}^K
- \frac{1}{M_{\rm P}^2 a^3} \mathcal{D}_t \left(\frac{a^3}{H}\dot{\varphi}^I \dot{\varphi}_L\right),\label{eq:massterm}
\end{align}
where the $\mathcal{R}^I_{\ JKL}$ is the field-space Riemann tensor. Here all relevant quantities such as $V_0$, $G^{IJ}$, $\Gamma^{I}_{\ JK}$, $\mathcal{R}^I_{\ JKL}$ are evaluated at background order. In contrast to Ref.~\cite{Cado:2023zbm}, which focused on the unitary gauge for the Higgs sector, the EoMs for the perturbations $Q^{\phi_2}$, $Q^{\phi_3}$ and $Q^{\phi_4}$ now depend on the $Z$ and $W$ bosons at linear order. Reference~\cite{Cado:2023zbm} solely focused on the production of the hypermagnetic fields \textit{at the end of inflation}; the reheating temperature was treated as a free parameter. Therefore, the choice of unitary gauge was not relevant for the estimation of baryon asymmetry. However, for the present study of preheating, as is clear from the last terms involving $Z$ and $W$ bosons in the respective Eqs.~\eqref{eqpertur:phi_2}, \eqref{eqpertur:phi_3} and~\eqref{eqpertur:phi_4}, the unitary gauge becomes ill-defined (specifically at zero crossings of the background fields). 

Finally, and for later convenience, we re-express the EoMs of the scalar field fluctuations in conformal time~$\tau$ such that line element becomes $ds^2 = a^2(\tau) \eta_{\mu\nu} dx^\mu dx^\nu$ with the rescaled variables $X^I(x^\mu) \equiv a(t) Q^I(x^\mu)$. Hence, performing the replacements $Z_0 \to Z_0 / a$, $W^\pm_0 \to W^\pm_0 / a$ and $\partial_0 \to \partial_\tau / a$, we find new EoMs
\begin{align}
\mathcal{D}_\tau^2 X^\phi &-\bigg[\nabla^2  - a^2 \left(\mathcal{M}^\phi_{\ \ \phi} -\frac{1}{6}R_E G^\phi_{\ \ \phi} \right)\bigg] X^\phi = 0,\label{eqpertur:Xphi}\\
\mathcal{D}_\tau^2 X^h &-\bigg[\nabla^2  - a^2 \left(\mathcal{M}^h_{\ \ h} -\frac{1}{6}R_E G^h_{\ \ h} \right)\bigg] X^h = 0,\label{eqpertur:Xh}\\
\mathcal{D}_\tau^2 X^{\phi_2} &-\bigg[\nabla^2  - a^2 \left(\mathcal{M}^{\phi_2}_{\ \ \phi_2} -\frac{1}{6}R_E G^{\phi_2}_{\ \ \phi_2} \right)\bigg] X^{\phi_2}+ a \; g_Z h_0 \left( \Upsilon  Z_0 - Z_0^\prime \right)
= 0,\label{eqpertur:Xphi_2}\\
\mathcal{D}_\tau^2 X^{\phi_3} &-\bigg[\nabla^2 - a^2 \left(\mathcal{M}^{\phi_3}_{\ \ \phi_3} -\frac{1}{6}R_E G^{\phi_3}_{\ \ \phi_3} \right)\bigg] X^{\phi_3}+ a \;  \frac{i e}{ \sqrt{2} s_W} \, h_0 \bigg[ \Upsilon  \left(W^-_0-W^+_0\right)-  \left(W^{- \prime}_0-W^{+ \prime}_0\right) \bigg]=0,\label{eqpertur:Xphi_3}\\
\mathcal{D}_\tau^2 X^{\phi_4} &-\bigg[\nabla^2  - a^2 \left(\mathcal{M}^{\phi_4}_{\ \ \phi_4} -\frac{1}{6}R_E G^{\phi_4}_{\ \ \phi_4} \right)\bigg] X^{\phi_4}+ a \;  \frac{i e}{ \sqrt{2} s_W} \, h_0 \bigg[ \Upsilon \left(i W^-_0+ i W^+_0\right)- \left(i W^{- \prime}_0+ i W^{+ \prime}_0\right)\bigg]=0,\label{eqpertur:Xphi_4}
\end{align}
where we defined
\begin{align}
\Upsilon (\tau) = \frac{\varphi'}{\sqrt{6}M_{\rm P}} -  \frac{a'}{a} -   \frac{h_0'}{h_0}.
\label{def:Upsilon}
\end{align}
In these equations, as in the following, we have introduced the shorthand notation $(')$ for derivatives with respect to conformal time $\tau$.
%------------------------------------------------------------------------------------
\subsection{Gauge choice and equations of motion of the gauge and Goldstone bosons}
%------------------------------------------------------------------------------------
We have already mentioned that the unitary gauge becomes ill-defined at the zero crossing of the background field $h_0$ (see also Refs.~\cite{Ema:2016dny,Sfakianakis:2018lzf}). We, therefore, choose the Coulomb gauge to study the dynamics of field fluctuations. In the Coulomb gauge, we have $\partial_i Z^i = 0$ and $\partial_i W^{\pm i} = 0$ and the Goldstone bosons are dynamical.
%---------------------------
\subsubsection{$Z$ bosons}
%---------------------------
We begin with the EoM for the $Z$ boson. The time (i.e.~$\beta = 0$) and space (i.e.~$\beta = i$) components of Eq.~\eqref{eq:Zeom} at linear perturbation order in conformal time are
\begin{align}
-\partial_i\left(\partial_i Z_0 -Z_i'\right) +   a^2\, e^{-\sqrt{\frac{2}{3}}\frac{\varphi}{M_{\rm P}}} &\bigg( \frac{g_Z^2}{4} h_0^2 Z_0 + \frac{g_Z}{2}
\left(\phi_2 h_0' - h_0 \phi_2'\right)\bigg)= 0,\label{eq:Zoem-timeconf} \\
-\partial_\tau \left(\partial_\tau Z_i -\partial_i Z_0 \right)
+\partial_j \left(\partial_j Z_i - \partial_i Z_j\right) 
&- a^2\, e^{-\sqrt{\frac{2}{3}}\frac{\varphi}{M_{\rm P}}} \bigg(\frac{g_Z^2}{4}   h_0^2 Z_i -  \frac{g_Z}{2} h_0 \partial_i\phi_2 \bigg)\nn\\
&+  \frac{4 M_{\rm P}^2}{\xi_R \Lambda^2} \partial_\tau \left( F(\varphi^I)e^{\sqrt{\frac{2}{3}}\, \frac{\varphi}{M_{\rm P}}} \right) \epsilon^{i jk}
\left(\partial_j Z_k - \partial_k Z_j\right) =0. \label{eq:Zoem-spaceconf}
\end{align}
One can move to momentum space by considering
\begin{align}
&Z_0(\tau,\vb{x}) = \int \frac{d^3k}{\left(2\pi\right)^3} \widetilde{Z}_0(\tau,\vb{k}) e^{-i \vb{k}\cdot\vb{x}},\hspace{1.5cm}
Z_i(\tau,\vb{x}) = \int \frac{d^3k}{\left(2\pi\right)^3} \widetilde{Z}_i(\tau,\vb{k}) e^{-i \vb{k}\cdot\vb{x}},\label{eq:Zmomdecomp}
\end{align}
where the $\widetilde{\vb{Z}}\equiv \left(\widetilde Z_i(\tau,\vec k)\right)$ field can be written in terms of transverse and longitudinal components as
\begin{align}
\widetilde{\vb{Z}}(\tau,\vb{k}) =  \sum_{\lambda=\pm,L} \widetilde{Z}^\lambda(\tau,\vb{k})\ \hat{\epsilon}_Z^\lambda(\vb{k}), \label{Z:pol}
\end{align}
with
\begin{align}
i\vb{k}\cdot\hat{\epsilon}_Z^\pm(\vb{k}) = 0, \hspace{7mm}i\vb{k}\cdot\hat{\epsilon}_Z^L(\vb{k}) = |\vb{k}|=k, \hspace{7mm}i\vb{k}\times\hat{\epsilon}_Z^\pm(\vb{k}) = \pm k \ \hat{\epsilon}_Z^\pm(\vb{k}), \hspace{7mm} \hat{\epsilon}_Z^\lambda(\vb{k})^\ast=   \hat{\epsilon}_Z^\lambda(-\vb{k}). \label{def:helical-basis-Z}
\end{align}
The Coulomb gauge condition for the $Z$ boson then implies
\begin{align}
\partial_i Z^i = \partial_i \left(\overline{g}_E^{ik}Z_k\right)= \overline{g}_E^{ik} \partial_i Z_k = \frac{1}{a^2} \partial_i Z_i =0,\label{gauge:Zpos}
\end{align}
which translates in the momentum space to
\begin{align}
k_i \widetilde{Z}_i = \eta_{ij} k_i \widetilde{Z}^j =  k_j \widetilde{Z}^j =  \vb{k}\cdot\widetilde{\vb{Z}}=0.\label{gauge:Zmomconf}
\end{align}
Inserting Eq.~\eqref{Z:pol} in Eq.~\eqref{gauge:Zmomconf}, we get $\widetilde{Z}^L = 0$ as a consequence of the gauge condition and Eq.~\eqref{def:helical-basis-Z}. Using the gauge condition of Eq.~\eqref{gauge:Zpos} in Eqs.~\eqref{eq:Zoem-timeconf}-\eqref{eq:Zoem-spaceconf} and going to momentum space, we find
\begin{align}
 & \widetilde{Z}_0 =e^{-\sqrt{\frac{2}{3}}\frac{\varphi}{M_{\rm P}}} \frac{g_Z}{2 \mathcal{K}_Z} 
 \left(h_0 \widetilde{\phi}_2'- \widetilde{\phi}_2 h_0'\right)\label{eq:gaugecond1}, \\
& \widetilde{Z}_i''+a^2 \mathcal{K}_Z  \widetilde{Z}_i +ik_i\left( \widetilde{Z}_0' +a^2\,e^{-\sqrt{\frac{2}{3}}\frac{\varphi}{M_P}}\frac{g_Z}{2}
h_0  \widetilde{\phi}_2\right)+i \; \frac{4 M_{\rm P}^2}{\xi_R \Lambda^2 } \ \partial_\tau \left( F(\varphi^I)e^{\sqrt{\frac{2}{3}}\, \frac{\varphi}{M_P}} \right) \epsilon^{ijm}(k_j \widetilde{Z}_m -k_m\widetilde{Z}_j  )  =0,\label{eq:Zmom} 
\end{align}
where we have defined
\begin{align}
\mathcal{K}_Z = \frac{k^2}{a^2}+ m_Z^2,  \hspace{2cm} m_Z^2=  \frac{g_Z^2}{4}  e^{-\sqrt{\frac{2}{3}}\frac{\varphi}{M_{\rm P}}} h_0^2. \label{def:K_Z}
\end{align}
The three-components of Eq.~\eqref{eq:Zmom} can then be brought to the form
\begin{align}
&\bigg[\widetilde{\vb{Z}}'' + a^2 \mathcal{K}_Z \widetilde{\vb{Z}} \bigg]
+ i \vb{k} \bigg\{  \widetilde{Z}_0' +a^2\,e^{-\sqrt{\frac{2}{3}}\frac{\varphi}{M_P}}\frac{g_Z}{2}
h_0  \widetilde{\phi}_2 \bigg\}+  \frac{8i  M_{\rm P}^2}{\xi_R \Lambda^2} \partial_\tau \left( F(\varphi^I)e^{\sqrt{\frac{2}{3}}\, \frac{\varphi}{M_P}} \right) 
\left(\vb{k}\cross \widetilde{\vb{Z}}\right)=\vb{0}.\label{eq:Zmom1}
\end{align}
This is a system of three differential equations. A linear combination of them is obtained by
multiplying  both sides of Eq.~\eqref{eq:Zmom1} by $i\vb{k}$ and utilizing Eq.~\eqref{gauge:Zmomconf}: all terms in square bracket go to zero since $i\vb{k}\cdot\hat{\epsilon}_Z^\pm(\vb{k}) = 0$ 
and $\widetilde{Z}^L =0$. The last term vanishes due to $\vb{k}\cdot\left(\vb{k}\cross \widetilde{\vb{Z}}\right)\equiv 0$. In the end, we obtain the equation 
\begin{align}
 \widetilde{Z}_0' = - a^2\,e^{-\sqrt{\frac{2}{3}}\frac{\varphi}{M_P}}\frac{g_Z}{2}
h_0  \widetilde{\phi}_2\label{eq:gaugecond2}.
\end{align}
Together, Eqs.~\eqref{eq:gaugecond1} and \eqref{eq:gaugecond2} imply a constraint for the scalar fields.
We can now use these Eqs.~\eqref{eq:gaugecond1} and~\eqref{eq:gaugecond2} alongside the gauge condition to remove the $Z$ fields from the EoM of $X^{\phi_2}$. This makes $\widetilde{\phi}_2$ dynamical in place of the longitudinal component of the $Z$ boson. Finally, multiplying  $\hat{\epsilon}_Z^\pm(\vb{k})$ to both sides of Eq.~\eqref{eq:Zmom1}, we obtain the EoM for the transverse modes (the two remaining equations)
\begin{align}
\partial^2_\tau \widetilde{Z}^\lambda+(\omega_Z^\lambda)^2\;\widetilde{Z}^\lambda = 0, \hspace{1cm }\mbox{with}~(\lambda = \pm),
\label{eq:Z-EoM-tranverse-conf}
\end{align}
where
\begin{align}
&(\omega_Z^\lambda(\tau,k))^2=a^2 \mathcal{K}_Z+\lambda k\;\dfrac{8  M_{\rm P}^2}{\xi_R \Lambda^2} \partial_\tau\left( F(\varphi^I)e^{\sqrt{\frac{2}{3}}\, \frac{\varphi}{M_{\rm P}}} \right)
= k^2+ a^2\,m_Z^2(\tau)+\zeta^\lambda(\tau,k),\label{eq:womegaZ}
\end{align}
where, we identify 
\begin{align}
 \zeta^\lambda(\tau,k)= \lambda k\;\dfrac{8 M_{\rm P}^2}{\xi_R \Lambda^2} \partial_\tau\left( F(\varphi^I)e^{\sqrt{\frac{2}{3}}\, \frac{\varphi}{M_{\rm P}}}\right).
 \label{def:zeta}
\end{align}
We shall return to the impact of different terms in $(\omega_Z^\lambda(\tau,k))^2$ shortly.

%---------------------------
\subsubsection{$W^\pm$ bosons}
%---------------------------
An identical consideration as for the $Z$ bosons leads for the
time and space components of Eq.~\eqref{eq:Wpmeom} to
\begin{align}
&-\partial_i\left(\partial_i W^\pm_0 -W^{\pm\prime}_i\right) +   a^2\, e^{-\sqrt{\frac{2}{3}}\frac{\varphi}{M_{\rm P}}} \bigg( \frac{e^2}{4 s_W^2} h_0^2 W^\pm_0 \pm \frac{ie}{2 \sqrt{2} s_W}
\bigg[(\phi_3\pm i \phi_4) h_0' - h_0 (\phi_3'\pm i \phi_4') \bigg]   \bigg)= 0,\label{eq:Wpoem-timeconf} \\
&-\partial_\tau \left(\partial_\tau W^\pm_i -\partial_i W^\pm_0 \right)+\partial_j \left(\partial_j W^\pm_i - \partial_i W^\pm_j\right)
- a^2\, e^{-\sqrt{\frac{2}{3}}\frac{\varphi}{M_{\rm P}}} \bigg( \frac{e^2}{2 s_W^2}  h_0^2 W^\pm_i \mp  \frac{i e}{2 \sqrt{2} s_W} h_0 (\partial_i\phi_3\pm i \partial_i \phi_4 ) \bigg)\nn\\
& \hspace{4cm}+  \frac{4 M_{\rm P}^2}{\xi_R \Lambda^2}\; \partial_\tau\left( F(\varphi^I)e^{\sqrt{\frac{2}{3}}\, \frac{\varphi}{M_{\rm P}}} \right) \epsilon^{i jk}
\left(\partial_j W^\pm_k - \partial_k W^\pm_j\right) =0. \label{eq:Wpoem-spaceconf}
\end{align}
We can again go in the momentum space, where the $\widetilde{\vb{W}}^\pm$ fields can be written in terms of transverse and longitudinal components as
\begin{align}
\widetilde{\vb{W}}^\pm(\tau,\vb{k}) =  \sum_{\lambda=\pm,L} \widetilde{W}^{\pm,\lambda}(\tau,\vb{k})\ \hat{\epsilon}_W^\lambda(\vb{k}), \label{W:pol}
\end{align}
with
\begin{align}
i\vb{k}\cdot\hat{\epsilon}_W^\pm(\vb{k}) = 0, \hspace{7mm}i\vb{k}\cdot\hat{\epsilon}_W^L(\vb{k}) = |\vb{k}|=k, \hspace{7mm}i\vb{k}\times\hat{\epsilon}_W^\pm(\vb{k}) = \pm k \ \hat{\epsilon}_W^\pm(\vb{k}), \hspace{7mm} \hat{\epsilon}_W^\lambda(\vb{k})^\ast=   \hat{\epsilon}_W^\lambda(-\vb{k}).\label{def:helical-basis-W}
\end{align}
The Coulomb gauge condition gives
\begin{align}
\partial_i W^{\pm i} = \partial_i \left(\overline{g}_E^{ik}W^\pm_k\right)= \overline{g}_E^{ik} \partial_i W^\pm_k = \frac{1}{a^2} \partial_i W^\pm_i =0\label{gauge:Wpos}
\end{align}
and it translates to
\begin{align}
k_i \widetilde{W}^\pm_i =  \eta_{ij} k_i \widetilde{W}^{\pm j} =  k_j \widetilde{W}^{\pm j} =  \vb{k}\cdot\widetilde{\vb{W}}^\pm=0.\label{gauge:Wmomconf}
\end{align}

Inserting Eq.~\eqref{W:pol} in Eq.~\eqref{gauge:Wmomconf}, we obtain again $\widetilde{W}^{\pm,L} = 0$ similar to the previous section. Using the gauge condition of Eq.~\eqref{gauge:Wpos} in Eq.~\eqref{eq:Wpoem-timeconf} and going to momentum space, we then find
\begin{align}
&\widetilde{W}^\pm_0 = \mp\; e^{-\sqrt{\frac{2}{3}}\frac{\varphi}{M_{\rm P}}}\frac{ie}{2 \sqrt{2} s_W \mathcal{K}_W }
\bigg[(\phi_3\pm i \phi_4) h_0' - h_0 (\phi_3'\pm i \phi_4') \bigg],\label{eq:constr1}\\
& \widetilde{W}_i^{\pm\prime\prime}+a^2 \mathcal{K}_W  \widetilde{W}^\pm_i + ik_i\left( \widetilde{W}^{\pm\prime}_0 \pm a^2\,e^{-\sqrt{\frac{2}{3}}\frac{\varphi}{M_P}}  \frac{i e}{2 \sqrt{2} s_W} h_0 (\widetilde{\phi}_3\pm i \widetilde{\phi}_4 )\right)  \nn \\
& \hspace{2cm}+ ik_i \; \frac{4 M_{\rm P}^2}{\xi_R \Lambda^2 } \ \partial_\tau \left( F(\varphi^I)e^{\sqrt{\frac{2}{3}}\, \frac{\varphi}{M_P}} \right) \epsilon^{ijm}(k_j \widetilde{W}_m^\pm -k_m\widetilde{W}_j^\pm  )  =0,\label{eq:Wpmom}
\end{align}
with
\begin{align}
 \mathcal{K}_W = \frac{k^2}{a^2}+ m_W^2, \hspace{2cm} m_W^2= \frac{e^2}{4 s_W^2} e^{-\sqrt{\frac{2}{3}}\frac{\varphi}{M_{\rm P}}}  h_0^2.
 \label{def:KW}
\end{align}
Likewise, the three components of Eqs.~\eqref{eq:Wpmom} read as
\begin{equation} \begin{aligned}
&\bigg[\widetilde{\vb{W}}^{\pm\prime\prime} + a^2 \mathcal{K}_W \widetilde{\vb{W}}^\pm \bigg]
+ i \vb{k} \bigg\{ \widetilde{W}^{\pm \prime}_0 \pm a^2\,e^{-\sqrt{\frac{2}{3}}\frac{\varphi}{M_P}}  \frac{i e}{2 \sqrt{2} s_W} h_0 (\widetilde{\phi}_3\pm i \widetilde{\phi}_4 ) \bigg\}  \\
& \hspace{4cm}+ \frac{8i  M_{\rm P}^2}{\xi_R \Lambda^2} \partial_\tau \left( F(\varphi^I)e^{\sqrt{\frac{2}{3}}\, \frac{\varphi}{M_P}} \right)
\left(\vb{k}\cross \widetilde{\vb{W}}^\pm\right)=\vb{0}.\label{eq:Wpmomthree}
\end{aligned} \end{equation}
Applying the same procedure as for the $Z$ boson, and using Eq.~\eqref{eq:Wpmomthree}, we find
\begin{align}
&\widetilde{W}^{\pm\prime}_0 =\mp\;a^2\,e^{-\sqrt{\frac{2}{3}}\frac{\varphi}{M_P}}  \frac{i e}{2 \sqrt{2} s_W} h_0 (\widetilde{\phi}_3\pm i \widetilde{\phi}_4 ) \label{eq:constr2}. 
\end{align}
Combing Eqs.~\eqref{eq:constr1} and~\eqref{eq:constr2}, we can remove two degrees of
freedom each from $W^\pm$ making the Goldstone $\phi_3$ and $\phi_4$ dynamical, leaving a constraint on the Goldstone fields $\widetilde \phi_3$ and $\widetilde\phi_4$. Finally, from Eq.~\eqref{eq:Wpmomthree} we get the EoM for the transverse modes
\begin{align}
\partial^2_\tau\widetilde{W}^{\pm,\lambda} +(\omega_W^\lambda)^2 \; \widetilde{W}^{\pm,\lambda} = 0~\mbox{with}~(\lambda = \pm).\label{eq:Wpmeom-conf}
\end{align}
where
\begin{align}
&(\omega_W^\lambda(\tau,k))^2=a^2 \mathcal{K}_W+\lambda k\;\dfrac{8 M_{\rm P}^2}{\xi_R \Lambda^2} \partial_\tau\left( F(\varphi^I)e^{\sqrt{\frac{2}{3}}\, \frac{\varphi}{M_{\rm P}}}\right)=k^2+ a^2\, m_W^2(\tau)+\zeta^\lambda(\tau,k). \label{eq:womegaW}
\end{align}

%-------------------------------------
\subsubsection{Goldstone bosons}
\label{sec:gauge-Goldstone}
%--------------------------------------
As discussed earlier, the Goldstone bosons are dynamical in the Coulomb gauge.
The gauge choices remove the longitudinal components of the $Z$ and $W^\pm$ and constraint equations for each fields.
In the case of the $Z$ boson, these equations, namely Eq.~\eqref{eq:gaugecond1} and Eq.~\eqref{eq:gaugecond2}, can be used to make the $\phi_2$ field dynamical. This is obtained as follows:
We first rewrite the Eq.~\eqref{eqpertur:Xphi_2} in momentum space as
\begin{align}
&\mathcal{D}_\tau^2 \widetilde{X}^{\phi_2} +\left(k^2  + a^2 m_{\mathrm{eff},(\phi_2)}^2  \right) \widetilde{X}^{\phi_2}+ a \;  g_Z h_0 \left( \Upsilon \widetilde{Z}_0 - \widetilde{Z}_0'\right)= 0,
\end{align}
where we used the gauge condition $\partial_i  \widetilde{Z}_i =0$ and defined
\begin{align}
m_{\mathrm{eff},(\phi_i)}^2 = \mathcal{M}^{\phi_i}_{\ \ \phi_i} -\frac{1}{6}R_E G^{\phi_i}_{\ \ \phi_i}, \hspace{2cm} (i=2,3,4).
\label{def:meff-Goldstones}
\end{align}
We then employ Eqs.~\eqref{eq:gaugecond1} and~\eqref{eq:gaugecond2} to get the EoM of the $\widetilde{X}^{\phi_2}$ as
\begin{align}
\mathcal{D}_\tau^2 \widetilde{X}^{\phi_2} +\mathcal{E}_{(\phi_2)}(\tau,k) 
\mathcal{D}_\tau\widetilde{X}^{\phi_2}+\omega^2_{(\phi_2)}(\tau,k) \widetilde{X}^{\phi_2}= 0, \label{eom:phi_2}
\end{align}
where
\begin{subequations} \begin{eqnarray}
\mathcal{E}_{(\phi_2)}(\tau,k)  &=& \ 2 \, \frac{m_Z^2}{\mathcal{K}_Z}\;\Upsilon, \\
\omega^2_{(\phi_2)}(\tau,k) &=& \ k^2  + a^2\left( m_{\mathrm{eff},(\phi_2)}^2 +  m_Z^2 \right) +\mathcal{E}_{(\phi_2)}\Upsilon.\label{eom:phi_2-omega2}
\end{eqnarray} \end{subequations}
Similarly, Eqs.~\eqref{eqpertur:Xphi_3} and \eqref{eqpertur:Xphi_4} lead to
\begin{subequations} \begin{eqnarray}
\mathcal{D}_\tau^2 \widetilde{X}^{\phi_3} +\mathcal{E}_{(\phi_3)}(\tau,k)
\mathcal{D}_\tau\widetilde{X}^{\phi_3}+\omega^2_{(\phi_3)}(\tau,k) \widetilde{X}^{\phi_3}&=& 0, \label{eom:phi_3}\\
\mathcal{D}_\tau^2 \widetilde{X}^{\phi_4} +\mathcal{E}_{(\phi_4)}(\tau,k)
\mathcal{D}_\tau\widetilde{X}^{\phi_4}+\omega^2_{(\phi_4)}(\tau,k) \widetilde{X}^{\phi_4}&=& 0, \label{eom:phi_4}
\end{eqnarray}  \label{eom:phi_3+4}  \end{subequations}
with
\begin{subequations} \begin{eqnarray}
\mathcal{E}_{(\phi_3)}(\tau,k)  &=& \mathcal{E}_{(\phi_4)}(\tau,k) = 2 \, \frac{m_W^2}{\mathcal{K}_W}  \; \Upsilon ,\\
\omega^2_{(\phi_3)}(\tau,k) &=& \ k^2  +a^2\left( m_{\mathrm{eff},(\phi_3)}^2 +  m_W^2 \right)+\mathcal{E}_{(\phi_3)} \Upsilon  \label{eom:phi_3-omega2},\\
\omega^2_{(\phi_4)}(\tau,k)& =& \ k^2  + a^2\left( m_{\mathrm{eff},(\phi_4)}^2 +  m_W^2 \right)+ \mathcal{E}_{(\phi_4)}\Upsilon\label{eom:phi_4-omega2}.
\end{eqnarray}\end{subequations}
It is clear that all gauge bosons are decoupled from Eqs.~\eqref{eom:phi_2} and \eqref{eom:phi_3+4}. Furthermore, as only $\phi$ and $h$ acquire background field values, the EoMs of the Goldstone bosons are decoupled not only from the EoMs of $\widetilde{X}^{\phi}$ and $\widetilde{X}^{h}$, they are also decoupled from each other.

%%%%%%%%%%%%%%%%%%%%%%%%%%%%%%%%%%%%%%%%%%%%%%%%%%%%%%%%%%%%%%%%%%%%%%%%%%%%%%%%%%%%%%
\section{Inflaton and Higgs quanta production}
\label{sec:Inf+Higgs-fluctuations}
%%%%%%%%%%%%%%%%%%%%%%%%%%%%%%%%%%%%%%%%%%%%%%%%%%%%%%%%%%%%%%%%%%%%%%%%%%%%%%%%%%%%%%
We now proceed with the quantization of the $\phi$ and $h$ fields, and the production of the respective particles. For this, we follow closely Refs.~\cite{Amin:2014eta,DeCross:2015uza}, which consider non-trivial field space manifolds relevant to our analysis. The $G_{IJ}$ matrix is diagonal and only depends on $\phi^1 = \phi$, whereas the $3\times 3$ lower block of $\mathcal{M}^I_J$ matrix involving the Goldstones is diagonal. Therefore, we can reduce the upper $2\times2$ block of $\mathcal{M}^I_J$ in our scenario to a two field model with $\phi$ and $h$. Quantization of the three Goldstone bosons is discussed separately in the following section alongside the massive gauge bosons. One may still have nonzero $\mathcal{M}^\phi_{~~h}$ and $\mathcal{M}_{~~\phi}^h$ if $h_0$ and $\xi_H$ are not vanishingly small. The second order action involving the inflaton and Higgs fluctuations $Q^I$ (with $I=\{1,2\}$) can be derived as in~\cite{Gong:2011uw,DeCross:2015uza}
\begin{align}
S^{(2)}_{(\phi h)} =& \int d^3x \ dt \ a^3
\bigg[-\frac{1}{2}\overline{g}^{\mu\nu}_{E} G_{IJ} \mathcal{D}_\mu Q^I\mathcal{D}_\nu Q^J-\frac{1}{2}\mathcal{M}_{IJ} Q^I Q^J
\bigg],\label{action:quadfluc}
\end{align}
where $G_{IJ}$, $\mathcal{M}_{IJ}$ are evaluated at background order and $\overline{g}^{\mu\nu}_{E}\equiv(-1,a^2(t), a^2(t), a^2(t))$ is the unperturbed spatially flat FLRW metric.
The latter action can be written in conformal time and with the rescaled variables $X^I(x^\mu)$ as
\begin{align}
S^{(2)}_{(\phi h)} =& \int d^3x \ d\tau \Biggl[-\frac{1}{2} \eta^{\mu\nu} G_{IJ} (\mathcal{D}_\mu X^I)(\mathcal{D}_\nu X^J)-\frac{1}{2}\mathscr{M}_{IJ} X^I X^J\Biggr], 
\label{action:S_X2-inf}
\end{align}
with
\begin{align}
\mathscr{M}_{IJ} =a^2\bigg(\mathcal{M}_{IJ} -\frac{1}{6} G_{IJ}R_E\bigg),~~\mbox{with}~~R_E = \frac{6a''}{a^3}.
\label{equ:mathcal-MIJ-def}
\end{align}
The energy momentum tensor for the field fluctuation is given for the linearized theory as
\begin{align}
T_{\mu\nu}^{(\phi h)} =\ & G_{IJ} (\mathcal{D}_\mu X^I)(\mathcal{D}_\nu X^J)
+ \eta_{\mu\nu} \Biggl[-\frac{1}{2} \eta^{\alpha\beta} G_{IJ} (\mathcal{D}_\alpha X^I)(\mathcal{D}_\beta X^J)-\frac{1}{2}\mathscr{M}_{IJ} X^I X^J
 \Biggr],\label{eq:infHiggflucstress}
\end{align}
with $T_{00}^{(\phi h)}$ denoting the associated energy density.

Transforming to momentum space, one can recast Eq.~(\ref{action:S_X2-inf}) into the form
\begin{align}
S_{(\phi h)} =&  \int d\tau \, \mathcal{L}_{(\phi h)} = \int \, d\tau \,\frac{d^3k}{(2\pi)^3}\, \ \bigg[\frac{1}{2} \left|\partial_\tau  \widetilde{X}^I\right|^2
- \frac{1}{2}\omega_{(I)}^2(\tau,k) \left|\widetilde{X}^I\right|^2 \bigg], \label{action:quanphih}
\end{align}
with
\begin{align}
\omega_{(I)}^2(\tau,k) = \left(k^2 + a^2 m_{\mathrm{eff},(I)}^2(\tau)  \right).
\label{def:omega2}
\end{align}
The effective masses are given by
\begin{align}
&m_{\mathrm{eff},(\phi)}^2(\tau)= \mathcal{M}_\phi^\phi - \frac{1}{6}R_E = \frac{1}{a^2}\; \mathscr{M}^\phi_{~~\phi} , \hspace{1.cm} m_{\mathrm{eff},(h)}^2(\tau)= \mathcal{M}_h^h - \frac{1}{6}R_E=\frac{1 }{a^2}\; \mathscr{M}^h_{~~h}.
\label{def:meff2}
\end{align}
We identify
\begin{subequations} \begin{eqnarray}
m_{1,(I)}^2 &=& G^{(I)J} (\mathcal{D}_{(I)}\mathcal{D}_J V_E),  \label{def:effective-masses-decomposition-1} \\
m_{2,(I)}^2 &=& - \mathcal{R}^{(I)}_{\ \ JK(I)} \dot{\varphi}^J \dot{\varphi}^K, \label{def:effective-masses-decomposition-2} \\
m_{3,(I)}^2 &=& - \frac{1}{M_{\rm P}^2 a^3} \mathcal{D}_t \left(\frac{a^3}{H}\dot{\varphi}^{(I)} \dot{\varphi}_{(I)}\right), \label{def:effective-masses-decomposition-3} \\
m_{4,(I)}^2&=&-\frac{R_E}{6},  \label{def:effective-masses-decomposition-4} 
\end{eqnarray}  \label{def:effective-masses-decomposition}   \end{subequations}
without summing over $(I)$, such that 
\begin{align}
m_{\mathrm{eff},(I)}^2= \sum_k m_{k,(I)}^2 .
\end{align}
One advantage of writing the action in Eq.~\eqref{action:quanphih} (and consequently also Eqs.~\eqref{eqpertur:Xphi} and \eqref{eqpertur:Xh}) in conformal time as opposed to cosmic time is the absence of terms linear in $\mathcal{D}_\tau \widetilde{X}^I$. Hence, the canonical momentum 
in momentum space is found as
\begin{align}
&\hat{\widetilde{\pi}}^{I}(\tau,\vb{k})=  \partial_\tau \hat{\widetilde{X}}^{I}(\tau,\vb{k}), \\
&~\mbox{with}~\bigg[ \hat{\widetilde{X}}^{I}(\tau,\vb{k}),\hat{\widetilde{\pi}}^{J}(\tau,\vb{q})\bigg]= i (2\pi)^3\delta^{IJ}\delta^{(3)}(\vb{k}+\vb{q}) \label{quant:phih},
\end{align}
where we have elevated the classical field fluctuations $\widetilde{X}^{I}$ to their respective quantized $\hat{\widetilde{X}}^{I}$ versions.

The quantized fluctuations $\hat{\widetilde{X}}^{I}(\tau,\vb{k})$ can be decomposed in momentum space as
\begin{align}
\hat{\widetilde{X}}^I(\tau, \vb{k})=  \sum_m \left[ u_m^I(\tau,k) \hat{a}_m(\vb{k}) + u_m^{I*}(\tau,k) \hat{a}^\dagger_m(-\vb{k})   \right],\label{eq:Xmuquantization}
\end{align}
where $m \in \{1,2\}$ for $\{\phi,h\}$. $u_m^I(\tau,k)$ corresponds to the associated mode functions of the creation and annihilation operators $\hat{a}_m(\vb{k})$ and $\hat{a}^\dagger_m(-\vb{k})$. These are defined as
\begin{align}
 \hat{a}_m(\vb{k})  \ket{0} = 0,\hspace{3 cm} \bra{0} \hat{a}^\dagger_m(\vb{k}) = 0,
\end{align}
and obey the usual commutator relationships
\begin{align}
\left[\hat{a}_m(\vb{k}), \hat{a}_n(\vb{q})\right] = \left[\hat{a}^\dagger_m(\vb{k}), \hat{a}^\dagger_n(\vb{q})\right] = 0, \hspace{2 cm} 
\left[\hat{a}_m(\vb{k}), \hat{a}^\dagger_n(\vb{q})\right] = (2\pi)^3 \delta_{mn} \delta^{(3)}(\vb{k}-\vb{q}).
\end{align}
Note that, we have $N=2$ second order differential equations for the $\widetilde{X}^{\phi}$ and $\widetilde{X}^{h}$, the parametrization  in Eq.~\eqref{eq:Xmuquantization} leads to $N^2$ complex mode functions $u_m^I(\tau,k)$, and hence $2 N^2$ real-valued scalar functions. However, these two fluctuations are coupled through the EoMs via $\mathcal{M}^I_{\ \ J}$. This implies $2N(N-1)$ constraints, leading to $2N^2-2N(N-1)=2N =4$ independent solutions. 
The mode functions can be parametrized as
\begin{align}
u_m^I(\tau,k) = t_{(m,I)}(\tau,k) e_m^I(\tau),
\end{align}
where the $t_{(m,I)}(\tau,k)$ are complex scalar functions and $e_m^I(\tau)$ are vielbeins of the field-space metric. Note, $x_{(m,I)}$ is not a field space vector, but the index $I$ in brackets denotes individual species and is not summed over. The vielbeins satisfy the following conditions
\begin{align}
\delta^{mn} e_m^I(\tau) e_n^J(\tau) = G^{IJ}(\tau), \label{eq:vielbein-conditon}
\end{align}
and are real functions. 
For any arbitrary vector $A^I$ in field space, we have,
\begin{equation} \begin{aligned}
A^m = e_I^m A^I,\hspace{2.5 cm}  A^I = e^I_m A^m,\\
e^m_I e_n^I = \delta^m_n, \hspace{2.5 cm} e^m_I e_m^J = \delta^I_J.
\end{aligned} \end{equation}
The covariant derivative of the vielbeins in terms of the spin connection reads 
\begin{align}
\mathcal{D}_I e^m_J = - \omega^{mn}_I e_{n J},
\end{align}
with $\omega^{mn}_I = -\omega^{nm}_I$ antisymmetric in the internal indices. Further, since the  $\omega^{mn}_I$ is antisymmetric, the covariant derivative with respect to conformal time vanishes
\begin{align}
\mathcal{D}_\tau  e^m_J = 0,
\end{align}
for all $m$ and $J$.

We take  $t_{(1,\phi)}(\tau,k)=v_{1k}(\tau)$, $t_{(2,\phi)}(\tau,k)=v_{2k}(\tau)$, $t_{(1,h)}(\tau,k)=y_{1k}(\tau)$
and $t_{(2,h)}(\tau,k)=y_{2k}(\tau)$ for the quantization of Eq.~\eqref{eq:Xmuquantization}. Hence, we get,
\begin{subequations} \begin{eqnarray}
&\hat{\widetilde{X}}^\phi =   \left[ \left(v_{1k}(\tau) e_1^\phi(\tau) \hat{a}_1(\vb{k}) + v_{2k}(\tau) e_2^\phi(\tau) \hat{a}_2(\vb{k})\right)
+ \left(v^*_{1k}(\tau) e_1^\phi(\tau) \hat{a}^\dagger_1(-\vb{k}) + v_{2k}^*(\tau) e_2^\phi(\tau) \hat{a}^\dagger_2(-\vb{k})\right)   \right],\label{eq:phiquant}\\
&\hat{\widetilde{X}}^h =  \left[ \left(y_{1k}(\tau) e_1^h(\tau) \hat{a}_1(\vb{k}) + y_{2k}(\tau) e_2^h(\tau) \hat{a}_2(\vb{k})\right)
+  \left(y^*_{1k}(\tau) e_1^h(\tau) \hat{a}^\dagger_1(-\vb{k}) + y^*_{2k}(\tau) e_2^h(\tau) \hat{a}^\dagger_2(-\vb{k})\right)  \right].\label{eq:hquant}
\end{eqnarray}  \end{subequations}
Going into momentum space and utilizing the quantization above, from the EoMs in Eq.~\eqref{eqpertur:Xphi} and \eqref{eqpertur:Xh} we get
\begin{subequations} \begin{eqnarray}
&v_{1k}'' +\omega^2_{(\phi)} \, v_{1k}\, e_1^\phi = -a^2 \mathcal{M}^\phi_{~~h} \ y_{1k} \,e_1^h,\label{eq:vk}\\
&v_{2k}'' +\omega^2_{(\phi)} \, v_{2k}\, e_2^\phi = -a^2 \mathcal{M}^\phi_{~~h} \ y_{2k} \,e_2^h,\label{eq:wk}\\
&y_{1k}'' +\omega^2_{(h)}\,y_{1k} \,e_1^h = -a^2 \mathcal{M}_{~~\phi}^h \ v_{1k} \,e_1^\phi,\label{eq:yk}\\
&y_{2k}'' +\omega^2_{(h)}\, y_{2k}\, e_2^h = -a^2 \mathcal{M}_{~~\phi}^h \ v_{2k} \,e_2^\phi,\label{eq:zk}
\end{eqnarray} \end{subequations}
with $\omega^2_{(I)} $ given by Eq.~\eqref{def:omega2}.

We are now equipped with necessary tools to derive the energy density.
The comoving vacuum averaged energy  density is defined as
\begin{align}
\rho_{(\phi h)}=\int d^3x \left\langle T_{00}^{(\phi h)} \right\rangle=  \int \frac{d^3k}{(2\pi)^3} \left\langle \rho_{k,(\phi h)}\right\rangle
= \int \frac{d^3k}{(2\pi)^3} \ \rho_{k,(\phi h)}^{\rm{vev}}, \label{eq:energydensity}
\end{align}
where we recall we are only considering $\phi^I\in\{\phi,h\}$ in this section.
Taking the $00$-component of $T_{00}^{(\phi h)}$ and expressing different fluctuation fields in momentum space via 
\begin{equation*}
F(t,\vec x)= \int \frac{d^3k}{\left(2\pi\right)^3} \widetilde{f}(t,\vb{k}) e^{-i \vb{k}\cdot\vb{x}},
\end{equation*}
we can perform one momentum integration via the Dirac delta function that appears after performing the position space integral. In $\rho_{k,(\phi h)}$, which is a quadratic function of fluctuations, one fluctuation is $\widetilde{X}^J(\tau, -\vb{k})$. The energy density spectra is therefore read
\begin{align}
\rho_{k,(\phi h)}=& \frac{1}{2} G_{IJ} \left(\mathcal{D}_\tau \widetilde{X}^I(\tau, \vb{k})\right)\left(\mathcal{D}_\tau \widetilde{X}^J(\tau, -\vb{k})\right) + \frac{1}{2}\left(k^2 G_{IJ} + \mathscr{M}_{IJ}\right) \widetilde{X}^I(\tau, \vb{k}) \widetilde{X}^J(\tau, -\vb{k}). \label{eq:energydensityfluc}
\end{align}
Inserting Eq.~\eqref{eq:Xmuquantization} in Eq.~\eqref{eq:energydensityfluc} we get energy density per mode $k$ of the quantized fluctuations as
\begin{equation}
\begin{split}
\rho_{k,(\phi h)}^{\rm{vev}}&= \left\langle \rho_{k,(\phi h)}\right\rangle= \frac{1}{2} \sum_{m,n}\bigg[\delta^{mn} \bigg( G_{IJ}t'_{(m,I)} {t^*}'_{(n,J)}  + \left(k^2 G_{IJ} + \mathscr{M}_{IJ}\right) t_{(m,I)} t_{(n,J)}^* \bigg) e_m^I e_n^J\bigg]\\
&= \rho_k^{(\phi)} + \rho_k^{(h)} +\rho_k^{\mathrm{int}},
\label{eq:energmodek}
\end{split}
\end{equation}
with
\begin{subequations} 
\begin{eqnarray}
\rho_k^{(\phi)} &=& \dfrac{1}{2} G_{\phi\phi}\left[(|v_{1k}'|^2 + \omega^2_{(\phi)} \, |v_{1k}|^2)e_1^\phi e_1^\phi + (|v_{2k}'|^2 +\omega^2_{(\phi)} \, |v_{2k}|^2)e_2^\phi e_2^\phi \right],\label{enphi_conf} \\
 \rho_k^{(h)} &=& \dfrac{1}{2} G_{hh}\left[(|y_{1k}'|^2 +  \omega^2_{(h)} \, |y_{1k}|^2)e_1^h e_1^h + (|y_{2k}'|^2 +\omega^2_{(h)} \, |y_{2k}|^2)e_2^h e_2^h \right],\label{enh_conf}\\
  \rho_k^{\mathrm{int}} &=& \dfrac{1}{2}( \mathscr{M}_{\phi h}+ \mathscr{M}_{h \phi} ) \left[ v_{1k} y_{1k}^* e_1^h e_1^\phi 
+ v_{2k} y_{2k}^* e_2^h e_2^\phi  \right],\label{eq:enephih}
\end{eqnarray}  
\end{subequations}
where we used Eq.~\eqref{def:omega2} and Eq.~\eqref{def:meff2}. 

As we already discussed, in this paper we primarily focus on the $R^2$-like regime and assume that the background value of $h_0$ is much smaller than $\varphi$, which essentially reduces our scenario to a single field attractor-like scenario. 
We consider three benchmark points (BP)~\cite{Cado:2023zbm} for our analysis, summarized in Tab.~\ref{parmeterchoices}. The BP$a$ is deep in the $R^2$-like regime with $\xi_H$ being negligibly small. We can therefore expect this scenario to behave practically like Starobinsky inflation. BP$b$ and BP$c$ parametrize mixed $R^2$-Higgs scenarios, with $\xi_H=1$ for BP$b$ and $\xi_H=10$ for BP$c$. For all three BPs, the corresponding scalar amplitude, spectral index and tensor-to-scalar ratio are in agreement with the Planck 2018 data~\cite{Planck:2018jri} within the 95\% confidence level (CL) interval~\cite{Cado:2023zbm}.
%%%%%%%%%%%%%%%%%%%
\begin{table}[!t]
\begin{tabular}{|c |c| c| c| c | c | c| c | c |c| c| c| c}
    \hline
	BP                  & $\xi_R$             &  $\xi_H$   &  $\varphi(t_{\text{in}})$ [$M_{\rm P}$]  & $h_0(t_{\text{in}})$  [$M_{\rm P}$] \\
   \hline
        $a$                   & $2.35\times 10^9$   &  $10^{-3}$ &  5.5                                  & $2\times10^{-4}$ \\
        $b$                   & $2.55\times 10^9$   &  $1$       &  5.5                                  & $8.94\times10^{-4}$ \\
        $c$                   & $2.2\times 10^9$   &   $10$        &  5.4                                  & $5.00\times10^{-3}$ \\
	\hline
	\end{tabular}
	\caption{Benchmark points chosen for our analysis. Scales are given in units of the Planck mass~$M_{\rm P}$. See text for details.}
	\label{parmeterchoices}
\end{table}
%%%%%%%%%%%%%%%%%%%
We have checked that the off-diagonal elements $\mathcal{M}^\phi_{~~h} \sim 0$ and $\mathcal{M}_{~~\phi}^h \sim 0$ and, hence, $\mathcal{M}^I_{~~J}$ is essentially diagonal for all three BPs. Consequently, the vielbeins also are diagonal, $e_2^\phi\sim 0$, $e_1^h \sim 0$ and $\widetilde{X}^\phi$ and $\widetilde{X}^h$ depend only on the scalar mode functions $v_{1k}(\tau)$ and $y_{2k}(\tau)$, respectively. Therefore, Eq.~\eqref{eq:vk} and Eq.~\eqref{eq:zk} essentially satisfy source-free EoMs while Eq.~\eqref{eq:wk} and Eq.~\eqref{eq:yk} vanish. We are left with
\begin{subequations} \begin{eqnarray}
&v_{1k}'' + \omega^2_{(\phi)}\,v_{1k} \simeq 0,\label{eq:inflatonfluc}\\
&y_{2k}'' + \omega^2_{(h)}\,y_{2k} \simeq 0,\label{eq:higgsfluc}
\end{eqnarray}  \label{eq:h+phi-fluc} \end{subequations}
and the energy densities for the inflaton and Higgs fluctuations per mode read
\begin{subequations} \begin{eqnarray}
\rho_k^{(\phi)} &=& \dfrac{1}{2} G_{\phi\phi}\left(|v'_{1k}|^2 +\omega^2_{(\phi)}   |v_{1k}|^2\right) e_1^\phi e_1^\phi = \dfrac{1}{2}\left(|v'_{1k}|^2 +\omega^2_{(\phi)}   |v_{1k}|^2\right) , \label{enphi_conf-simp}\\
 \rho_k^{(h)} &=& \dfrac{1}{2} G_{hh}\left(|y'_{2k}|^2 +\omega^2_{(h)}  |y_{2k}|^2 \right)e_2^h e_2^h = \dfrac{1}{2}\left(|y'_{2k}|^2 +\omega^2_{(h)}  |y_{2k}|^2 \right), \label{enh_conf-simp}\\
 \rho_k^{\mathrm{int}}& =& \mathcal{O}(h^2)\sim 0.  \label{en-phi+h_conf-simp}
\end{eqnarray}  \end{subequations}
Note that equation of motion of the mode functions are decoupled for all three BPs.
It is worth noting that, in the single field like regime $v_{1k}$ corresponds to the mode function for the adiabatic mode while $y_{2k}$ corresponds to
isocurvature mode~\cite{Cado:2023zbm,DeCross:2016cbs}.

%%%%%%%%%%%%%%%%%%%
\begin{figure}[h]
\centering
\hspace{1.2cm}\includegraphics[height= 3.6cm]{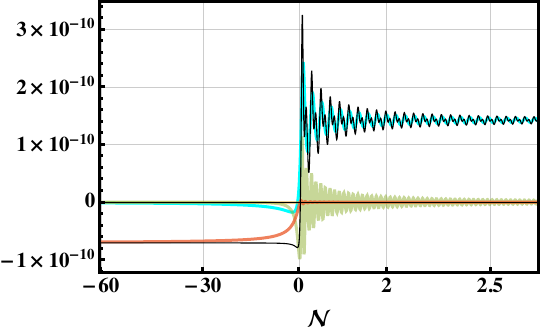}\hspace{2mm}
\includegraphics[height= 3.6cm]{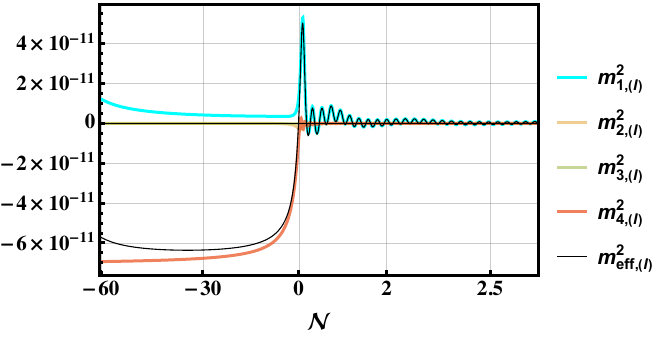}
\includegraphics[height= 3.6cm]{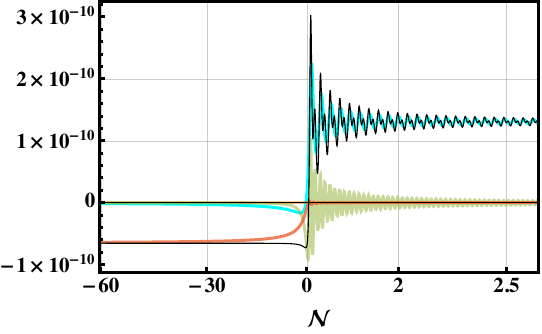}\hspace{2mm}
\includegraphics[height= 3.6cm]{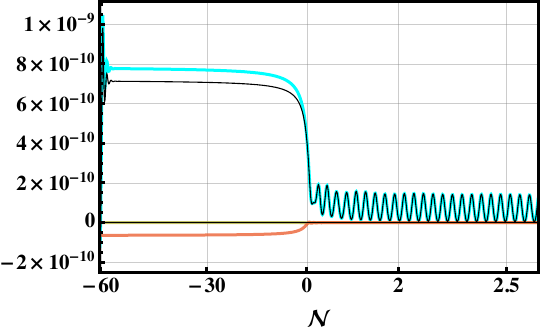} \\
\includegraphics[height= 3.6cm]{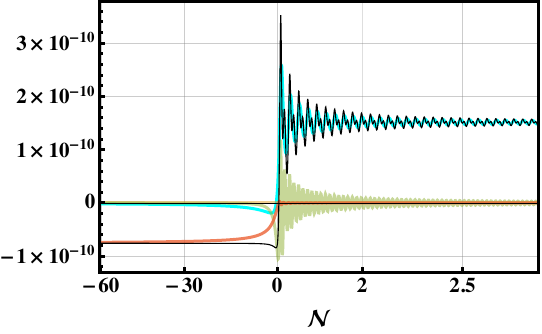} \hspace{2mm}
\includegraphics[height= 3.6cm]{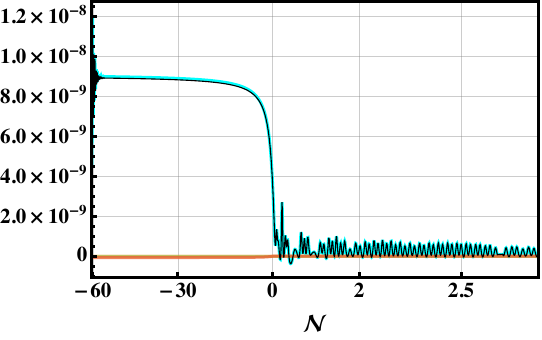}
\caption{The effective masses $m_{\mathrm{eff},(I)}^2$ (black) for $\phi$ (left) and $h$ (right) as in Eq.~\eqref{def:meff2} and their respective contributions \eqref{def:effective-masses-decomposition} for BP$a$ (upper row), BP$b$ (middle row) and BP$c$ (lower row). See text for details.}
\label{meff_plot}
\end{figure}
%%%%%%%%%%%%%%%%%%%

 Let us take a closer look at the different contributions to $m_{\mathrm{eff,(\phi)}}^2$ and $m_{\mathrm{eff},(h)}^2$ as displayed in Fig.~\ref{meff_plot} for all three BPs. The dominant contribution to $m_{\mathrm{eff,(\phi)}}^2$ for all three BPs before the end of inflation arises from $m_{4,(I)}^2$, which turns
 $m_{\mathrm{eff,(\phi)}}^2$ large and negative. After inflation, $m_{1\mathrm{,(\phi)}}^2$ provides the largest but positive contributions to $m_{\mathrm{eff,(\phi)}}^2$ 
for all three BPs. The same behavior is observed for the Higgs field for BP$a$, the dominant contribution to $m_{\mathrm{eff,}(h)}^2$  before the end of the inflation stems again from $m_{4,(h)}^2$. However, for BP$b$ and BP$c$, $m_{1\mathrm{,(h)}}^2$  overpowers $m_{4\mathrm{,(h)}}^2$. After the end of inflation, 
$m_{1\mathrm{,(h)}}^2$ constitutes the largest contribution to $m_{\mathrm{eff,}(h)}^2$ for all three BPs, which oscillates around the minimum. This oscillation, makes $m_{\mathrm{eff,}(h)}^2$ 
negative periodically with largest amplitude for BP$c$, due to the comparably larger value of $\xi_H$ for BP$c$. We note that if $m_{\mathrm{eff},(I)}^2 < 0$, modes with $k^2/a^2 < \left|m_{\mathrm{eff},(I)}^2 \right|$ will experience tachyonic instability, which may lead 
to exponential growth in energy density. In the case of a tachyonic regime, we shall use the definition
\begin{align}
\omega_{(I)}^2(\tau,k) = \left(k^2 + \left|a^2 m_{\mathrm{eff},(I)}^2(\tau) \right|  \right),
\label{def:omega2-tachyonic}
\end{align}
instead of Eq.~\eqref{def:omega2} when computing the energy density~\cite{Felder:2001kt}. This is because the standard definition of the occupation number, which was extensively used in modeling preheating, is only valid for $m_{\mathrm{eff},(I)}^2 \geqslant 0$. We shall return to the impact of $m_{\mathrm{eff},(I)}^2$ on preheating shortly.

The energy densities in Eqs.~\eqref{eq:energy_den-phi+h} are not vacuum-subtracted. To identify the latter, we first define the Bunch-Davies (BD) vacuum as
\begin{subequations} \begin{eqnarray}
 v_{1k} = c^\phi_1 \left(1 -\frac{i}{k \tau}\right)e^{- i k \tau} + c^\phi_2 \left(1 +\frac{i}{k \tau}\right) e^{ i k \tau},\label{eq:vacchoiphi}\\
 y_{2k} = c^h_1 \left(1 -\frac{i}{k \tau}\right)e^{- i k \tau}  + c^h_2 \left(1 +\frac{i}{k \tau}\right) e^{ i k \tau}\label{eq:vacchoih}.
\end{eqnarray} \label{eq:vacchoi-h+phi}  \end{subequations}
The normalization $v_{1k} v^{*\prime}_{1k} - v^\prime_{1k} v^*_{1k} = i$ and $y_{2k} y^{*\prime}_{2k} - y^\prime_{2k} y^*_{2k} = i$ yields constraints
\begin{subequations} 
\label{eq:const1+2} 
%\begin{equation}
\begin{align}
|c^\phi_1|^2 - |c^\phi_2|^2 &= \frac{1}{2k},\label{eq:const1}\\
|c^h_1|^2 - |c^h_2|^2 &= \frac{1}{2k}\label{eq:const2}.
\end{align}
%\end{equation}
\end{subequations}
Here, variations in the mode functions $v_{1k}$ and $y_{2k}$ could be accompanied by respective annihilation operators
such that $\widetilde{X}^\phi$ and $\widetilde{X}^h$ remains unchanged. Each such solution corresponds to a different vacuum, however,
we may require that the vacuum state $\ket{0}$ is the minimum energy state (ground state) of the Hamiltonian.

The Hamiltonian for the inflaton and Higgs fluctuations is written as
\begin{align}
 \mathcal{\hat{H}}&=  \int \frac{d^3 k}{(2\pi)^3} \left\langle T_{00}^{(\phi h)} \right\rangle \nn\\
%                   &= \int \frac{d^3 k}{(2\pi)^3} \left[ G_{IJ}  (\partial_\tau\hat{\widetilde{X}}^{I}  )(\partial_\tau\hat{\widetilde{X}}^{J}) +\frac{1}{2} \eta^{\mu\nu} G_{IJ} (\mathcal{D}_\mu \hat{\widetilde{X}}^{I})(\mathcal{D}_\nu \hat{\widetilde{X}}^{J})+\frac{1}{2}\mathscr{M}_{IJ} \hat{\widetilde{X}}^{I} \hat{\widetilde{X}}^{J}\right],\nn \\
                  &= \frac{1}{2}\int \frac{d^3 k}{(2\pi)^3}  \left[\left(|v'_{1k}|^2 +\omega^2_{(\phi)} |v_{1k}|^2\right)    \left(\hat{a}^\dagger_1(\vb{k})\hat{a}_1(\vb{k})+\delta^3(\vb{0})\right) + \left(
                  |y'_{2k}|^2 + \omega^2_{(h)}  |y_{2k}|^2 \right) \left(\hat{a}^\dagger_2(\vb{k})\hat{a}_2(\vb{k})+\delta^3(\vb{0})\right) \right].
\end{align}
The vacuum expectation value of the Hamiltonian is
\begin{align}
\left\langle\mathcal{\hat{H}}\right\rangle = \int \frac{d^3 k}{(2\pi)^3}  \left[\rho_k^{(\phi)} \delta^3(\vb{0}) +\rho_k^{(h)} \delta^3(\vb{0}) \right],
\end{align}
where $\delta^3(\vb{0})$ is divergent and arises due to integrating an infinite volume as usual. At sufficiently early times i.e. $\tau \to - \infty$, the vacuum choice of
Eqs.~\eqref{eq:vacchoi-h+phi} leads to
\begin{subequations} \begin{eqnarray}
 v_{1k} = c^\phi_1 e^{- i k \tau} + c^\phi_2 e^{ i k \tau},\label{eq:vacphi}\\
 y_{2k} = c^h_1 e^{- i k \tau} + c^h_2 e^{ i k \tau}\label{eq:vach},
\end{eqnarray}  \end{subequations}
where along with the constraints Eqs.~\eqref{eq:const1+2}, $\left\langle\mathcal{\hat{H}}\right\rangle$ minimized if
\begin{align}
 c^\phi_1 =\frac{1}{\sqrt{2k}},\hspace{1.5cm} c^\phi_2 =0,\hspace{1.5cm} c^h_1 =\frac{1}{\sqrt{2k}},\hspace{1.5cm} c^h_2 =0.
\end{align}
Therefore, the desired vacuum solutions (i.e.~the so-called BD vacuum) at early times become
\begin{align}
v_{1k}^{\rm BD} = \frac{1}{\sqrt{2k}} e^{-i k \tau},\hspace{2cm}  y_{2k}^{\rm BD} = \frac{1}{\sqrt{2k}} e^{-i k \tau}.
\end{align}
The physical BD-vacuum energy is identified as
\begin{subequations} \begin{align}
\rho_{(\phi)}^{\rm BD} &=  \frac{1}{2a^4}\int \frac{d^3 k}{(2\pi)^3} \left(|{v^{\rm BD}_{1k}}'|^2 +\omega^2_{(\phi)} |v^{\rm BD}_{1k}|^2\right),\\
\rho_{(h)}^{\rm BD} &=  \frac{1}{2a^4}\int \frac{d^3 k}{(2\pi)^3} \left(|{y^{\rm BD}_{2k}}'|^2 +\omega^2_{(h)} |y^{\rm BD}_{2k}|^2\right).
\end{align}  \end{subequations}
At sufficiently early times and for large modes we have ${k^2}/{a^2} \gg |m^2_{\mathrm{eff},(I)}(t)|$. Hence $\omega^2_{(I)} \simeq k^2 $, and the vacuum energy becomes for both fields
\begin{align} \rho_{(I)}^{\rm BD}  &=\int dk \; \frac{k^2}{2 \pi^2 a^4} \rho_{k,(I)}^{\rm BD} = \frac{1}{a^4}\int dk \;  \frac{k^3}{4\pi^2}  . \label{def:rho-BD_phi+h}
\end{align}
The energy density of the inflaton and Higgs quanta is then obtained by removing the BD vacuum from the classical solution as
\begin{subequations} \label{eq:energy_den-quantum} 
 \begin{eqnarray}
\rho^q_{(\phi)} &= \rho_{(\phi)}- \rho_{(\phi)}^{\rm BD},\label{eq:quantumphi}  \\
 \rho^q_{(h)} &= \rho_{(h)}-\rho_{(h)}^{\rm BD}.\label{eq:quantumh}
\end{eqnarray}  \end{subequations}
It is, however, computationally less challenging to solve Eqs.~\eqref{eq:h+phi-fluc}
in cosmic time. Therefore, we rewrite them in cosmic time as
\begin{subequations} 
 \label{eq:h+phi-flucT} 
\begin{align}
&\ddot{v}_{1k} +  H \dot{v}_{1k} + \left(\dfrac{k^2}{a^2}+ m^2_{\mathrm{eff},(\phi)}(t) \right) v_{1k} \simeq 0,\label{eq:inflatonflucT}\\
&\ddot{y}_{2k} +  H \dot{y}_{2k} + \left(\dfrac{k^2}{a^2}+ m^2_{\mathrm{eff},(h)}(t) \right) y_{2k} \simeq 0 \label{eq:higgsflucT}.
\end{align} 
\end{subequations}
Utilizing, Eq.~\eqref{eq:energydensity}, and Eqs.~\eqref{en-phi+h_conf-simp}, we can find the corresponding (physical) energy densities
for $\phi$ and $h$ fluctuations in cosmic time
\begin{subequations} 
\label{eq:energy_den-phi+h}  
\begin{align}
\rho_{(\phi)}= \dfrac{\left\langle \rho_c^{\phi}(x^\mu) \right\rangle}{a^4}  %= \frac{1}{a^4}\int \frac{d^3k}{(2\pi)^3} \rho_k^{\mathrm{fluc},\rm{vev}}
&= \dfrac{1}{a^2} \displaystyle\int \dfrac{k^2}{4\pi^2} dk \left[|\dot{v}_{1k}|^2 + \left(\dfrac{k^2}{a^2}+ \left| m^2_{\mathrm{eff},(\phi)}(t) \right| \right) |v_{1k}|^2\right], \label{eq:energy_denphi}\\
\rho_{(h)}= \dfrac{\left\langle \rho_c^{h}(x^\mu) \right\rangle}{a^4}  %= \frac{1}{a^4}\int \frac{d^3k}{(2\pi)^3} \rho_k^{\mathrm{fluc},\rm{vev}}
&= \dfrac{1}{a^2} \displaystyle\int \dfrac{k^2}{4\pi^2}  dk \left[|\dot{y}_{2k}|^2 + \left(\frac{k^2}{a^2}+ \left| m^2_{\mathrm{eff},(h)}(t)\right| \right) |y_{2k}|^2 \right].
\label{eq:energy_denh}
\end{align} 
\end{subequations}
To solve the Eqs.~\eqref{eq:h+phi-flucT} in cosmic time  we use the BD initial condition
and initialize all relevant modes about $\sim 5$ $e$-foldings before the end of inflation, $\mathcal N\sim -5$
\begin{align}
\lim_{t \to - \infty}v_{1k}(k,t) = \lim_{t \to - \infty}y_{2k}(k,t) = \frac{ e^{-\frac{i k t}{a}}}{\sqrt{2 k}}, \hspace{1.cm}
\lim_{t \to - \infty}\dot{v}_{1k}(k,t)=\lim_{t \to - \infty}\dot{y}_{2k}(k,t) = - \frac{i}{a}  \sqrt{\frac{k}{2}} \; e^{-\frac{i k t}{a}}.\label{eq:ini}
\end{align}
After solving Eqs.~\eqref{eq:h+phi-flucT} and evaluating $\rho_{(\phi)}$ and $\rho_{(h)}$ in cosmic time, one can plug them in Eqs.~\eqref{eq:energy_den-quantum} to find respective energy densities.

\begin{figure}[h]
\centering
\hspace{1.2cm}\includegraphics[height= 3.6cm]{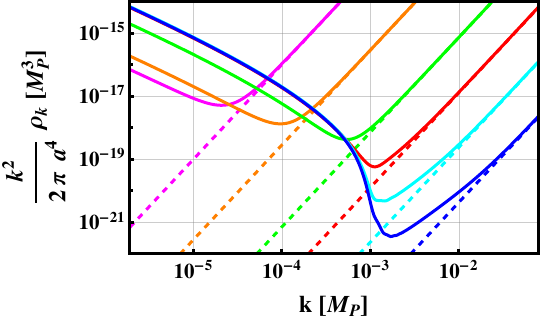} \hspace{1cm}
\includegraphics[height= 3.6cm]{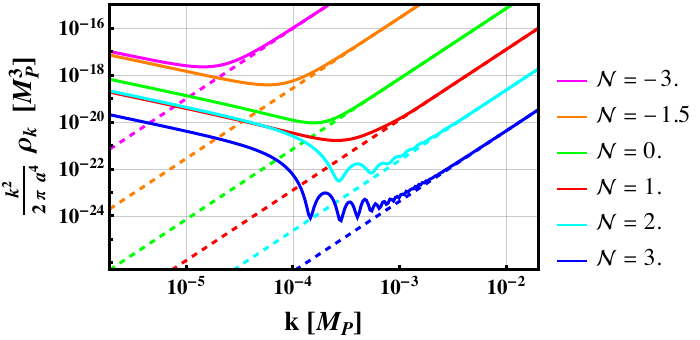} \\
\includegraphics[height= 3.6cm]{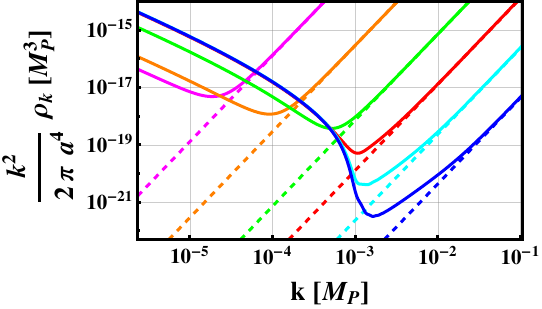}\hspace{6mm}
\includegraphics[height= 3.6cm]{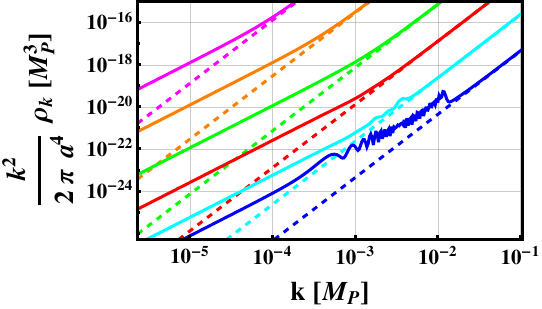} \\
\includegraphics[height= 3.6cm]{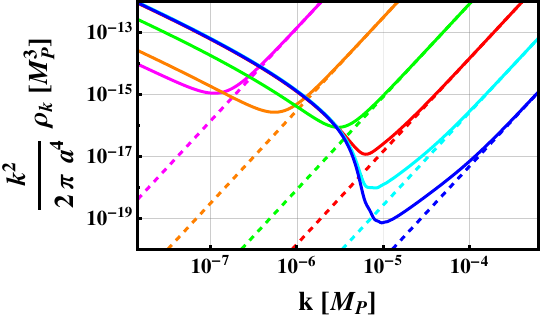}\hspace{8mm}
\includegraphics[height= 3.6cm]{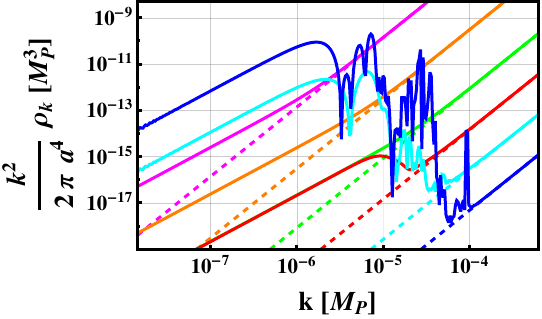}
\caption{Spectra of $\frac{k^2}{2 \pi^2 a^4}\rho_{k,(I)}$ (solid) and $\frac{k^2}{2 \pi^2 a^4}\rho_{k,(I)}^{\rm BD}$ (dashed)
 for $\phi$ (left) and $h$ (right) as given in Eqs.~\eqref{def:rho-BD_phi+h} and \eqref{eq:energy_den-quantum} for BP$a$ (upper row), BP$b$ (middle row) and BP$c$ (lower row).}
\label{plot:phi+h-spectra}
\end{figure}

In Fig.~\ref{plot:phi+h-spectra},  we plot the spectra (solid) of the energy densities for $\phi$ and $h$ alongside the BD spectra (dashed) for all three BPs for different $\mathcal{N}$. The figure allows us to identify the upper limit of the $k$, i.e., the corresponding value where the bare spectra match the BD ones. Once the upper limit is identified, we evaluate the quantum energy densities for $\phi$ and $h$ via Eqs.~\eqref{eq:energy_den-quantum}. In practice, we generate all spectra with wave number close to the unitarity cut-off $k_{\rm UV}$ (which is $M_{\rm P}$ in case of $R^2$-Higgs inflation~\cite{Ema:2017rqn,Gorbunov:2018llf}). However, to find the quantum energy density from Eqs.~\eqref{eq:energy_den-quantum}, we utilize an adaptive numerical code that considers only those modes for which the relative error between $\rho_k$ and $\rho_k^{\rm BD}$ is about 10\% with $\rho_k >\rho_k^{\rm BD}$, ensuring vacuum is subtracted properly. The lower limit of the relevant modes in Eqs.~\eqref{eq:energy_den-quantum}, on the other hand, is chosen from different dynamics.
Since thermalization during (p)reheating proceeds through particle interactions, the relevant modes are those which reside within the horizon at the time of consideration. Modes that are super-horizon are so-called frozen-in and cannot take part in such processes \cite{Sfakianakis:2018lzf}.
In our numerical analysis, this is done via the same adaptive code that only includes modes that have large enough physical wave-numbers to be inside the horizon at the time that we are considering. Further, for solving Eqs.~\eqref{eq:h+phi-flucT}, we have ensured that all modes are initialized deep inside the horizon such that the dynamics at the onset of preheating is captured.

\begin{figure}[h]
\centering
\includegraphics[width=.32 \textwidth]{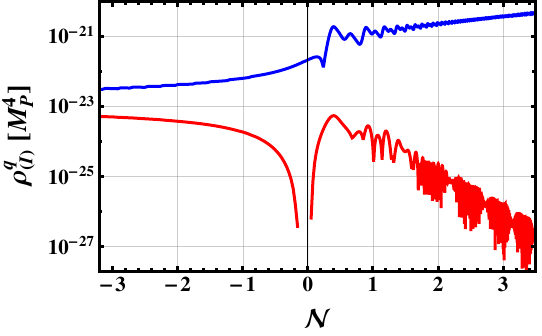}
\includegraphics[width=.32 \textwidth]{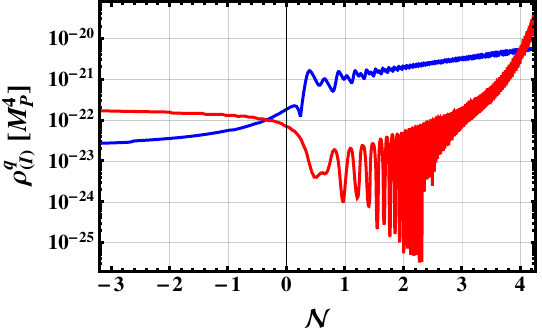}
\includegraphics[width=.32 \textwidth]{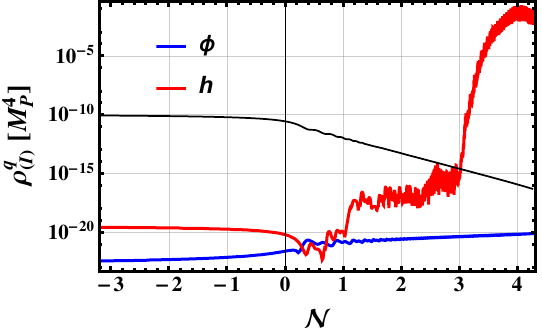}
\caption{The energy densities $\rho^q_{(\phi)}$ (blue) and $\rho^q_{(h)}$ (red) as defined in Eqs.~\eqref{eq:energy_den-quantum}
for the three benchmark points. The black line displays the background energy density $\rho_{\mathrm{inf}}$.\label{plot:inflaton-energy-density}}
\label{fig:energyphih}
\end{figure}

We show the energy densities $\rho^q_{(\phi)}$ (blue), $\rho^q_{(h)}$ (red) and the background energy density $\rho_{\rm inf}$ (black) in Fig.~\ref{fig:energyphih}.
For BP$a$ and BP$b$, both $\rho^q_{(\phi)}$ and  $\rho^q_{(h)}$ are much smaller than $\rho_{\rm inf}$, which is well above the plotted range and not displayed. The $\rho^q_{(\phi)}$ remains practically unchanged between all the BPs primarily because of the value of $\xi_R$, which is practically the same for all three BPs. Around the end of inflation, $\rho^q_{(\phi)}$ receives a tachyonic (exponential) amplification, which can be easily understood as $m_{\mathrm{eff,(\phi)}}^2 < 0$ in Fig.~\ref{meff_plot}. This is due to $m_{4\mathrm{,(\phi)}}^2$ dominating over all other terms in $m_{\mathrm{eff,(\phi)}}^2$ before the end of inflation. After inflation, $m_{1\mathrm{,(\phi)}}^2$ dominates and oscillates but never goes 
below zero. We then find the production of inflaton quanta will not preheat the Universe.

%%%%%%%%%%%%%%%%%%%%%%%%%%%%%%%%%%%%%%%%%%%%%%%%%%%%%%%%%%%%%%%%%%%%%%%%%%%%%%%%%%%%%%
%%%%%%%%%%%%%%%%%%%%%%%%%%%%%%%%%%%%%%%%%%%%%%%%%%%%%%%%%%%%%%%%%%%%%%%%%%%%%%%%%%%%%%

The situation changes for $\rho^q_{(h)}$ due to the range of $\xi_H$ values for different BPs. For BP$a$, $m_{\mathrm{eff},(h)}^2$ remains negative before the end of inflation due to
negative $m_{4\mathrm{,(h)}}^2$, however, after the end of inflation $m_{1\mathrm{,(h)}}^2$ dominates and becomes periodically negative due to zero crossing of $h_0$ condensate, as can be seen from Fig.~\ref{meff_plot}. But this is not sufficient to drive tachyonic growth of $\rho^q_{(h)}$ due to a small amplitude, rather we observe a damping of $\rho^q_{(h)}$ as displayed in Fig.~\ref{fig:energyphih}. This is chiefly due to the smallness of $\xi_H\sim 10^{-3}$ for  BP$a$, which is essentially deep inside $R^2$-like regime. In contrast, for BP$b$ and BP$c$, $m_{\mathrm{eff},(h)}^2$ is large positive for $\mathcal{N} <0$ but oscillates for $\mathcal{N} \gtrsim 0$. For BP$b$, after the end of inflation, $m_{\mathrm{eff},(h)}^2$ does not go below zero in the plotted range $\mathcal{N} \lesssim 3$ (see Fig.~\ref{meff_plot}). Thus, $\rho^q_{(h)}$ does not undergo tachyonic growth for BP$b$ and $\rho^q_{(h)}$ still remains much smaller than $\rho_{\rm inf}$. 
As a result, for BP$b$, Higgs production will lead to inefficient and incomplete preheating. In contrast, for BP$c$, $m_{\mathrm{eff},(h)}^2$ turns negative for $\mathcal{N} \gtrsim 0$, but the amplitude becomes small as we approach $\mathcal{N} \gtrsim 1.5$. Therefore, in the initial stage of preheating, $\rho^q_{(h)}$ experiences tachyonic growth, but parametric resonance takes over for the later part of the preheating. We find successful preheating for BP$c$ which is completed at $\mathcal{N}\sim 3$. Here we understand ``completion'' of preheating as the point in the time evolution when $\rho^q_{(h)}$ becomes equal to the background energy density $\rho_{\rm inf}$. A more conservative approach is adopted by Ref.~\cite{Sfakianakis:2018lzf} where the authors understand completion of preheating as $\rho^q_{(h)}\sim 0.1 \rho_{\rm inf}$ (the linear analysis is not reliable when $\rho^q_{(h)}$ approaches $\rho_{\rm inf}$). 
Our results for baryogenesis are not significantly impacted by the choice between these conventions. We also remark that the growth in $\rho^q_{(h)}$ beyond $\mathcal{N}\sim 3$ for BP$c$ is indicative of the breakdown of our linear order estimation. This growth is expected to be shut off once the decay of the produced particles, backreaction and rescattering effects are taken into account. We shall discuss this in more detail in Sec.~\ref{sec:reheating}.

%%%%%%%%%%%%%%%%%%%%%%%%%%%%%%%%%%%%%%%%%%%%%%%%%%%%%%%%%
\section{Production of  $Z$, $W$ and Goldstone bosons}
\label{sec:GWZ-production}
%%%%%%%%%%%%%%%%%%%%%%%%%%%%%%%%%%%%%%%%%%%%%%%%%%%%%%%%%
%----------------------------------------------
In this section, we focus on a detailed discussion of particle production of gauge and Goldstone bosons.

\subsection{Equations of motion and quantization}
%----------------------------------------------

 We consider the second-order action involving Goldstone and gauge bosons
\begin{equation} \begin{aligned}
S^{(2)}_{\mathrm{G}}
=& \int d^3x \ dt \ a^3
\Biggl[-\frac{1}{2}\overline{g}^{\mu\nu}_{E} G_{IJ} \mathcal{D}_\mu Q^I\mathcal{D}_\nu Q^J-\frac{1}{2}\mathcal{M}_{IJ} Q^I Q^J  \\
&- e^{-\sqrt{\frac{2}{3}}\frac{\varphi}{M_{\rm P}}} \overline{g}^{\mu\nu}_{E}\Biggl\{  \left(Q^J \partial_\mu h_0 -  h_0 \partial_\mu Q^J\right) \left(\delta^3_J \frac{g_Z}{2} Z_\nu + \frac{i e}{2 \sqrt{2} s_W} \bigg[  (\delta^4_J +i\delta^5_J ) W^-_\nu-(\delta^4_J -i\delta^5_J ) W^+_\nu \bigg] \right) \\
&+\frac{g_Z^2}{8}h_0^2 Z_\mu Z_\nu + \frac{e^2}{4 s_W^2} h_0^2  W^+_\mu W^-_\nu\Biggr\}
 -  \dfrac{1}{4} \overline{g}_E^{\mu\rho} \overline{g}_E^{\nu\sigma} F_{Z\mu\nu}F_{Z\rho\sigma}   
 - \dfrac{1}{2} \overline{g}_E^{\mu\rho} \overline{g}_E^{\nu\sigma} F^+_{W\mu\nu}F^-_{W\rho\sigma}   \\
 &- \frac{  M_{\rm P}^2}{\xi_R\Lambda^2}F(\varphi^I)e^{\sqrt{\frac{2}{3}}\, \frac{\varphi}{M_{\rm P}}}\frac{\epsilon^{\mu\nu\rho\sigma}}{a^3} F_{Z\mu\nu}F_{Z\rho\sigma}
- \frac{ 2 M_{\rm P}^2}{\xi_R\Lambda^2}F(\varphi^I)e^{\sqrt{\frac{2}{3}}\, \frac{\varphi}{M_{\rm P}}}\frac{\epsilon^{\mu\nu\rho\sigma}}{a^3} F^+_{W\mu\nu}F^-_{W\rho\sigma}  \Biggr],
\label{action:WZgol}
\end{aligned} \end{equation}
where $I,J=\{3,4,5\}$ everywhere, except inside $F(\varphi^I)$ where $\varphi^I \in \{\varphi,h_0\}$.
In conformal time, the action becomes
\begin{align}
S^{(2)}_{\mathrm{G}}
=& \int d^3x \ d\tau \Biggl[-\frac{1}{2} \eta^{\mu\nu} G_{IJ} (\mathcal{D}_\mu X^I)(\mathcal{D}_\nu X^J)-\frac{1}{2}\mathscr{M}_{IJ} X^I X^J
\nn\\
&- a \, e^{-\sqrt{\frac{2}{3}}\frac{\varphi}{M_{\rm P}}}  \, \eta^{00} \left( X^J (\partial_\tau h_0)-h_0 (\partial_\tau X^J)  + \frac{\partial_\tau a}{a} h_0 X^J \right) 
 \left( \delta^3_J \frac{g_Z}{2} Z_0 + \frac{i e}{2 \sqrt{2} s_W} \bigg[  (\delta^4_J +i\delta^5_J ) W^-_0-(\delta^4_J -i\delta^5_J ) W^+_0 \bigg] \right)  \nn\\
 &-  a^2 \, e^{-\sqrt{\frac{2}{3}}\frac{\varphi}{M_{\rm P}}}\eta^{\mu\nu} \left(\frac{g_Z^2}{8}h_0^2 Z_\mu Z_\nu + \frac{e^2}{4 s_W^2} h_0^2  W^+_\mu W^-_\nu\right) -  \dfrac{1}{4} \eta^{\mu\rho} \eta^{\nu\sigma} F_{Z\mu\nu}F_{Z\rho\sigma}  - \dfrac{1}{2} \eta^{\mu\rho} \eta^{\nu\sigma} F^+_{W\mu\nu}F^-_{W\rho\sigma}\label{eq:2ndorderaction}\\
 &- \frac{  M_{\rm P}^2}{\xi_R\Lambda^2}F(\varphi^I)e^{\sqrt{\frac{2}{3}}\, \frac{\varphi}{M_{\rm P}}}\epsilon^{\mu\nu\rho\sigma} F_{Z\mu\nu}F_{Z\rho\sigma}
- \frac{ 2 M_{\rm P}^2}{\xi_R\Lambda^2}F(\varphi^I)e^{\sqrt{\frac{2}{3}}\, \frac{\varphi}{M_{\rm P}}}\epsilon^{\mu\nu\rho\sigma} F^+_{W\mu\nu}F^-_{W\rho\sigma}\Biggr].
\nn
\end{align}
We first proceed to quantize the $Z$ and $W$ bosons. With Eq.~\eqref{eq:2ndorderaction}, and aided
by Eqs.~\eqref{eq:Z-EoM-tranverse-conf} and~\eqref{eq:Wpmeom-conf}, the quadratic actions for the respective fields can be written as
%%%%%%
\begin{equation} \begin{aligned}
S^\lambda_{W,Z} =& \int d\tau \,\mathcal{L}^\lambda_{W,Z} = \int \, d\tau \,\frac{d^3k}{(2\pi)^3}\, \Bigg[\bigg(\frac{1}{2} |\partial_\tau  \widetilde{Z}^\lambda|^2
- \frac{1}{2}(\omega_Z^\lambda(\tau,k))^2 |\widetilde{Z}^\lambda|^2 \bigg) \\
&+ \bigg(\frac{1}{2} |\partial_\tau  \widetilde{W}^{+,\lambda}|^2
- \frac{1}{2}(\omega_W^\lambda(\tau,k))^2|\widetilde{W}^{+,\lambda}|^2 \bigg)+\bigg(\frac{1}{2} |\partial_\tau  \widetilde{W}^{-,\lambda}|^2
- \frac{1}{2}(\omega_W^\lambda(\tau,k))^2|\widetilde{W}^{-,\lambda}|^2 \bigg)\Bigg].
\end{aligned} \end{equation}
%%%%%%
The canonical momenta for the $Z$ and $W$ in Fourier space are therefore
%%%%%%
\begin{subequations} \begin{eqnarray}
&\hat{\widetilde{\pi}}^\lambda_Z(\tau,\vb{k})=  \partial_\tau \hat{\widetilde{Z}}^{\lambda}(\tau,\vb{k}),\\
&\hat{\widetilde{\pi}}^\lambda_{W^\pm}(\tau,\vb{k})=  \partial_\tau \hat{\widetilde{W}}^{\mp,\lambda}(\tau,\vb{k}),
\end{eqnarray}  \end{subequations}
%%%%%%
with the commutation relations
%%%%%%
\begin{subequations} \begin{eqnarray}
&\bigg[ \hat{\widetilde{Z}}^\lambda(\tau,\vb{k}), \hat{\widetilde{\pi}}^{\lambda\prime}_Z(\tau,\vb{q})\bigg]=
i (2\pi)^3 \delta^{\lambda\lambda\prime}\delta(\vb{k}+\vb{q}),\\
&\bigg[ \hat{\widetilde{W}}^{\pm,\lambda}(\tau,\vb{k}), \hat{\widetilde{\pi}}^{\lambda\prime}_{W^\pm}(\tau,\vb{q})\bigg]=
 i (2\pi)^3 \delta^{\lambda\lambda\prime}\delta(\vb{k}+\vb{q}),
\end{eqnarray}  \end{subequations}
%%%%%%
where the field operators are given as
%%%%%%
\begin{subequations} \label{eq:ZWquant} \begin{eqnarray}
\hat{\widetilde{Z}}^\lambda &=  z^\lambda_{k}(\tau) \hat{a}_Z^\lambda(\vb{k}) + z^{\lambda*}_{k}(\tau) \hat{a}_Z^{\lambda\dagger}(\vb{-k}),\hspace{1cm}(\lambda = \pm)\label{eq:Zquant}\\
\hat{\widetilde{W}}^{+,\lambda} &= w^\lambda_{k}(\tau) \hat{a}_W^\lambda(\vb{k}) + w^{\lambda*}_{k}(\tau)\hat{b}_W^{\lambda\dagger}(\vb{-k}) , \hspace{1cm}(\lambda = \pm) \label{eq:Wquant}
\end{eqnarray}  \end{subequations}
%%%%%%
where $\hat{a}_W^\dagger$, $\hat{b}_W^{\dagger}$ are the creation operators for $W^+$ and $W^-$, respectively. These obey the usual commutation relations
%%%%%%
\begin{subequations} \begin{eqnarray}
 \bigg[\hat{a}_Z^\lambda(\vb{k}),\hat{a}_Z^{\lambda\prime}(\vb{q})\bigg] = 0,\hspace{1cm}\bigg[\hat{a}_Z^{\dagger\lambda}(\vb{k}),\hat{a}_Z^{\dagger\lambda\prime}(\vb{q})\bigg] =0,\hspace{1cm}
 \bigg[\hat{a}_Z^\lambda(\vb{k}),\hat{a}_Z^{\lambda'\dagger}(\vb{q})\bigg] = (2\pi)^3\delta^{\lambda\lambda\prime} \delta^3(\vb{k}-\vb{q}),\\
  \bigg[\hat{a}_W^\lambda(\vb{k}),\hat{a}_W^{\lambda\prime}(\vb{q})\bigg] = 0,\hspace{1cm}\bigg[\hat{a}_W^{\dagger\lambda}(\vb{k}),\hat{a}_W^{\dagger\lambda\prime}(\vb{q})\bigg] =0,\hspace{1cm}
 \bigg[\hat{a}_W^\lambda(\vb{k}),\hat{a}_W^{\lambda'\dagger}(\vb{q})\bigg] = (2\pi)^3 \delta^{\lambda\lambda\prime} \delta^3(\vb{k}-\vb{q}).
\end{eqnarray}  \end{subequations}
%%%%%%
Similar commutation relations hold for $\hat{b}_W^{\lambda}$ while all commutation relationships between $\hat{a}_W$, $\hat{b}_W$ vanish. 
Inserting Eqs.~\eqref{eq:ZWquant} in the Eq.~\eqref{eq:Z-EoM-tranverse-conf} and Eq.~\eqref{eq:Wpmeom-conf}, the mode equations of the $W$ and $Z$ fields can be found as
%%%%%%
\begin{subequations}\label{eq:modeeomWZ} \begin{eqnarray}
&{z^{\lambda}_{k}}''+ (\omega_Z^\lambda)^2\, z^\lambda_{k} = 0 \hspace{1. cm} (\lambda = \pm), \label{eq:modeeomZ}\\
&{w^{\lambda}_{k}}''+ (\omega_W^\lambda)^2 \, w^\lambda_{k} = 0  \hspace{1. cm} (\lambda = \pm). \label{eq:modeeomW}
\end{eqnarray}  \end{subequations}
%%%%%%
%
The quantization of the Goldstone bosons is a bit more involved. This is due to the presence of friction 
term $\mathcal{D}_\tau \widetilde{X}^{I}$ (with $I=\phi_2,\phi_3,\phi_4$) in the respective EoMs. 
We start by recasting Eqs.~\eqref{eom:phi_2}, \eqref{eom:phi_3} and \eqref{eom:phi_4} into the form
\begin{align}
S_{(\phi_i)} =& \int d\tau \, L_{(\phi_i)} = \int \, d\tau \,\frac{d^3k}{(2\pi)^3}\, \ \Delta_{(I)} \bigg[\frac{1}{2} \left|\partial_\tau  X^I\right|^2
- \frac{1}{2}\omega_{(I)}^2(\tau,k) \left|X^I\right|^2 \bigg],
\end{align}
with
\begin{align}
\Delta_{(I)} = \exp{\int_{-\infty}^{\tau}\mathcal{E}_{(I)}(\tau',k)\,d\tau'}.
\end{align}
$\mathcal{E}_{(I)}(\tau,k)$ and $\omega^2_{(I)}(\tau,k)$ are provided in Sec.~\ref{sec:gauge-Goldstone}.
As we deal with non-canonical kinetic terms, we apply the quantization procedure detailed in Ref.~\cite{Lozanov:2016pac}. We can quantize the fields as in before as
%%%%%%
\begin{subequations}\label{eq:phiiquant} \begin{eqnarray}
&\hat{\widetilde{X}}^{\phi_2} =   \bigg[s_{k}(\tau) e^{\phi_2}(\tau) \hat{a}_3(\vb{k})  + s^*_k(\tau) e^{\phi_2}(\tau) \hat{a}^\dagger_3(-\vb{k})   \bigg],\label{eq:phi2quant}\\
&\hat{\widetilde{X}}^{\phi_3} =   \bigg[q_{k}(\tau) e^{\phi_3}(\tau) \hat{a}_4(\vb{k})  + q^*_k(\tau) e^{\phi_3}(\tau) \hat{a}^\dagger_4(-\vb{k})   \bigg],\label{eq:phi3quant}\\
&\hat{\widetilde{X}}^{\phi_4} =   \bigg[r_{k}(\tau) e^{\phi_4}(\tau) \hat{a}_5(\vb{k})  + r^*_k(\tau) e^{\phi_4}(\tau) \hat{a}^\dagger_5(-\vb{k})   \bigg],\label{eq:phi4quant}
\end{eqnarray}  \end{subequations}
%%%%%%
and the canonical momentum can be found as
%%%%%%
\begin{align}
\hat{\widetilde{\pi}}^{I}(\tau,\vb{k}) = \Delta_{(I)} \partial_\tau \hat{\widetilde{X}}^{I}(\tau,\vb{k}),
\end{align}
with
\begin{align}
\bigg[ \hat{\widetilde{X}}^{I}(\tau,\vb{k}),\hat{\widetilde{\pi}}^{J}(\tau,\vb{q})\bigg]= \frac{i}{\Delta_{(I)}}(2\pi)^3\delta^{IJ}\delta(\vb{k}+\vb{q}),
\end{align}
%%%%%%
and similar commutation relations between creation and annihilation operators as in the previous 
section. From Eqs.~\eqref{eom:phi_2}, \eqref{eom:phi_3} and \eqref{eom:phi_4}, we obtain the EoMs of the mode functions as
%%%%%%
\begin{subequations} \label{eom:sqr_k}  \begin{eqnarray}
&s_k'' +\mathcal{E}_{(\phi_2)}(\tau,k) \ s_k' + \omega^2_{(\phi_2)}(\tau,k)\, s_k  = 0, \label{eom:s_k} \\
&q_k'' +\mathcal{E}_{(\phi_3)}(\tau,k) \ q_k' + \omega^2_{(\phi_3)}(\tau,k)\, q_k  = 0, \label{eom:q_k}\\
&r_k'' +\mathcal{E}_{(\phi_4)}(\tau,k) \ r_k' +\omega^2_{(\phi_4)}(\tau,k)\, r_k  = 0. \label{eom:r_k}
\end{eqnarray} \end{subequations}
%%%%%%

We have checked numerically that the constraints imposed by Eqs.~(\ref{eq:gaugecond1}) and (\ref{eq:gaugecond2}) on the Goldstone boson $\widetilde\phi_2$, and (\ref{eq:Wpmom}) and (\ref{eq:constr2}) on the Goldstones $\widetilde\phi_3$ and $\widetilde\phi_4$, are consistent with the EoM of the corresponding mode functions Eqs.~(\ref{eom:phi_2}) and (\ref{eom:phi_3})-(\ref{eom:phi_4}), respectively.

%%%%%%%%%%%%%%%%%%%
\subsection{Energy density}
%%%%%%%%%%%%%%%%%%%
We can define the physical energy density associated with the Goldstone and gauge bosons as 
\begin{align}
\rho_{\mathrm{G}}= \frac{1}{a^4}\int d^3x \left\langle T_{00}^{\mathrm{G}} \right\rangle= \frac{1}{a^4} \int \frac{d^3k}{(2\pi)^3} \left\langle \rho_{k,\mathrm{G}}\right\rangle
= \frac{1}{a^4}\int \frac{d^3k}{(2\pi)^3} \rho_{k,\mathrm{G}}^{\rm{vev}}, \label{eq:energydensityQG}
\end{align}
where the energy-momentum tensor can be derived from Eq.~\eqref{eq:2ndorderaction} as 
\begin{align}
T_{\mu\nu}^{\mathrm{G}} = & \ G_{IJ} (\mathcal{D}_\mu X^I)(\mathcal{D}_\nu X^J)
+ 2 a \, e^{-\sqrt{\frac{2}{3}}\frac{\varphi}{M_{\rm P}}}  \ \delta_\mu^0  \delta_\nu^0   \left( X^J (\partial_\tau h_0)-h_0 (\partial_\tau X^J)  + \frac{\partial_\tau a}{a} h_0 X^J \right) 
 \nn\\
& 
 \left( \delta^3_J \frac{g_Z}{2} Z_0 + \frac{i e}{2 \sqrt{2} s_W} \bigg[  (\delta^4_J +i\delta^5_J ) W^-_0-(\delta^4_J -i\delta^5_J ) W^+_0 \bigg] \right) +  2a^2 \, e^{-\sqrt{\frac{2}{3}}\frac{\varphi}{M_{\rm P}}}  \left(\frac{g_Z^2}{8}h_0^2 Z_\mu Z_\nu + \frac{e^2}{4 s_W^2} h_0^2  W^+_\mu W^-_\nu\right)
 \nn\\
&+\eta^{\rho\sigma} F_{Z\mu\rho}F_{Z\nu\sigma} +  2 \eta^{\rho\sigma} F^+_{W\mu\rho}F^-_{W\nu\sigma} + \eta_{\mu\nu} \Biggl[-\frac{1}{2} \eta^{\alpha\beta} G_{IJ} (\mathcal{D}_\alpha X^I)(\mathcal{D}_\beta X^J)-\frac{1}{2}\mathscr{M}_{IJ} X^I X^J
\nn\\
&+a \, e^{-\sqrt{\frac{2}{3}}\frac{\varphi}{M_{\rm P}}}  \left( X^J (\partial_\tau h_0)-h_0 (\partial_\tau X^J)  + \frac{\partial_\tau a}{a} h_0 X^J \right) 
 \left( \delta^3_J \frac{g_Z}{2} Z_0 + \frac{i e}{2 \sqrt{2} s_W} \bigg[  (\delta^4_J +i\delta^5_J ) W^-_0-(\delta^4_J -i\delta^5_J ) W^+_0 \bigg] \right)  \nn\\
 &-  a^2 \, e^{-\sqrt{\frac{2}{3}}\frac{\varphi}{M_{\rm P}}}\eta^{\alpha\beta} \left(\frac{g_Z^2}{8}h_0^2 Z_\alpha Z_\beta + \frac{e^2}{4 s_W^2} h_0^2  W^+_\alpha W^-_\beta\right) -  \dfrac{1}{4} \eta^{\alpha\rho} \eta^{\beta\sigma} F_{Z\alpha\beta}F_{Z\rho\sigma}  - \dfrac{1}{2} \eta^{\alpha\rho} \eta^{\beta\sigma} F^+_{W\alpha\beta}F^-_{W\rho\sigma}\Biggr].
\label{eq:gauge+Gold-Tmunu}
\end{align}
We can also find the energy density per Fourier mode of the of scalar field fluctuations in momentum space
\begin{align}
\rho_{k,\mathrm{G}}  =  & \ \frac{1}{2} G_{IJ} (\mathcal{D}_\tau \widetilde{X}^I)(\mathcal{D}_\tau \widetilde{X}^J) + \frac{1}{2}\left(k^2 G_{IJ} + \mathscr{M}_{IJ}\right) \widetilde{X}^I \widetilde{X}^J - \frac{1}{2} \ e^{-2\sqrt{\frac{2}{3}}\frac{\varphi}{M_{\rm P}}}  \bigg(h_0'  \widetilde{X}^I + h_0 \frac{a'}{a}  \widetilde{X}^I- h_0 \partial_\tau\widetilde{X}^I\bigg) \nn\\
 & \bigg(h_0' \widetilde{X}^J + h_0 \frac{a'}{a} \widetilde{X}^J- h_0 \partial_\tau\widetilde{X}^J  \bigg)\Biggl[  \frac{g_Z^2}{4 \mathcal{K}_Z} \delta^3_I \delta^3_J  
 +\frac{e^2}{4 s_W^2\mathcal{K}_W}
\big(\delta^4_I \delta^4_J+\delta^5_I \delta^5_J\big)  \Biggr] +\frac{1}{2}\widetilde{Z}'_i\widetilde{Z}'_i +\widetilde{W}^{+\prime}_i\widetilde{W}^{-\prime}_i\nn\\
&
+ a^2 \left[ \frac{\mathcal{K}_Z}{2} \widetilde{Z}_i \widetilde{Z}_i+\mathcal{K}_W\widetilde{W}^+_i \widetilde{W}^-_i \right]
+\mathcal{O}(\widetilde{X}^3), 
\label{eq:gaugegolen}
\end{align}
where we have used Eqs.~\eqref{eq:gaugecond1} and \eqref{gauge:Wmomconf}, as well as the gauge
conditions described in Eqs.~\eqref{gauge:Zmomconf} and \eqref{gauge:Wmomconf}. As the different fields 
do not mix, we can decompose $ \rho_{k,\mathrm{G}}^{\rm{vev}}$ into gauge and Goldstone energy densities
\begin{align}
 \rho_{k,\mathrm{G}}^{\rm{vev}}=  \rho_k^{Z} + \rho_k^{W}  + \rho_k^{(\phi_i)} .
\label{eq:energydecomposition1}
\end{align}
We are now ready to consider the $Z$ and $W$ boson cases before turning to the Goldstone bosons.

%-----------------------------------------------
\subsubsection{$Z$ and $W$ bosons}\label{sec:ZandW}
%-------------------------------------------------
From Eq.~\eqref{eq:gaugegolen}, we see that the energy densities for $W$ and $Z$ bosons are separated out as
\begin{subequations} \label{eq:WZenergy-density}  \begin{eqnarray}
&\rho_k^{Z} =  \displaystyle\sum_{\lambda=\pm}\left[\left|z^{\lambda\prime}_{k}\right|^2+a^2 \mathcal{K}_Z\left|z^{\lambda}_{k}\right|^2 \right], \label{eq:Zenergy-density}\\
&\rho_k^{W}  = 2 \displaystyle\sum_{\lambda=\pm}\left( \left|w^{\lambda\prime}_{k}\right|^2+ a^2 \mathcal{K}_W\left|w^{\lambda}_{k}\right|^2 \right).\label{eq:Wenergy-density}
\end{eqnarray} \end{subequations}
To find the respective vacuum-subtracted energy densities, we have to minimize the associated Hamiltonian 
%%%%%%
\begin{equation} \begin{aligned}
 \mathcal{\hat{H}}_{W,Z}   =  \int d^3x \left\langle T^{\mathrm{G}}_{00,\ WZ} \right\rangle
 & =\frac{1}{2}\int \frac{d^3 k}{(2\pi)^3}  \sum_{\lambda=\pm}  \bigg[\left(|z_{k}^{\lambda\prime}|^2 +(\omega_Z^\lambda)^2 |z_{k}^{\lambda}|^2\right)
   \left(\hat{a}_Z^{\lambda\dagger}(\vb{k}) \hat{a}_Z^{\lambda}(\vb{k})+\delta^3(\vb{0})\right)\\
   &+ 2 \left(|w_{k}^{\lambda\prime}|^2 +(\omega_W^\lambda)^2 |w_{k}^{\lambda}|^2\right)
   \left(\hat{a}_W^{\lambda\dagger}(\vb{k}) \hat{a}_W^{\lambda}(\vb{k})+\delta^3(\vb{0})\right) \bigg],
\end{aligned} \end{equation}
%%%%%%
($T^{\mathrm{G}}_{00,\ WZ}$ is the energy associated with $Z$ and $W$). As before, $\mathcal{\hat{H}}_{W,Z}$ can be minimized by the BD vacuum solution
\begin{align}
z_{k,{\rm BD}}^{\lambda}  = \frac{1}{\sqrt{2k}} e^{-i k \tau},\hspace{1cm}  w_{k,{\rm BD}}^{\lambda} = \frac{1}{\sqrt{2k}} e^{-i k \tau}.
\end{align}
The corresponding energy densities, obtained from Eqs.~\eqref{eq:WZenergy-density}, are
%%%%%%
\begin{subequations} \begin{eqnarray}
\rho_{Z}^{\rm BD} &= \displaystyle\int dk \frac{k^2}{2 \pi^2 a^4}  \; \rho_{k,Z}^{\rm BD}  =\frac{1}{a^4}\int dk \; \frac{k^3}{\pi^2}, \\
\rho_{W}^{\rm BD} &=\displaystyle\int dk \frac{k^2}{2 \pi^2 a^4} \; \rho_{k,W}^{\rm BD}  = \frac{1}{a^4}\int dk \; \frac{2k^3}{\pi^2},
\end{eqnarray}  \end{subequations}
%%%%%%
where, at sufficiently early times and for large modes, $(\omega_{Z,W}^\lambda)^2 \to k^2$.
The quantum gauge energy density is obtained by removing the BD vacuum from the classical solution as
%%%%%%
\begin{subequations} \begin{eqnarray}
\rho^q_{Z} &= \rho_{Z}-\rho_{Z}^{\rm BD} \label{def:quantum-energy-density-Z}, \\
 \rho^q_{W} &= \rho_{W}-\rho_{W}^{\rm BD}.\label{def:quantum-energy-density-W}
\end{eqnarray}  \end{subequations}
%%%%%%
When finding the energy densities, we solve Eqs.~\eqref{eq:modeeomZ} and~\eqref{eq:modeeomW} in cosmic time.

%%%%%%%%%%%%%%%%%%%
\begin{figure}[h]
\begin{center}
\includegraphics[width=.32 \textwidth]{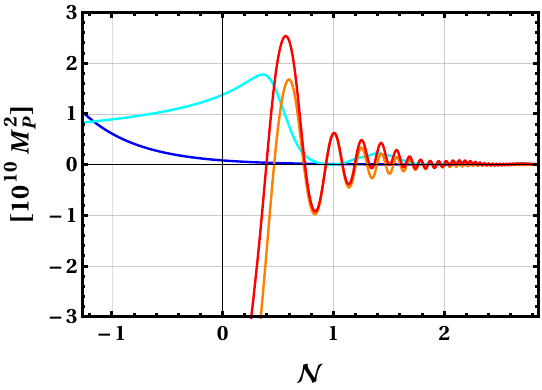}
\includegraphics[width=.32 \textwidth]{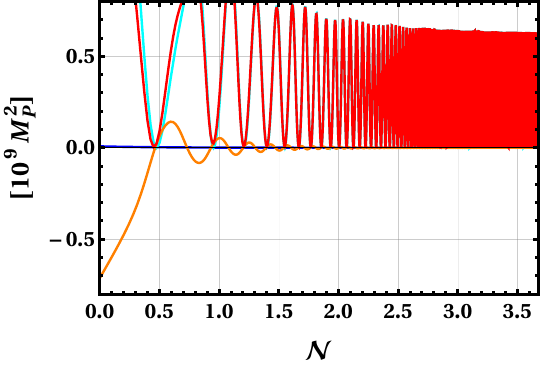}
\includegraphics[width=.32 \textwidth]{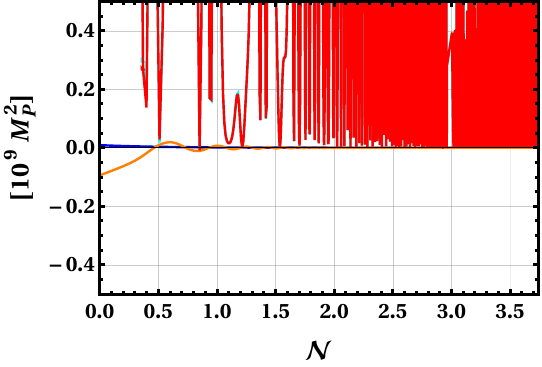}
\includegraphics[width=.32 \textwidth]{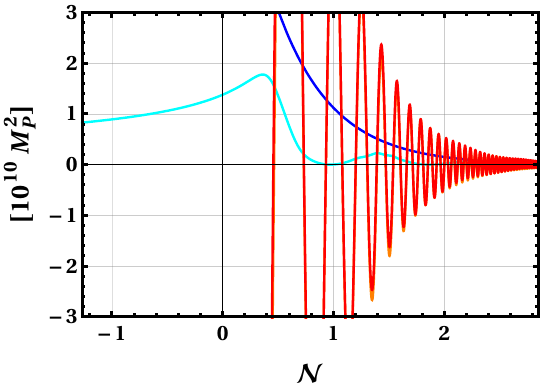}
\includegraphics[width=.32 \textwidth]{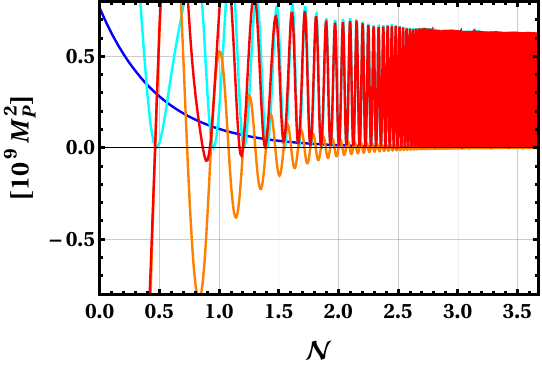}
\includegraphics[width=.32 \textwidth]{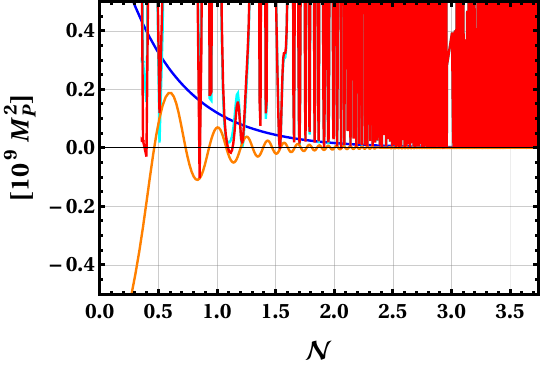}
\end{center}
\caption{Relative strength of different terms of $({\omega_{Z}^{+}})^2$ in Eq.~\eqref{eq:womegaZ} for BP$a$ (left panel), BP$b$ (middle panel) and BP$c$ (right panel).
The upper panel conforms $k = a(t_{\rm end})H(t_{\rm end})$ while the lower panel is for
$k = 10 \, a(t_{\rm end})H(t_{\rm end})$. The blue, cyan, orange, and red lines in each figure 
display $k^2/a^2$, $m_Z^2$, $\zeta^+$ and a combination of $m_Z^2$ plus $\zeta^+$. We set $\Lambda =2\times10^{-5}~M_{\rm P}$ throughout.}
\label{plot:omegaZW}
\end{figure}
%%%%%%%%%%%%%%%%%%%

In Fig.~\ref{plot:omegaZW}, we display the different contributions to ${\omega_{Z}^{+}}^2$, 
Eq.~\eqref{eq:womegaZ}, for all three BPs. For BP$a$ (first column of Fig.~\ref{plot:omegaZW}), 
as the non-minimal coupling $\xi_H$ is small, the impact of $m_Z^2$ is suppressed compared 
to $\zeta^+$ (see Eq.\eqref{eq:womegaZ}). This is visible from the cyan and orange solid lines, 
respectively. However, for BP$b$ and BP$c$, the larger non-minimal coupling renders $m_Z^2$ more 
dominant compared to $\zeta^\lambda$. This can be seen from the second and third columns of Fig.~\ref{plot:omegaZW}. 
${\omega_{Z}^{-}}^2$ shows a similar behavior, except for a sign change of the CS term. We also find that the different contributions of
${\omega_{W}^{\lambda}}^2$ follow a similar pattern, and we choose to not repeat these here.

%%%%%%%%%%%%%%%%%%%
\begin{figure}[h]
\centering
\includegraphics[width=.32 \textwidth]{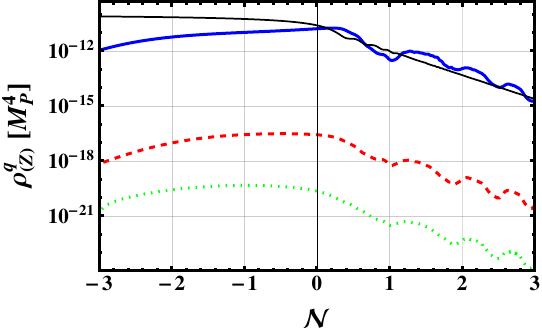}
\includegraphics[width=.32 \textwidth]{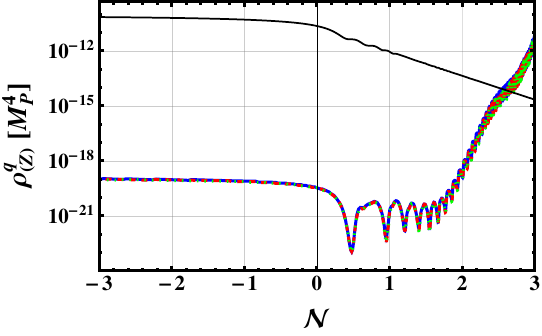}
\includegraphics[width=.32 \textwidth]{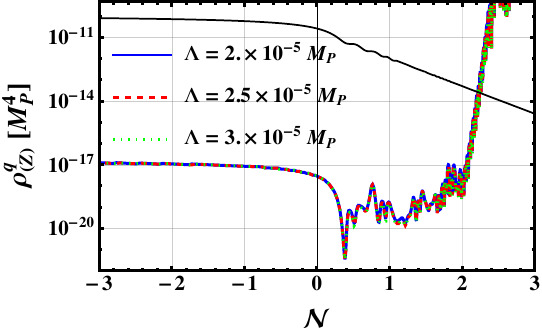}
\includegraphics[width=.32 \textwidth]{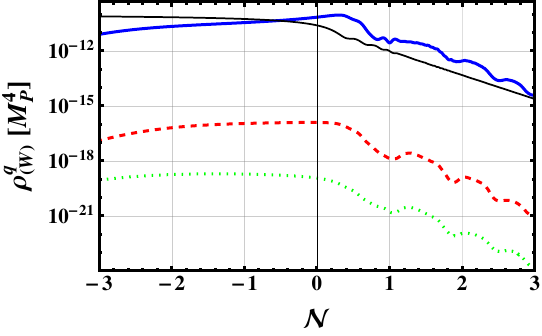}
\includegraphics[width=.32 \textwidth]{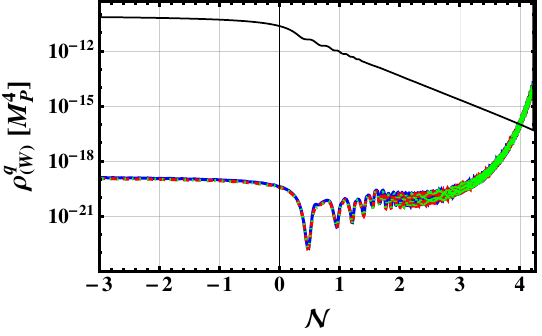}
\includegraphics[width=.32 \textwidth]{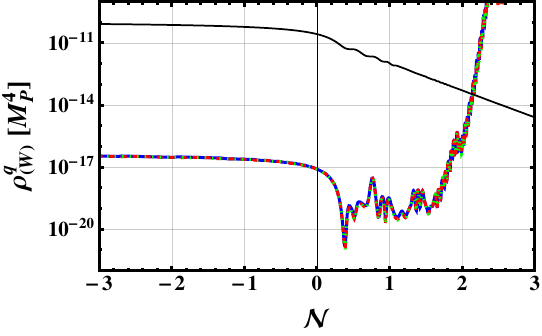}
\caption{The energy densities $\rho^q_{Z}$ (upper row) and $\rho^q_{W}$ (lower row) for various values of $\Lambda$ for the benchmark points BP$a$ (left column), BP$b$ (middle column) and BP$c$ (right column). The black line displays the background energy density $\rho_{\mathrm{inf}}$.
\label{plot:WZenergydensity}}
\end{figure}
%%%%%%%%%%%%%%%%%%%

The behavior of $({\omega_{Z}^{+}})^2$ of Fig.~\ref{plot:omegaZW} has severe implications for the energy densities of the $Z$ and $W$ bosons, Fig.~\ref{plot:WZenergydensity}. For BP$a$, as the $\zeta^\lambda$ term dominantes, $\rho^q_{Z}$ and $\rho^q_{W}$ both scale as $1/\Lambda$. However, for BP$b$ and BP$c$, as the non-minimal coupling becomes larger, the $m_Z^2$ term overpowers $\zeta^\lambda$. We, therefore, see parametric resonance taking over. We find that $Z$ boson production can preheat the Universe within 2 $e$-foldings after the end of inflation for both BP$b$ and BP$c$. For $W$ production, the preheating is completed within $\mathcal{N}\approx 4$ ($\mathcal{N}\approx2$) for BP$b$ (BP$c$). We remark that gauge preheating is also possible for BP$a$ if $\Lambda \lesssim 2\times10^{-5}~M_{\rm P}$, however, we shall see below that such values of $\Lambda$ will overproduce hypermagnetic fields and imply an overproduction of the observed baryon asymmetry. Note that the produced $Z$ boson $W$ boson can decay into SM matter. We will return to this in Sec.~\ref{sec:reheating}.

%%%%%%%%%%%%%%%%%%%
\subsubsection{Goldstone bosons}
%%%%%%%%%%%%%%%%%%%
The Goldstone fields do not mix. Therefore, we can further decompose VEV energy density $\rho_k^{(\phi_i)}$ from Eq.~\eqref{eq:energydecomposition1} into the $\phi_i$ VEV energy density as
\begin{align}
\rho_k^{(\phi_i)}= \rho_k^{(\phi_2)} + \rho_k^{(\phi_3)} + \rho_k^{(\phi_4)}
\end{align}
with
\begin{subequations}
\begin{align}
\rho_k^{(\phi_2)}\; = \;   \frac{1}{2} \Biggl\{ \left(1-\frac{m_Z^2}{\mathcal{K}_Z} \right)\left|s_k^\prime\right|^2
& +  \Biggl[k^2+a^2 m_{\mathrm{eff},(\phi_2)}^2 - \frac{m_Z^2}{\mathcal{K}_Z} \;\Upsilon^2\Biggl] \left|s_k\right|^2
- \frac{m_Z^2}{\mathcal{K}_Z} \;\Upsilon \; (s'_k s^\ast_k +s^{\ast\prime}_k s_k )
\Biggr\}\label{eq:golphi2},\\
%%%%%%%%%%%%%%%
\rho_k^{(\phi_3)} \; = \; \frac{1}{2} \Biggl\{  \left(1-\frac{m_W^2}{\mathcal{K}_W} \right) \left|q_k^\prime\right|^2&+ \Biggl[k^2+a^2 m_{\mathrm{eff},(\phi_3)}^2 - \frac{m_W^2}{\mathcal{K}_W} \;\Upsilon^2 \Biggl] \left|q_k\right|^2
- \frac{m_W^2}{\mathcal{K}_W} \;\Upsilon\;  (q'_k q^\ast_k +q^{\ast\prime}_k q_k )\Biggr\}  \label{eq:golphi3},\\
%%%%%%%%%%%%%%%
\rho_k^{(\phi_4)} \; = \; \frac{1}{2} \Biggl\{\left(1-\frac{m_W^2}{\mathcal{K}_W} \right)\left|r_k^\prime\right|^2 &+ \Biggl[k^2+a^2 m_{\mathrm{eff},(\phi_4)}^2 - \frac{m_W^2}{\mathcal{K}_W} \;\Upsilon^2\Biggl] \left|r_k\right|^2
-\frac{m_W^2}{\mathcal{K}_W} \;\Upsilon \; (r'_k r^\ast_k +r^{\ast\prime}_k r_k )\Biggr\} ,
\label{eq:golphi4}
\end{align}
\end{subequations}
where $\Upsilon$ was defined in Eq.~\eqref{def:Upsilon}.

\begin{figure}[!t]
\centering
\includegraphics[height = 4.8 cm]{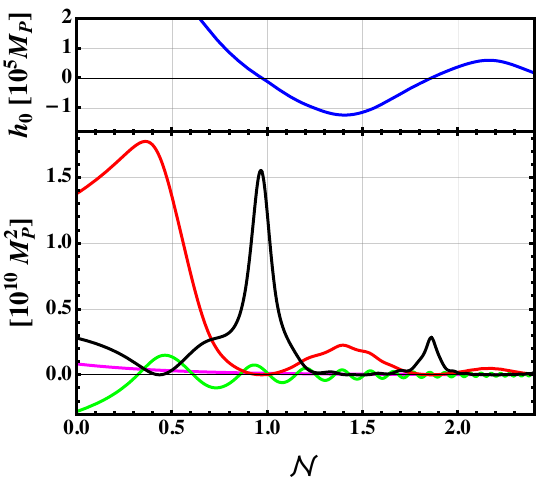}\hspace{2mm}
\includegraphics[height = 4.8 cm]{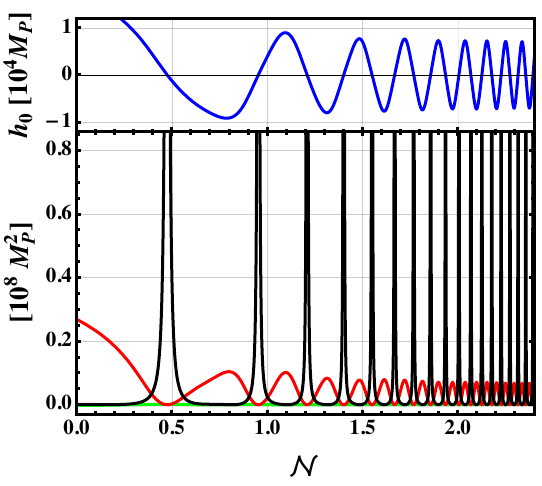}\hspace{2mm}
\includegraphics[height = 4.8 cm]{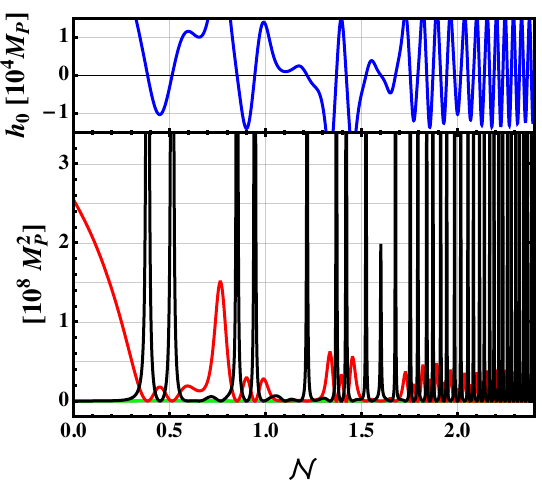}
\caption{Different terms in $\omega^2_{(\phi_2)}(\tau,k) / a^2$. On the bottom panels, the magenta, green, red and black lines correspond to  $k^2/a^2$, $m_{\mathrm{eff},(\phi_2)}^2$, $m_Z^2$ and $\mathcal{E}_{(\phi_2)}\Upsilon / a^2$, see Eq.~\eqref{eom:phi_2-omega2}, for all three BPs. Here as a reference we set $k=a(t_{\rm end}) H(t_{\rm end})$. We can see that the spikes correspond to the zero-crossings of $h_0$ (blue lines) displayed on the upper panels.
See text for detailed explanation. 
\label{plot:spikes}}
\end{figure}

Before solving Eqs.~\eqref{eom:sqr_k} to find the energy densities, let us briefly discuss the different contributions to $\omega^2_{(I)}$ and their impact on preheating. For this purpose, we take $\phi_2$ as a representative field for the Goldstone bosons, and we checked that $\phi_3$ and $\phi_4$ show a similar behavior (with $m_Z^2$ replaced by $m_W^2$). In Fig.~\ref{plot:spikes}, we have plotted the different contributions to $\omega^2_{(\phi_2)}(\tau,k)$ for $k=a(t_{\rm end}) H(t_{\rm end})$ for illustration.  It is clear, around the end of inflation, $m_Z^2$ dominates over all other terms but soon after, the last term associated with $\mathcal{E}_{(\phi_2)}$ dominates every time $h_0$ crosses zero,
resulting in spike-like structures. The spikes have the amplitude
\begin{align}
\left.\omega^2_{(\phi_{2})} \right|_{h_0=0}=k^2+a^2 \left[ m_{\mathrm{eff},(\phi_{2})}^2 + \frac{g_Z^2}{2}\frac{a^2}{k^2} e^{-\sqrt{\frac{2}{3}}\frac{\varphi}{M_{\rm P}}} h_0^{\prime 2} \right],
\label{eq:spike-amplitude}
\end{align}
which is well below the unitarity cut-off scale.
For comparison, we have also plotted $k^2/a^2$ for the $k=a(t_{\rm end}) H(t_{\rm end})$ in magenta in the lower panels and the evolution of $h_0$ in blue in the upper panels. It is clear that the spikes are smaller for BP$a$, which is deep in the $R^2$-like regime with $\xi_H \ll 1$. However, they increase significantly for BP$b$ and BP$c$ due to a comparably larger $\xi_H$. As we will see shortly, such spikes induce a growth of the corresponding modes leading to preheating of the Goldstone bosons (see Refs.~~\cite{Ema:2016dny,He:2018mgb,Sfakianakis:2018lzf,Bezrukov:2019ylq,Bezrukov:2020txg,He:2020ivk} for similar discussions) without violating unitarity.

%%%%%%%%%%%%%%%%%%%
%%%%%%%%%%%%%%%%%%%

To solve Eqs.~\eqref{eom:sqr_k}, we make the change of variable $\bar{s}_k=  \sqrt{\Delta_{(\phi_2)} } \,s_k$ to write
\begin{subequations} \label{eom:sqr_kscaled}  \begin{eqnarray}
&\bar{s}''_k(\tau,k) + \bar{\omega}^2_{(\phi_2)} \bar{s}_k(\tau,k)  = 0, \label{eom:s_kscaled} \\
&\bar{q}''_k(\tau,k) + \bar{\omega}^2_{(\phi_3)} \bar{q}_k(\tau,k)  = 0, \label{eom:q_kscaled}\\
&\bar{r}''_k(\tau,k) + \bar{\omega}^2_{(\phi_4)} \bar{r}_k(\tau,k)  = 0, \label{eom:r_kscaled}
\end{eqnarray} \end{subequations}
where
\begin{align}
\bar{\omega}^2_{(I)} =\omega^2_{(I)} -\dfrac{\mathcal{E}_{(I)}'}{2}-\dfrac{ \mathcal{E}^2_{(I)}}{4}.
\end{align}
At sufficiently early times, when all modes of interest are deep inside the horizon (i.e. $k \gg a H$), the frequencies become $\bar{\omega}^2_{(I)} \to k^2$ and the solutions of Eqs.~\eqref{eom:sqr_kscaled} reduce to plane waves
\begin{align}
s_{k}  = \frac{e^{-i k \tau}}{\sqrt{2k\Delta_{(\phi_2)}}} ,\hspace{1.5cm}  q_{k} = \frac{e^{-i k \tau}}{\sqrt{2k\Delta_{(\phi_3)}}} ,\hspace{1.5cm}  r_{k}  = \frac{e^{-i k \tau}}{\sqrt{2k\Delta_{(\phi_4)}}}. \label{gold_ini}
\end{align}
Hence, by taking Eq.~\eqref{gold_ini} as the initial conditions, Eqs.~\eqref{eom:sqr_kscaled} will enable us to find the evolution of the relevant modes from sub-horizon to super-horizon scales. Here, unlike the BD solution of all other fields, the appearance of the $\Delta_{(I)}$s is due to the presence of additional friction terms in Eqs.~\eqref{eom:sqr_k} as shown above. At early times, for the relevant modes, we simultaneously have ${k^2}/{a^2} \gg |m^2_{\mathrm{eff},(I)}(t)|$ and $m_{Z,W}^2/\mathcal{K}_{Z,W} \to 0$. As before, Eq.~\eqref{gold_ini} will minimize the associated Hamiltonian. 

The energy density associated with vacuum for the Goldstone modes reads
\begin{align}
 \rho_{(\phi_i)}^{\rm BD}=\frac{1}{4\pi^2  a^4} \;  \int dk \; \frac{k^3}{\Delta_{(\phi_i)}(k,\tau)}.
 \end{align}
The quantum Goldstone energy densities are obtained by removing the BD vacua from the respective classical solutions as
\begin{align}
\rho^q_{(\phi_i)} &= \rho_{(\phi_i)}- \rho_{(\phi_i)}^{\rm BD},~~~~\mbox{ with}~i =  2,3,4.\label{def:quantum-energy-density-phiG}
\end{align}
Note that we have solved the respective EoMs in cosmic time. 

%%%%%%%%%%%%%%%%%%%
\begin{figure}[!t]
\centering
\includegraphics[width=.32 \textwidth]{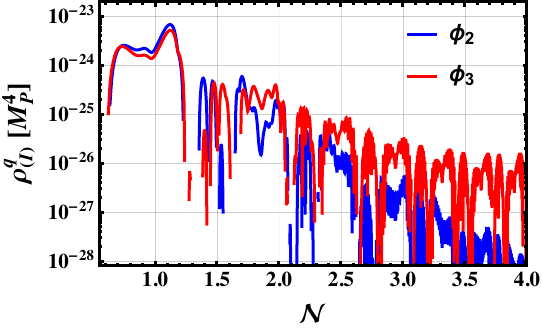}
\includegraphics[width=.32 \textwidth]{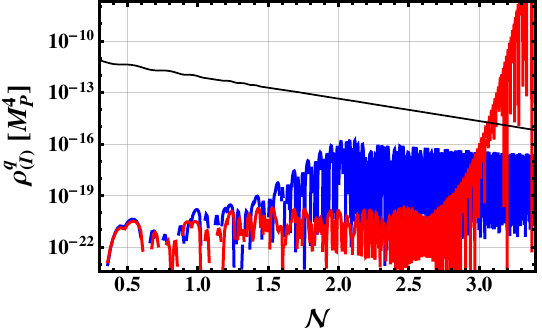}
\includegraphics[width=.32 \textwidth]{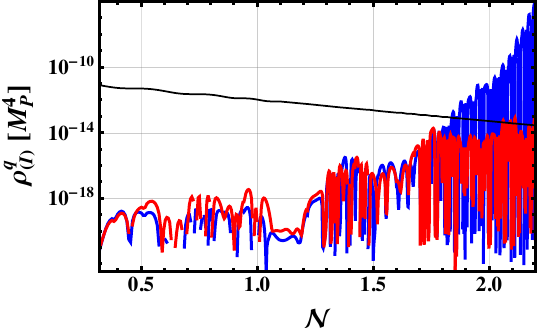}
\caption{The energy densities $\rho^q_{(\phi_2)}$ and $\rho^q_{(\phi_3)}$ for the benchmark points BP$a$ through $c$ from left to right. We recall that $\rho^q_{(\phi_3)}=\rho^q_{(\phi_4)}$. The black line displays the background energy density $\rho_{\mathrm{inf}}$.
\label{plot:rhoG}}
\end{figure}
%%%%%%%%%%%%%%%%%%%

We show the $\rho^q_{(\phi_2)}$ and $\rho^q_{(\phi_3)}$ as function of $\mathcal{N}$ in Fig.~\ref{plot:rhoG} in blue and red, respectively. 
We recall that $\omega^2_{(\phi_3)}=\omega^2_{(\phi_4)}$ and $\rho^q_{(\phi_3)}=\rho^q_{(\phi_4)}$.
We find that for all of our BPs no particles are produced before the end of inflation. However, the spike behavior in $\omega^2_{(\phi_2)}$ and $\omega^2_{(\phi_3)}$ induces a growth of the energy densities for BP$b$ and BP$c$. We find that preheating is possible for both BP$b$ and BP$c$. In case of BP$b$, the preheating is complete for $\mathcal{N} \approx 3$ ($\mathcal{N} \approx 1.9$ for the BP$c$). However, there is a subtlety. For BP$b$, the preheating is completed via the $\phi_3$ field, whereas, for BP$c$ it is completed by $\phi_2$. This is primarily due to $m_Z > m_W$ and the evolution of $h_0$. For both these BPs, initially $\rho^q_{(\phi_2)}$ rises faster due to the heaviness of the $Z$ boson while $\rho^q_{(\phi_3)}$ remains constant. At a later time, the ``spike forest'' becomes denser which overpowers the mild mass difference between $Z$ and $W$. Therefore, the delayed growth of $\rho^q_{(\phi_3)}$ plateaus when the spikes are more abundant. The exponential growth of $\rho^q_{(\phi_3)}$ for BP$c$ is not shown in the right panel of Fig.~\ref{plot:rhoG} because the initial growth of $\rho^q_{(\phi_2)}$ is sufficiently high enough to preheat the Universe. We stress again that our linearized result here does not include decay, which we shall discuss in Sec.~\ref{sec:reheating}.

%%%%%%%%%%%%%%%%%%%%%%%%%%%%%%%%%%%%%%%%%%%%%%%%%%%
\section{Production of the electromagnetic field}
\label{sec:EM-production}
%%%%%%%%%%%%%%%%%%%%%%%%%%%%%%%%%%%%%%%%%%%%%%%%%%%
Following the same steps as in the previous sections, we now consider the production of electromagnetic (EM) field.
The time component ($\beta = 0$) of Eq.~\eqref{eq:photon} at linear order in the perturbations is
\begin{align}
-\frac{1}{a^2}\partial_i\left(\partial_i A_0 -\partial_0 A_i\right) = 0,\label{eq:0A1}
\end{align}
and the spatial components ($\beta = i$) are
\begin{equation} \begin{aligned}
- \frac{\dot{a}}{a}\left(\partial_0 A_i - \partial_i A_0 \right)- \partial_0& \left(\partial_0 A_i -\partial_i A_0 \right)
+\frac{1}{a^2}\partial_j \left(\partial_j A_i - \partial_i A_j\right)
\\
&+  \frac{4 M_{\rm P}^2}{\xi_R \Lambda^2 a} \partial_0 \left( F(\varphi^I)e^{\sqrt{\frac{2}{3}}\, \frac{\varphi}{M_{\rm P}}} \right)  \epsilon^{i jk}
\left(\partial_j A_k - \partial_k A_j\right) =0.\label{eq:Aspace}
\end{aligned} \end{equation}
We move to conformal time $d\tau = dt/a(t)$ such that the line element becomes $ds^2 = a^2(\tau) \eta_{\mu\nu} dx^\mu dx^\nu$. Hence, performing the replacements $A_0 \to A_0 / a$, $\partial_0 \to \partial_\tau / a$, we find that the above two equations written in comoving coordinates are
\begin{align}
&-\partial_i\left(\partial_i A_0 -A_i'\right) = 0,\label{eq:Aoem-timeconf} \\
&-\partial_0 \left(\partial_0 A_i -\partial_i A_0 \right)
+\partial_j \left(\partial_j A_i - \partial_i A_j\right) 
+  \frac{4 M_{\rm P}^2}{\xi_R \Lambda^2} \partial_\tau \left( F(\varphi^I)e^{\sqrt{\frac{2}{3}}\, \frac{\varphi}{M_{\rm P}}} \right) \epsilon^{i jk}
\left(\partial_j A_k - \partial_k A_j\right) =0. \label{eq:Aoem-spaceconf}
\end{align}
As before, we can transform to momentum space by
\begin{align}
&A_0(x^\mu) = \int \frac{d^3k}{\left(2\pi\right)^{3}} \widetilde{A}_0(\tau,\vb{k}) e^{-i \vb{k}\cdot\vb{x}},\hspace{1.5cm}
A_i(x^\mu) = \int \frac{d^3k}{\left(2\pi\right)^{3}} \widetilde{A}_i(\tau,\vb{k}) e^{-i \vb{k}\cdot\vb{x}},\label{eq:Amomdecomp}
\end{align}
where the $\widetilde{\vb{A}}$ field can be written in terms of transverse and longitudinal components as
\begin{align}
\widetilde{\vb{A}}(\tau,\vb{k}) =  \sum_{\lambda=\pm,L} \widetilde{A}^{\lambda}(\tau,\vb{k})\ \hat{\epsilon}_A^\lambda(\vb{k}), \label{A:pol}
\end{align}
with
\begin{align}
i\vb{k}\cdot\hat{\epsilon}_A^\pm(\vb{k}) = 0, \hspace{7mm}i\vb{k}\cdot\hat{\epsilon}_A^L(\vb{k}) = |\vb{k}|=k, \hspace{7mm}i\vb{k}\times\hat{\epsilon}_A^\pm(\vb{k}) = \pm k \ \hat{\epsilon}_A^\pm(\vb{k}), \hspace{7mm} \hat{\epsilon}_A^\lambda(\vb{k})^\ast=   \hat{\epsilon}_A^\lambda(-\vb{k}).\label{def:helical-basis-A}
\end{align}
We are free to choose any gauge for the EM field regardless of our choice of gauge for the massive gauge bosons. Choosing the Coulomb gauge for the EM field, we have $\partial_j A^j = \frac{1}{a^2} \partial_j A_j = 0$. This reduces in momentum space to $i\vb{k}\cdot\widetilde{\vb A}=0$, and removes one degree of freedom from the EM field. Utilizing Eq.~\eqref{def:helical-basis-A}, the gauge condition $i\vb{k}\cdot\widetilde{\vb A}=0$ translates to $\widetilde{A}^L(t,\vb{k})=0$, i.e. the longitudinal component of the photon vanishes. Further, inserting the gauge condition $\widetilde{A}^L(t,\vb{k})=0$  in Eq.~\eqref{eq:0A1}, we find that $A_0 = 0$, so that the photon is left with two independent (transverse) degrees of freedom. This is unchanged by the presence of $F\widetilde F$ term as expected. Notice the stark difference between the photon and the massive gauge bosons discussed earlier. In the former case the constraint equations from the gauge condition render the Goldstone bosons $\phi_2$, $\phi_3$ and $\phi_3$ dynamical, while for the photon only two transverse degrees of freedom are dynamical. 

In momentum space and conformal time, Eq.~\eqref{eq:Aoem-spaceconf} reads
\begin{align}
&\bigg[\widetilde{\vb{A}}'' + k^2 \widetilde{\vb{A}} \bigg]
+  \frac{8i M_{\rm P}^2}{\xi_R \Lambda^2} \partial_\tau \left( F(\varphi^I)e^{\sqrt{\frac{2}{3}}\, \frac{\varphi}{M_{\rm P}}} \right) 
\left(\vb{k}\cross \widetilde{\vb{A}}\right)=\vb{0}.\label{eq:Amom1}
\end{align}
The EoM for the transverse components becomes
\begin{align}
\partial^2_\tau \widetilde{A}^\lambda+ (\omega_A^\lambda)^2 \widetilde{A}^\lambda = 0, \hspace{2cm} (\lambda = \pm)
\label{eq:A-EoM-tranverse-conf}
\end{align}
with
\begin{align}
(\omega_A^\lambda(\tau,k))^2= k^2  + \zeta^\lambda(\tau,k), \label{eq:photontransmode}
\end{align}
where $ \zeta^\lambda(\tau,k)$ is given by Eq.~\eqref{def:zeta},
as well as $\widetilde{A}^L =0$.

\begin{figure}[h]
\centering
\includegraphics[height = 3.4 cm]{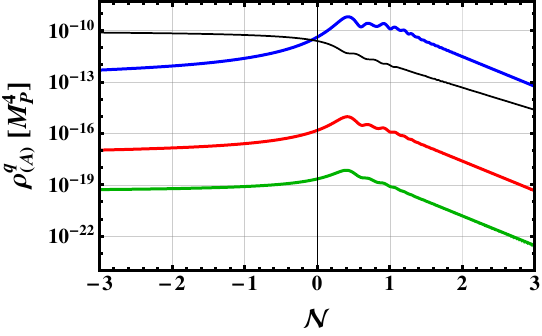}
\includegraphics[height = 3.4 cm]{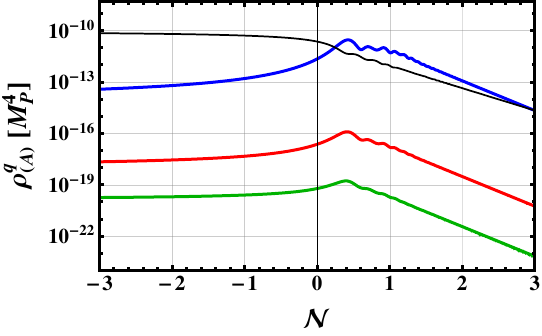}
\includegraphics[height = 3.4 cm]{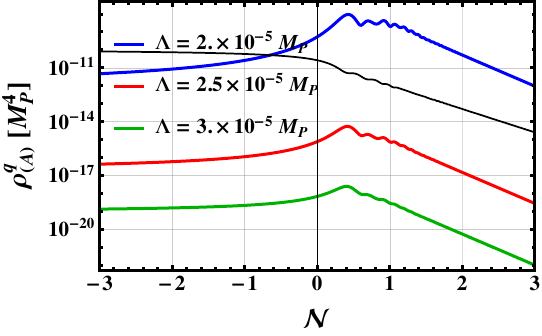}
\caption{The energy densities $\rho^q_{A}$ for various values of $\Lambda$ and all the benchmark points. The black line displays the background energy density $\rho_{\mathrm{inf}}$.
\label{plot:rhoA-All-Lambda}}
\end{figure}

In order to quantize the EM fields, we first integrate the photon part of the Lagrangian by parts to get the action quadratic in the fields
\begin{align}
S^\lambda_A = \int d\tau \,\mathcal{L}^\lambda_A = \int \, d\tau \,\frac{d^3k}{(2\pi)^3}\, \bigg[\frac{1}{2} |\partial_\tau  \widetilde{A}^\lambda|^2
- \frac{1}{2}(\omega_A^\lambda)^2  |\widetilde{A}^\lambda|^2\bigg].
\end{align}
The canonical momentum of the transverse modes are 
\begin{align}
\hat{\pi}^\lambda_A(\tau,\vb{x})=\frac{\partial \mathcal{L}^\lambda_A}{\partial\left(\partial_\tau  \hat{A}^\lambda(\tau,\vb{x})\right)}=  \partial_\tau \hat{A}^{\lambda}(\tau,\vb{x}),
\end{align}
with the commutation relation expressed as
\begin{align}
\bigg[ \hat{A}^\lambda(\tau,\vb{x}), \hat{\pi}^\lambda_A(\tau,\vb{y})\bigg]= i \delta^{\lambda\lambda\prime}\delta(\vb{x}-\vb{y}). \hspace{1cm}(\lambda = \pm)
\end{align}
In momentum space these expressions become
\begin{align}
& \hat{\widetilde{\pi}}^\lambda_A(\tau,\vb{k})=  \partial_\tau \hat{\widetilde{A}}^{\lambda}(\tau,\vb{k}), \\
&\bigg[ \hat{\widetilde{A}}^\lambda(\tau,\vb{k}), \hat{\widetilde{\pi}}^\lambda_A(\tau,\vb{q})\bigg]= i (2\pi)^3 \delta^{\lambda\lambda\prime}\delta(\vb{k}+\vb{q}) \hspace{1cm}(\lambda = \pm).
\end{align}
The field operator $\hat{\widetilde{A}}^\lambda(\tau,\vb{k})$ can be written as
creation and annihilation operators
\begin{align}
\hat{\widetilde{A}}^\lambda(\tau,\vb{k}) =  u^\lambda_{k}(\tau) \hat{a}_A^\lambda(\vb{k}) + u^{\lambda*}_{k}(\tau) \hat{a}_A^{\lambda\dagger}(\vb{-k})\hspace{1cm}(\lambda = \pm)\label{eq:Aquant}
\end{align}
that obey
\begin{align}
 \bigg[\hat{a}_A^\lambda(\vb{k}),\hat{a}_A^{\lambda\prime}(\vb{q})\bigg] = 0,\hspace{1cm}\bigg[\hat{a}_A^{\dagger\lambda}(\vb{k}),\hat{a}_A^{\dagger\lambda\prime}(\vb{q})\bigg] =0,\hspace{1cm}
 \bigg[\hat{a}_A^\lambda(\vb{k}),\hat{a}_A^{\lambda'\dagger}(\vb{q})\bigg] = (2\pi)^3\delta^{\lambda\lambda\prime} \delta^3(\vb{k}-\vb{q}).
 \end{align}
Inserting Eq.~\eqref{eq:Aquant} in Eq.~\eqref{eq:A-EoM-tranverse-conf}, the mode equations of $A$ can be found as
\begin{align}
&{u^{\lambda}_{k}}''+ (\omega_A^\lambda)^2\, u^\lambda_{k} = 0 \qquad (\lambda = \pm). \label{eq:modeeomA}
\end{align}
As in the last section, we can define the physical energy density as 
\begin{align}
\rho_{A}= \frac{1}{a^4}\int d^3x \left\langle T_{00}^{A} \right\rangle= \frac{1}{a^4} \int \frac{d^3k}{(2\pi)^3} \left\langle \rho_{k,A}\right\rangle
= \frac{1}{a^4}\int dk \, \frac{k^2}{2\pi^2} \,\rho_{k,A}^{\rm{vev}}, 
\label{eq:energydensityQA}
\end{align}
where the EM energy-momentum tensor can be derived from the action \eqref{eq:actionfinal}
\begin{align}
S^{(2)}_A = \int d^3x \ d\tau \left[-  \dfrac{1}{4} \, \eta^{\mu\rho} \eta^{\nu\sigma} F_{A\mu\nu}F_{A\rho\sigma}-\frac{M_{\rm P}^2}{\xi_R\Lambda^2}
F(\varphi^I)e^{\sqrt{\frac{2}{3}}\frac{\varphi}{M_{\rm P}}}\, \epsilon^{\mu\nu\rho\sigma}  F_{A\mu\nu} F_{A\rho\sigma}\right]
\end{align}
as
\begin{align}
T_{\mu\nu}^{A} = \eta_{\mu\nu}\left[-  \dfrac{1}{4} \, \eta^{\alpha\rho} \eta^{\beta\sigma} F_{A\alpha\beta}F_{A\rho\sigma}\right]+ \eta^{\rho\sigma} F_{A\mu\rho}F_{A\nu\sigma}.
\end{align}
Moving to momentum space and the helical basis \eqref{def:helical-basis-A}, we find that the EM energy density is
\begin{align}
\rho_{k,A}^{\rm{vev}} =  \sum_{\lambda=\pm}\left[\frac{1}{2} \left|u^{\lambda\prime}_{k}\right|^2+\frac{k^2}{2} \,\left|u^{\lambda}_{k}\right|^2 \right],
\end{align}
where we used Eq.~\eqref{eq:Aquant} and the gauge conditions $A_0=0$ and $\partial_iA_i=0$.
The first term is the electric component, and the second is the magnetic one. The BD vacuum solution as before is
\begin{align}
u_{k,{\rm BD}}^{\lambda}  = \frac{e^{-i k \tau} }{\sqrt{2k}} ,\hspace{1cm} u_{k,{\rm BD}}^{\lambda\prime}  = \sqrt{\frac{k}{2}} \; e^{-i k \tau}
\label{def:vacuum-Amodes}
\end{align}
with energy density
\begin{align}
\rho_{A}^{\rm BD}  =\int dk \;  \frac{k^2}{2 \pi^2 a^4} \rho_{k,A}^{\rm BD} =\frac{1}{a^4}\int dk \; \frac{k^3}{2\pi^2}.
\end{align}
As in the other cases, the quantum EM energy density is obtained by removing the BD vacuum from the classical solution as
%%%%%%
\begin{align}
\rho^q_{A} &= \rho_{A}-\rho_{A}^{\rm BD} \label{def:quantum-energy-density-A}.
\end{align}
We display the result in Fig.~\ref{plot:rhoA-All-Lambda} for all the BPs. Preheating from photon production seems then possible only for low values of the scale $\Lambda$. We will not consider this case in this paper as for such low values of the cutoff the rate of baryon asymmetry production is too large. See also discussion in Sec.~\ref{sec:disc}.

%%%%%%%%%%%%%%%%%%%
%%%%%%%%%%%%%%%%%%%

%%%%%%%%%%%%%%%%%%%
\begin{figure}[!t]
\centering
\includegraphics[height = 3.4 cm]{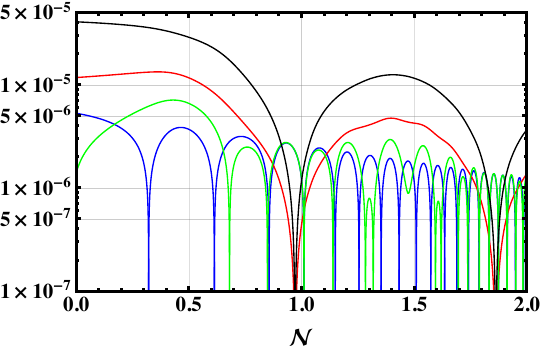}\includegraphics[height = 3.4 cm]{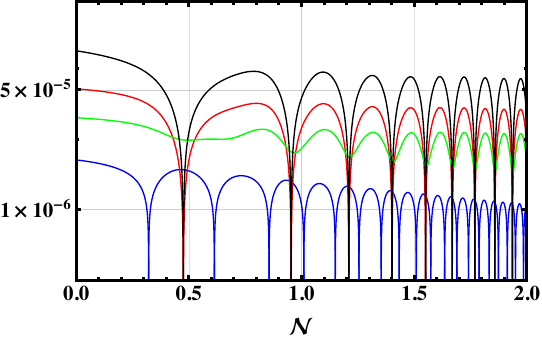}
\includegraphics[height = 3.4 cm]{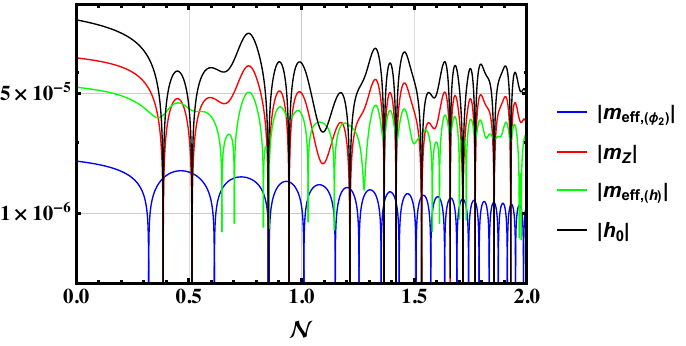}
\caption{Comparison of different masses for the three BPs after end of inflation.
\label{plot:decay-masses-ratio}}
\end{figure}
%%%%%%%%%%%%%%%%%%%

% %%%%%%%%%%%%%%%%%%%%%%%%%%%%%%%%%%%%%%%%%
%\section{Preheating and baryogenesis}
%\label{sec:results}
% %%%%%%%%%%%%%%%%%%%%%%%%%%%%%%%%%%%%%%%%%
\section{Reheating temperature}
\label{sec:reheating}
Our discussion in the preceding sections regarding preheating for different fields did not include effects such as decay and scattering of the produced particles.
These interactions are, of course, not summarized in the action of Eq.~\eqref{eq:actionfinal} since they are beyond the linearized approximation we adopted throughout this paper.
Nonetheless, these nonlinear effects may indeed dampen the strength of preheating and, in some cases may completely shut down preheating for certain species. We leave out a detailed estimation for separate work but briefly discuss their qualitative impact, in particular for those fields which display the capability of preheating the Universe.

Let us begin with the decay of the produced particles. For the case of Higgs and inflaton quanta, both Higgs and Goldstone bosons
can preheat for BP$c$ and BP$b$ but the Goldstones preheat faster. The produced Higgs as well as the
Goldstone particles may decay into SM fermions and gauge bosons. For pure Higgs inflation, it has been found that the decay of
Higgs particles into gauge bosons $ZZ,\ WW$ is kinematically disallowed~\cite{Sfakianakis:2018lzf}.
Our model is quite similar in this regard. In Fig.~\ref{plot:decay-masses-ratio}, we plot how  $|m_{{\rm eff}, (h)}|$, $|m_{\rm eff, (\phi_{2})}|$ and  $|m_Z|$ evolve in comparison to
$h_0$. It is clear that $|m_Z|$ is heavier than both $|m_{{\rm eff}, (h)}|$ and $|m_{\rm eff, (\phi_{2})}|$ for most times during preheating except when $h_0 = 0$. For the case of the Higgs in our qualitative discussion, we find that the duration when $|m_Z| < |m_{{\rm eff}, (h)}|$ is too small to deplete the energy density of the Higgs. The same holds for $|m_{\rm eff, (\phi_{3})}|$ ($|m_{\rm eff, (\phi_{2})}|$) for BP$b$ (BP$c$), where the effective masses of the respective gauge bosons are larger than those of Goldstone bosons such that decays into gauge bosons are not allowed.
The situation is different for the case of decays to fermions as discussed in Ref.~\cite{Sfakianakis:2018lzf}. For lighter fermions, the decays of the Higgs and Goldstone quanta into fermions are kinematically allowed. However, one needs a decay rate much greater than the Hubble expansion rate for the decay to efficiently deplete the energy densities. This is only possible for the heaviest fermions due to the largeness of their Yukawa couplings. However, even in this instance, as in Ref.~\cite{Sfakianakis:2018lzf}, we find the duration is too small to deplete the produced particles in our back-reactionless analysis.

\begin{figure}[!t]
\centering
\includegraphics[width=.305 \textwidth]{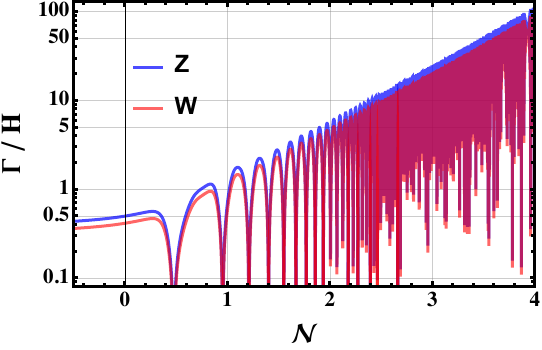}
\includegraphics[width=.32 \textwidth]{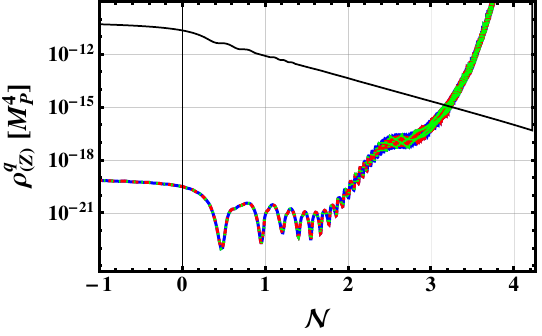}
\includegraphics[width=.32 \textwidth]{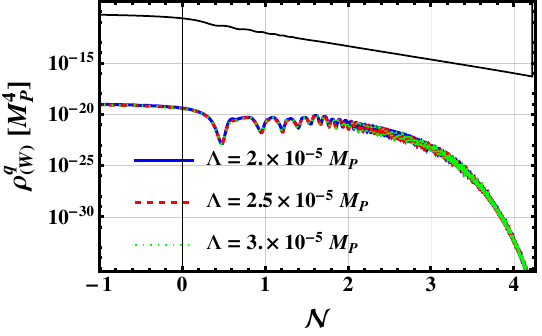}
\includegraphics[width=.32 \textwidth]{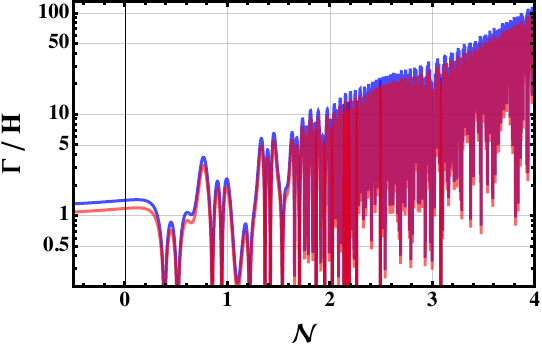}
\includegraphics[width=.32 \textwidth]{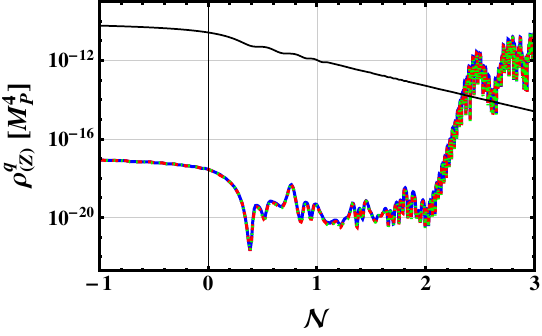}
\includegraphics[width=.32 \textwidth]{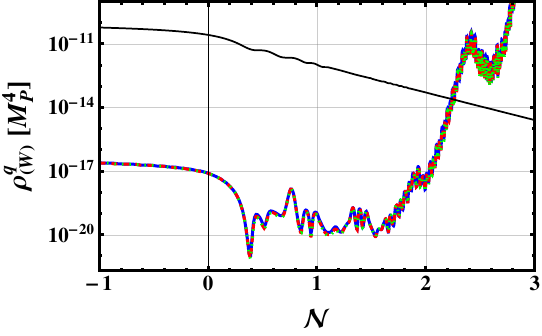}
\caption{The $\Gamma_{Z}/H$ and $\Gamma_{W}/H$ shown, in the left panels, in blue and red respectively for BP$b$ (top) and BP$c$ (bottom). The energy densities of the $Z$ (middle) and $W$ (right) taking into account the particle decay for the BP$b$ (top) and BP$c$ (bottom), see Eq.~\eqref{def:decay-factor}. The black line displays the background energy densities $\rho_{\mathrm{inf}}$ for the respective BPs.
\label{plot:decay-ZW}}
\end{figure}

The decays of the $Z$ and $W$ bosons into fermions may lead to a significant reduction of the energy densities. Similarly, the $Z$ boson can decay to two scalar bosons. To illustrate the impact of these decays, we consider $Z$ and $W$ boson decays into fermions~\cite{Garcia-Bellido:2008ycs,Bezrukov:2008ut}\footnote{Note here that $Z$ and $W$ decay rates include decay to all fermions and averaged over all polarization. We ignore the polarization averaging effect and utilize these expressions for the transverse modes.}
\begin{align}
\Gamma_Z(\tau) & = \frac{g_Z^2}{8\pi^2 \cos^2{\theta_W}} \, m_Z(\tau) \left( \frac{7}{2}- \frac{11}{3}\sin^2{\theta_W} + \frac{49}{9}\sin^4{\theta_W} \right), \\
\Gamma_W(\tau) & = \frac{3g^2}{16\pi} m_W(\tau),
\end{align}
where the time-dependent $Z$ and $W$ masses have been defined in Eqs.~\eqref{def:K_Z} and \eqref{def:KW}.
These decays may deplete the density of the produced gauge quanta if $\Gamma_{Z,W}/H \gg 1$. While these decay rates can be directly incorporated into the respective mode equations to estimate their impact, for simplicity we follow the approximate expression for the modified energy densities as in~\cite{Sfakianakis:2018lzf}
\begin{align}
\rho^q(\tau) \to \rho^q (\tau) \,\exp{-\int_{\tau_0}^{\tau} d\tau' \, \Gamma(\tau') }
\label{def:decay-factor}
\end{align}
where $\tau_0$ is the time when $\Gamma_{Z,W}/H$ becomes $\gg 1$ for the respective particles. The modified energy densities of the $W$ and $Z$ bosons are shown in Fig.~\ref{plot:decay-ZW}.
It is clear that the completion of the $Z$ preheating takes longer for the BPs, in comparison with the results shown in Fig.~\ref{plot:WZenergydensity}. For the $W$ boson, the decay may completely shut off preheating for BP$b$; completion takes longer for BP$c$.
The case of BP$a$ is more involved as for small $\Lambda$ there will be an explosive production of all the gauge fields. However, it was shown that the gauge fields will trigger the production of fermion anti-fermion pairs in the electromagnetic plasma that will strongly reduce the energy density, see e.g.~Refs.~\cite{Domcke:2018eki,Cado:2022pxk}. This backreaction (dubbed the Schwinger effect) may jeopardize any gauge preheating. Hence, we will conservatively consider that there is no preheating for the BP$a$ benchmark point.

% %%%%%%%%%%%%%%%%%%%%%%%%%%%%%%%%%%%%%%%%%
% %%%%%%%%%%%%%%%%%%%%%%%%%%%%%%%%%%%%%%%%%

\begin{figure}[!t]
\includegraphics[height=.21 \textwidth]{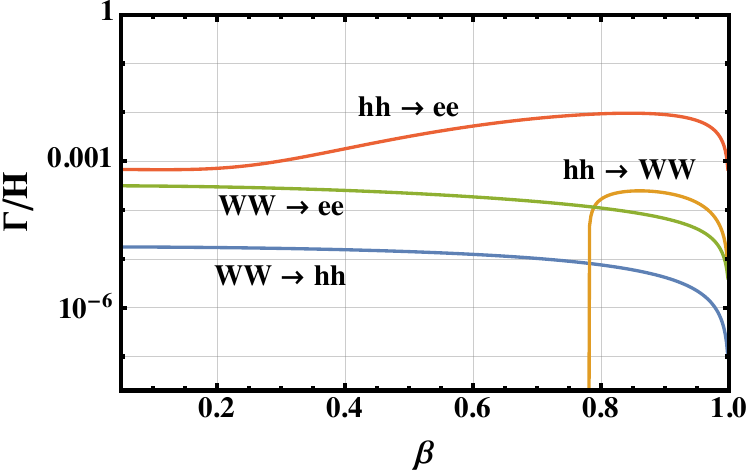}\hspace{5mm}
\includegraphics[height=.21 \textwidth]{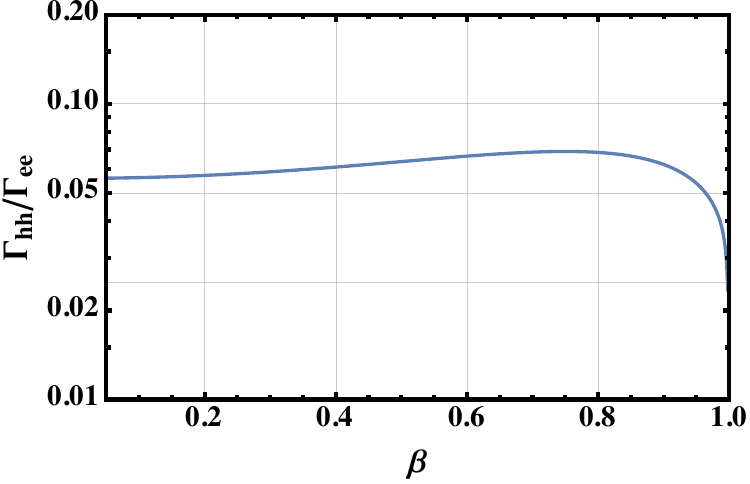}
\caption{Representative conversion rates $WW\to ee$, $WW\to hh$, and $hh\to ee$ as a function of the Hubble constant and the $W/h$ velocity $\beta$ for a typical epoch at the end of inflation characterized by $h_0\simeq 10^{-5}$$M_P$, $m_{\text{eff}, (h)} \sim  2.5 \times 10^{-5}$$M_P$, $m_W \sim  4 \times 10^{-5}$$M_P$, from Fig.~\ref{plot:decay-masses-ratio}. (Right) comparison of $WW\to hh$ and $WW\to ee$ conversion as a function of the $W$ velocity.
\label{plot:wwcomp}}
\end{figure}

Finally, while a more detailed study is required for the consideration of nonlinear effects, we turn to a qualitative discussion of rescattering and its potential relevance for preheating. In Fig.~\ref{plot:wwcomp}, we combine the effective mass information at a representative time of Fig.~\ref{plot:decay-masses-ratio} into a computation of representative $2\to 2$ scattering processes. The rate of particle conversion can be approximated as
\begin{equation}
\label{eq:conversion}
\Gamma = n \sigma \beta,
\end{equation}
in natural units where $\beta$ is the velocity of a representative $W$ in the plasma. We can further estimate the number density as
\begin{equation}
n_i\simeq {\rho_i\over m_i}
\end{equation}
for a particle species $i$. The energy densities for, e.g., the gauge bosons are collected in Sec.~\ref{sec:ZandW}. We obtain the cross section in Eq.~\eqref{eq:conversion} keeping the full mass, background field, and centre-of-mass dependencies. There is interesting phenomenology toward the end of inflation; $W$ particles can quickly convert into fermions and Higgs bosons, and vice versa. If the low-mass particles are sufficiently relativistic, they can convert back to vector bosons as indicated in Fig.~\ref{plot:wwcomp} (left). This is also true for fermions with sufficient energy very close to unity to transfer kinetic energy to heavy particle creation (these processes are not shown) as well as for any other crossed process shown in Fig.~\ref{plot:wwcomp}. Compared to the change of occupation number resulting from the particle decay discussed above, however, we see that particle conversion turns out to be insignificant and will not quantitatively impact the preheating implications that are derived from the particle decay in isolation. Again this is consistent with the findings of Ref.~\cite{Sfakianakis:2018lzf}.

%%%%%%%%%%%
%%%%%%%%%%%

%%%%%%%%%%%
\begin{table}[t!]
\begin{tabular}{|c |c| c| c| c | c | c| c | c |c| c| c| c}
    \hline
	BP  & preheating field(s)    & $\mathcal{N}_{\rm rh}$ &  $a_{\rm rh}$   & $\rho_{\rm rh} \; [M_{\rm P}^4]$ &  $T_{\rm rh}$ [GeV]\\
   \hline
        $a$ & --  & --&  -- &  --& -- \\
        $b$ & $\phi_3,\phi_4$  & 3.05  &  21 & $5 \times 10^{-14}$  & $5 \times 10^{14}$ \\
        $c$ & $\phi_2$ & 1.83  &   6.2   &  $10^{-13}$   &  $6 \times 10^{14}$ \\
	\hline
	\end{tabular}
	\caption{Preheating summary in the benchmark points chosen for our analysis. Values given are approximate.}
	\label{table:reheating}
\end{table}
%%%%%%%%%%%

In the following, for definiteness, we take the completion of preheating exactly when $\rho_{\mathrm{inf}}=\rho^{(q)}_{(X)}$ (with $X$ is any fields), corresponding to a cosmic time $t_{\rm rh}$,
i.e.~$a_{\rm rh}$, i.e.~$\mathcal N_{\rm rh}$.
The energy density at the time $a_{\rm rh}$, namely $\rho_{\rm rh}$ is therefore identified with the thermal bath energy density
\begin{align}
\rho_{\mathrm{inf}} (a_{\rm rh})\equiv\rho_{\rm rh}=\frac{g_{\rm rh} \pi^2}{30} \; T_{\rm rh}^4,
\end{align}
from which we can extract the (p)reheating temperature $T_{\rm rh}$ relevant for baryogenesis.
Note, as we stressed before, our linearized results here neglect back-reaction of the excited modes onto the background condensates; this limitation should be kept in mind.
%Therefore, our result has some uncertainties.
We summarize which fields can preheat the Universe individually and table the corresponding value $T_{\rm rh}$ in Table~\ref{table:reheating}. The impact of $T_{\rm rh}$
on the geometric baryogenesis will be discussed shortly.

% %%%%%%%%%%%%%%%%%%%%%%%%%%%%%%%%%%%%%%%%%
\section{Baryogenesis}
\label{sec:baryogensis}
% %%%%%%%%%%%%%%%%%%%%%%%%%%%%%%%%%%%%%%%%%
The baryon asymmetry of the Universe is characterized, in its entropic version, by the parameter
\begin{align} 
\eta_B = \frac{n_b-n_{\bar{b}}}{s} ,
\label{def:etaB}
\end{align}
where $n_b-n_{\bar{b}}$ is the difference between the baryon and anti-baryon number density and $s$ the comoving entropy density of the SM plasma. The best fit of CMB anisotropy puts the contraint~\cite{Planck:2018vyg}
\begin{align} 
\eta_B = (8.70\pm 0.11) \times 10^{-11} \hspace{1.cm} (95\text{\% CL}).
\end{align}
Besides and for completeness, the observed abundances of all the Big Bang Nucleosynthesis (BBN) isotopes today coincide within the range value~\cite{ParticleDataGroup:2022pth}
\begin{align} 
 8.2\times 10^{-11}  \leqslant \eta_B \leqslant 9.2 \times 10^{-11}\hspace{1.cm} (95\text{\% CL}),
 \end{align}
as all the light element abundances depend on $\eta_B$, compatible with the CMB measurement.

The SM Higgs mass measurement of $125~\text{GeV}$ favors a smooth electroweak crossover at temperatures around $180~\text{GeV} \gtrsim T \gtrsim 130~\text{GeV}$. At first glance, this might jeopardize an electroweak baryogenesis scenario as the Sakharov conditions impose that baryon number and C/CP-violating processes occur in a non-equilibrium environment~\cite{Sakharov:1967dj}. However, by carefully 
analyzing the transport equations for all SM species during the EWPT, it was shown in Refs.~\cite{Kamada:2016eeb,Kamada:2016cnb}
that the difference between chirality sources and sphaleron washout yields an out-of-equilibrium configuration even for the crossover;
the chiral anomaly of the SM provides a baryon+lepton violating process, which is then sufficient to generate the BAU.
The anomaly expresses the fact that the $B+L$ charges, the $U(1)_Y$ helicity, and the weak sphaleron are connected as 
\begin{align} 
\Delta N_B = \Delta N_L = N_g \left(  \Delta N_{\rm CS} - \dfrac{g'^2}{16\pi^2}\; \Delta \mathcal{H}_Y\right),
\label{def:anomaly:vessel}
\end{align}
where the factor $N_g=3$ is the number of fermion generations. Under the thermal fluctuation of the $SU(2)_L$ gauge fields, the Chern-Simons number $N_{\rm CS}$ is diffusive, resulting in the rapid washout of both lepton $N_L$ and baryon $N_B$ numbers. In contrast, %On the contrary, 
a helical primordial magnetic field acts as a source, and a net baryon asymmetry can remain after the EW phase transition. These two observations open the possibility of a baryogenesis mechanism within the SM electroweak theory although physics beyond the SM is needed to provide a strong enough CP violation at a higher-dimensional operator level. Indeed, the SM CP-violating term from the CKM matrix phase is too small to induce a significant baryon asymmetry at a low energy scale. In our scenario, the dim-6 interaction term $\epsilon^{\mu\nu\rho\sigma}B_{\mu\nu} B_{\rho\sigma} R_J$ fulfills this role. 
%
%In any case, the proper modelling of the epoch 160~GeV~$\gtrsim T \gtrsim 130$~GeV is critical to an accurate prediction of the relic BAU. 

The proper modelling of the epoch $160~\text{GeV}$~$\gtrsim T \gtrsim 130~\text{GeV}$ is critical for an accurate prediction of the relic BAU. 
We will rely on a mechanism that introduces a time-dependent (temperature-dependent) weak mixing angle $\theta_W(T)$ which enters an additional source of the baryon number into the kinetic equation, see Refs.~\cite{Kamada:2016eeb,Kamada:2016cnb}.
The angle behavior is confirmed by analytic calculations~\cite{Kajantie:1996qd}, and numerical lattice simulations~\cite{DOnofrio:2015gop}. 
We follow Refs.~\cite{Kamada:2016cnb,Jimenez:2017cdr} and model it with a smooth step function
\begin{align}
\cos^2 \theta_W  =  \frac{g^2}{g_Z^2}  +  \frac{g'^2}{2g_Z^2}\left(1 + \tanh \left[\frac{T - T_{\rm step}}{ \Delta T} \right] \right) ,
 \label{SmoothStepFunction}
 \end{align}
which, for $155~\text{GeV}  \lesssim T_{\rm step}  \lesssim 160~\text{GeV} $ and $5~\text{GeV} \lesssim \Delta T  \lesssim 20~\text{GeV} $, describes reasonably well the analytical and lattice results for the temperature dependence.
%
%Consequently, it is possible to generate the observed BAU from a helical hypermagnetic field that was generated prior the EW crossover.
We will now present the main lines of this mechanism and refer the reader to Refs.~\cite{Kamada:2016eeb,Kamada:2016cnb,Domcke:2019mnd,Cado:2021bia,Cado:2022evn,Cado:2023zbm,Cado:2023jty} and references therein for further background details.

The Boltzmann equation for the baryon-to-entropy ratio $\eta_B$ reads \cite{Kamada:2016cnb}
\begin{align}
\frac{d \eta_B}{d x}\, =\, -\frac{111}{34} \gamma_{W, \rm sph} \, \eta_B \, + \, \frac{3 g_Z^2}{16 \pi^2} \, \sin(2 \theta_W) \, \frac{d \theta_W}{d x}  \, \frac{\mathcal{H}_Y}{s} ,\label{BoltzmannEquation}
\end{align}
where $x = T/H(T)$ and $\mathcal{H}_Y$ is the hypermagnetic helicity that is initially present.
Furthermore, $\gamma_{W, \rm sph}=6\,\Gamma_{W \rm sph}/T^4$ is the dimensionless transport coefficient for the EW sphaleron, which, for temperatures $T <  161~\text{GeV}$, is found from lattice simulations to be~\cite{DOnofrio:2014rug}
\begin{align}
\gamma_{W \rm sph} \, \simeq \,  {\rm exp} \left(-147.7 + 107.9 \; \frac{T}{130~\text{GeV}}\right)  . 
\end{align}
 
The Boltzmann equation \eqref{BoltzmannEquation} has been numerically solved in \cite{Kamada:2016cnb} and the baryon-to-entropy ratio $\eta_B$ was found to become frozen, i.e.~$d\eta_B / dx=0$, at a temperature $T \simeq 135~\text{GeV}$. As expected, this is close to $T \simeq  130~\text{GeV}$ at which EW sphalerons freeze out. 
Setting the RHS of Eq.~\eqref{BoltzmannEquation} to zero and solving for $\eta_B$ yields
\begin{align}
\eta_B = \frac{255}{592} \frac{g_Z^2}{\pi^3\sqrt{10g_\ast}} \, \frac{\mathcal{H}_Y}{(T_{\rm rh} a_{\rm rh})^3} \left.  \frac{ f_{\theta_W}}{M_{\rm P}} \frac{T}{\gamma_{W\rm sph}}  \right|_{T=135~\text{GeV}},
\label{eq:etaB-technic}
\end{align}
where we used that 
\begin{align}
 s= \frac{2\pi^2 }{45} \; g_{\ast} (T_{\rm rh} a_{\rm rh})^3, \hspace{1cm}  g_{\ast}  = 106.75.
 \end{align}
The parameter $f_{\theta_W} $ encodes all the details on the EWPT dynamics with significant uncertainties
\begin{align}
f_{\theta_W}  = -\sin (2 \theta_W) \, \dfrac{d\theta_W}{d\log T}\bigg\rvert_{T=135~\text{GeV}}, \hspace{1.cm} 5.6 \times 10^{-4} \lesssim f_{\theta_W}  \lesssim 0.32.
\label{eq:ftheta}
\end{align}

\begin{figure}[!t]
\centering
\includegraphics[height = 3.4 cm]{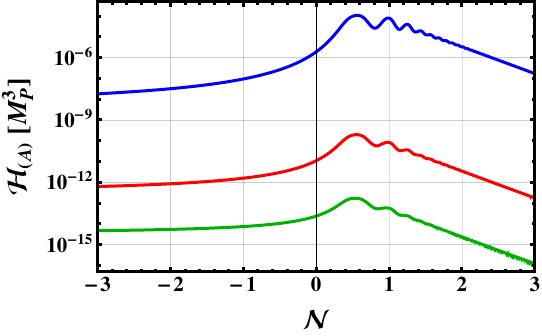}
\includegraphics[height = 3.4 cm]{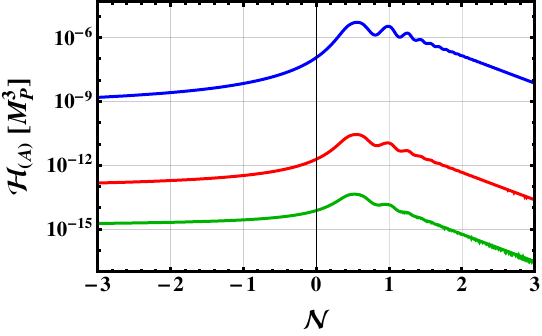}
\includegraphics[height = 3.4 cm]{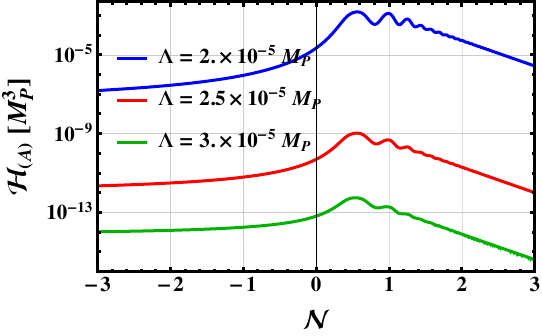}
\caption{The helicity $\mathcal{H}_{A}$ for various values of $\Lambda$ and all the benchmark points BP$a$, BP$b$ and BP$c$, from left to right.
\label{plot:HelA-All-Lambda}}
\end{figure}

Provided that the magnetic induction prevails over the dissipation effects in the plasma between reheating and the EWPT (see hereafter), we can estimate the hypermagnetic helicity at the start of the EWPT as
\begin{align}
\mathcal{H}_Y =\mathcal{H}_{A} (a_{\rm rh}) \; \cos^2{\theta_W},
\label{eq:hyperhelicity-to-EMhelicity}
\end{align}
where $\mathcal{H}_{A}$ is the helicity of the EM field defined as
\begin{align}
\mathcal{H}_{A} =  \frac{1}{a^3}\int dk \, \frac{k^3}{2\pi^2} \sum_{\lambda=\pm} \lambda \left|u^{\lambda}_{k}\right|^2 
\label{def:helicity-A}
\end{align}
that depends on time (or alternatively on the scale factor $a$ or the $e$-folding number $\mathcal N$) and on $\Lambda$, see Fig.~\ref{plot:HelA-All-Lambda}.
The $Z$ boson contribution vanishes from $\mathcal{H}_Y $ in Eq.~\eqref{eq:hyperhelicity-to-EMhelicity} because the massive fields are screened or decay away quickly compared to the time scale on which the baryon asymmetry evolves \cite{Kamada:2016cnb}.
Because the BD solution \eqref{def:vacuum-Amodes} is the same for both helicities, i.e. $u_{k,{\rm BD}}^+=u_{k,{\rm BD}}^-$, the BD vacuum contribution to the helicity vanishes, hence no vacuum subtraction is needed.

We can read the values of $T_{\rm rh}$, $\rho_{\rm rh}$ and $a_{\rm rh}$ from Tab.~\ref{table:reheating}. BP$a$, for which preheating is not evident and the relevant quantities have to be approximated, has been discussed in detail in our previous work, see Ref.~\cite{Cado:2023zbm}. In this work, we base the baryogenesis mechanism on the preheating results detailed in the previous section, and we will, therefore, mainly focus on BP$b$ and BP$c$. Of course, in all these cases, a detailed calculation of the perturbative reheating in the $R^2$-Higgs inflation model is needed to further improve on our findings.

%%%%%%%%%%%%%%%%%%%
%%%%%%%%%%%%%%%%%%%

\begin{figure}[!t]
\centering
\includegraphics[height = 4.3 cm]{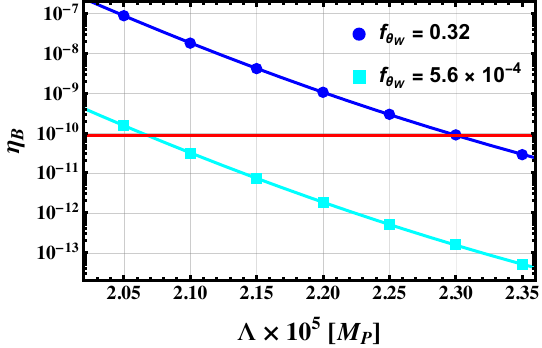} \hspace{1cm}
\includegraphics[height = 4.3 cm]{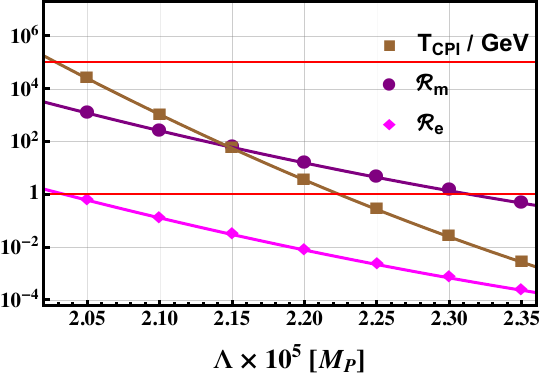} \\ \vspace{5mm}
\includegraphics[height = 4.3 cm]{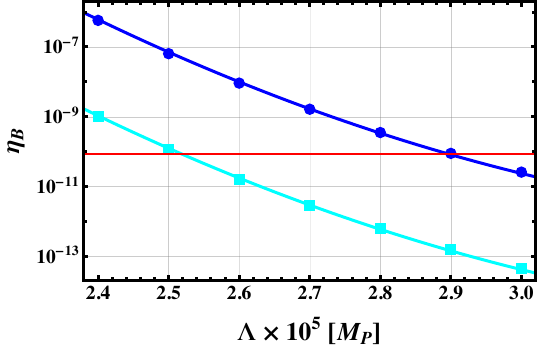} \hspace{1cm}
\includegraphics[height = 4.3 cm]{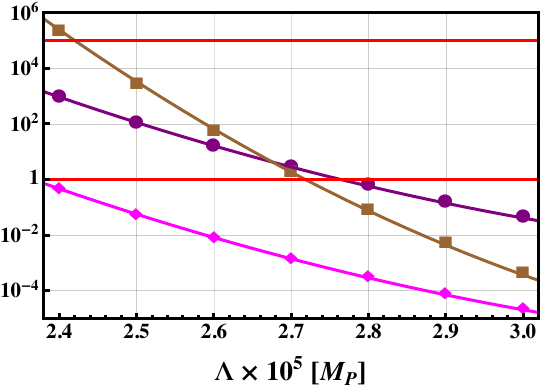} 
\caption{Baryogenesis parameter space for BP$b$ (top) and BP$c$ (bottom). On the left panels we show the asymmetry parameter in function of $\Lambda$ and $f_{\theta_W}$. The red line must be in between the light and dark blue curves to meet the observational constraint $\eta_B = 8.7 \times 10^{-11}$. On the right panels we display the two contraints, $\mathcal{R}_m > 1$ and $T_{\rm CPI}<10^5~\text{GeV}$ in function of $\Lambda$. To meet these contraints, both curves must be in between the horizontal red lines. We also display quantity $\mathcal{R}_e$ on which there is no constraint, see text tor detail.
\label{plot:baryogenesis}}
\end{figure}

The relation \eqref{eq:hyperhelicity-to-EMhelicity} holds only when the helicity is conserved between reheating and the electroweak crossover.
To guarantee that the magnetic induction dominates over dissipation in the plasma, we must require that the magnetic Reynolds number $\mathcal{R}_m$ evaluated at reheating is bigger than unity as
\begin{align}
\mathcal{R}_m \approx 2\,\alpha_Y \frac{c_\sigma}{c_\nu} \; \frac{\rho_{B_Y}^q}{\rho_{\rm rh}} \;\left(\frac{\ell_{B_Y}^q T_{\rm rh}}{a_{\rm rh}} \right)^2> 1 ,
\label{eq:magnetic-Reynolds-viscous}
\end{align}
where $c_\sigma \approx 4.5$ and $c_\nu \approx 0.01$ are respectively the conductivity  and the kinematic viscosity factors of the plasma \cite{Baym:1997gq,Arnold:2000dr} and $\alpha_Y = g^{\prime 2} / 4\pi$.
The former equation (as well as Eq.~\eqref{constraint-TCPI} hereafter) is valid only for $\mathcal{R}_e<1$, where 
\begin{align} \mathcal{R}_e \approx  \frac{2\,\alpha_Y^4}{c_\nu^2} \;  \log(\alpha_Y^{-1})^2  \;  \frac{\rho_{B_Y}^q}{\rho_{\rm rh}} \;\left(\frac{\ell_{B_Y}^q T_{\rm rh}}{a_{\rm rh}} \right)^2
\label{eq:electric-Reynolds-viscous}
\end{align}
is the electric Reynolds number. In Fig.~\ref{plot:baryogenesis} we show that this regime applies for the relevant values of $\Lambda$ in BP$b$ and BP$c$.
In the last two expressions, $\rho_{B_Y}^q$ is the quantum hypermagnetic energy density that can be computed as
\begin{align}
\rho_{B_Y}^q = \rho_{A,B}^q (a_{\rm rh}) \;  \cos^2{\theta_W} ,
\end{align}
with the EM magnetic energy
\begin{align}
\rho_{A,B} = \frac{1}{a^4}  \int d k \, \frac{k^4}{4\pi^2} \sum_{\lambda=\pm} \left|u^{\lambda}_{k}\right|^2, \hspace{1 cm} \rho_{A,B}^q= \rho_{A,B} - \rho_{A,B}^{\rm BD}, \hspace{1 cm} \rho_{A,B}^{\rm BD} =\frac{ \rho_{A}^{\rm BD}}{2},
\end{align}
and $\ell_{B_Y}^q$ is the hypermagnetic characteristic size given by 
\begin{align}
\ell_{B_Y}^q  = \frac{2\pi}{\rho_{A,B}^q\,a^3} \left[  \int d k \, \frac{k^3 }{4\pi^2} \sum_{\lambda=\pm} \left|u^{\lambda}_{k}\right|^2 -  \int d k \, \frac{k^2 }{4\pi^2}   \right] ,
\end{align}
where we performed a vacuum subtraction.
Note that $\ell_{B_Y}^q=\ell_{B}^q$.

%%%%%%%%%%%%%%%%%%%
%%%%%%%%%%%%%%%%%%%

The last constraint arises from the CP-odd term present in the magnetohydrodynamics description of the plasma. As the energy configuration in the gauge sector is more favorable than in the fermion sector \cite{Joyce:1997uy}, a helicity cancellation is induced because of the fermion asymmetry back-transformation into helical gauge fields with opposite sign. This phenomenon is called chiral plasma instability~(CPI).
Thus, one must ensure that all fermion asymmetry created alongside the helical field during inflation is erased by the action of the weak sphaleron for $10^{12}~\text{GeV}\gtrsim T\gtrsim 130~\text{GeV}$. 
Hence to preserve the helicity in the gauge sector, before the CPI can happen, we must require that $T_{\rm CPI} \lesssim 10^5~\text{GeV}$, where \cite{Joyce:1997uy, Domcke:2019mnd}
\begin{align} 
T_{\rm CPI}\approx \frac{4 \alpha_Y^5}{\pi^4 c_\sigma}  \log(\alpha_Y^{-1}) \left( \frac{2133}{481}  \right)^2 \; \frac{ \mathcal{H}_Y^2}{H(t_{\rm end}) \, T_{\rm rh}^4 \, \sqrt{a_{\rm rh}^9 \,a(t_{\rm end})}}.
\label{constraint-TCPI} 
\end{align}
Using Eqs.~\eqref{eq:etaB-technic}, \eqref{eq:magnetic-Reynolds-viscous}, \eqref{constraint-TCPI} and the values from Table~\ref{table:reheating}, we display in Fig.~\ref{plot:baryogenesis} the baryogenesis parameter space. The following ranges on $\Lambda$ meet all the constraints and hence yield a successful BAU:
\begin{equation} \begin{aligned}
2.07 \times 10^{-5} M_{\rm P} \ \lesssim \ \Lambda \ \lesssim \ 2.30 \times 10^{-5} M_{\rm P} \hspace{1cm} \text{for BP}b, \\
2.52 \times 10^{-5} M_{\rm P} \ \lesssim \ \Lambda \ \lesssim \ 2.76 \times 10^{-5} M_{\rm P} \hspace{1cm}  \text{for BP}c.
\end{aligned} \end{equation}
We observe that the smaller the reheating temperature, the smaller the coupling $\Lambda$ needs to be to achieve the BAU. This is in agreement with our result for the BP$a$ in Ref.~\cite{Cado:2023zbm} although there the reheating temperature was left as a free parameter.

% %%%%%%%%%%%%%%%%%%%%%%%%%%%%%%%%%%%%%%%%%
\section{Summary and Outlook}
\label{sec:disc}
In this work, we studied the implications of the preheating on gravity assisted baryogenesis in $R^2$-Higgs inflation, namely how preheating can impact on baryogenesis at the electroweak crossover from the production of helical hypermagnetic fields. To this end, we adopted the doubly-covariant formalism for both inflationary dynamics and gauge field production. We derived the equations of motion and energy densities for the inflaton, Higgs background fields, and relevant perturbations at linear order. This includes the inflationary fields, the $W^\pm$, $Z$ bosons, the photon, and the three Goldstone fields. The Coulomb gauge was used, as the unitary gauge becomes ill-defined at Higgs zero-crossings. Hence, the Goldstone bosons remained dynamical in our discussion. The preheating is governed by the field-space manifold, the dynamics of the background condensates, the respective effective masses and the coupled metric perturbations.

We primarily focused on $R^2$  and mildly mixed $R^2$-Higgs-like regimes, however expressions and the formalism can be applied to other regimes of $\xi_R$ and $\xi_H$.
We highlight different phenomenological possibilities by identifying three benchmark points: a deep $R^2$-like scenario with $\xi_H \approx 10^{-3}$ (BP$a$), $\xi_H \approx 1$ (BP$b$) and a mixed $R^2$-Higgs scenario with $\xi_H \approx 10$ (BP$c$) (see Tab.~\ref{parmeterchoices}). We find that the Higgs quanta and transverse modes of the $W$ boson can preheat the Universe for BP$c$ for $\xi_H \approx 10$, while the $Z$ boson can provide successful preheating for both BP$b$ and BP$c$. The Goldstone sector can also preheat for BP$b$ and BP$c$.
We find that for both BP$b$ and BP$c$, the Goldstone bosons can preheat the Universe faster than any other field: at $\mathcal N \approx \,3 \,$ and $\mathcal N \approx \,1.8 \,$, respectively (see Fig.~\ref{plot:rhoG}). In all cases, preheating never happens for BP$a$ unless $\Lambda$ is small. We remark that our results for preheating are in good agreement with previous studies in this model~\cite{He:2018mgb,He:2020ivk}, however, we find that $\xi_H \approx 1$ is also sufficient for preheating of the $\phi_3$ field at $\mathcal N \approx \,3$. This is caused predominantly by the spikes in $\omega^2_{(I)}$ due to the presence of the $\mathcal{E}_{(I)}$ term for the Goldstones as also discussed in Ref.~\cite{Ema:2016dny}.

We find that the value of $\Lambda$, required for baryogenesis does depend on $\xi_H$ indirectly via the reheating temperature.
We identify a window around $\Lambda \sim 2.2 \, (2.6) \times 10^{-5}\, M_{\rm P}$ for $\xi_H \approx 1 \, (10)$ where the observed baryon asymmetry of the Universe can be achieved. 
As preheating happens earlier for larger $\xi_H$, leading to larger reheating temperature, a larger value of $\Lambda$ is required for successful baryogenesis. However, our analysis also reveals that sufficiently small values of $\Lambda$ could lead to gauge preheating as found in Fig.~\ref{plot:WZenergydensity} and Fig.~\ref{plot:rhoA-All-Lambda} for BP$a$. Nevertheless, based on earlier findings from Ref.~\cite{Cado:2023zbm} and pending further investigation into fermion-gauge boson backreaction, we expect the inclusion of the Schwinger effect to significantly impact our results. Importantly, incorporating the Schwinger effect is unlikely to alter the successful sourcing of baryon number density (see also Refs.~\cite{Cado:2022pxk, Cado:2023zbm}).
Moreover, the relevance of the Schwinger effect dramatically depends on the value of the fermion masses at high values of the background Higgs field, and so on the mechanism for the generation of fermion masses at high scales. For instance a particular Froggatt-Nielsen mechanism was presented in Ref.~\cite{Cado:2023gan} where there is no Schwinger effect for the Standard Model at inflationary scales, while reproducing the fermion spectrum at electroweak scales. In that case, small values of $\Lambda$ are disfavored as they tend to overproduce the baryon asymmetry.
 
Our work systematically builds upon previous studies~\cite{Cado:2022evn,Cado:2023gan,Cado:2023zbm} by explicitly computing, for the first time, the reheating time, temperature, and energy without taking them as effective parameters in the model. One remaining uncertainty regarding the baryogenesis mechanism, which we leave for future work, concerns the specific dynamics of the electroweak crossover, particularly the evolution of the weak mixing angle from zero to its low-energy value. In the present analysis, we ensured that the helicity generated during reheating is preserved until the electroweak scale by carefully considering the plasma's Reynolds number and verifying that the chiral plasma instability does not impact our results. We also remark that for our preheating dynamics, we adopted a naive effective approach related to how decays impact the produced quanta instead of incorporating them in the EoMs directly. Furthermore, we have also not discussed how the produced particles backreact on both the background condensates (see Ref.~\cite{Bezrukov:2020txg}). In addition, we have ignored fermions from the picture to a large extent; they can impact significantly via the Schwinger effect as discussed in~\cite{Cado:2022evn,Cado:2023gan,Cado:2023zbm,Cado:2022pxk}. This requires a first principle derivation of all EoMs retaining terms beyond the linear order in this doubly covariant formalism including fermions and gauge bosons. This is beyond the scope of the current work and we leave a dedicated analysis of non-linear effects for future work.

%%%%%%%%%%%%%%%%%%%%%%%%%%%%%%%%%%%%%%%%%%
\subsection*{Acknowledgments}
We thank Evangelos Sfakianakis for helpful discussions.
YC acknowledges funding support from the Initiative Physique des Infinis (IPI), a research training program of the Idex SUPER at Sorbonne Universit\'{e}. 
CE is supported by the UK Science and Technology Facilities Council (STFC) under grant ST/X000605/1, the Leverhulme Trust under Research Project RPG-2021-031 and Research Fellowship RF-2024-300$\backslash$9, and by the Institute for Particle Physics Phenomenology Associateship Scheme.
The work of MQ is supported by the grant PID2023-146686NB-C31 funded by MICIU/AEI/10.13039/501100011033/ and by FEDER, EU. 
IFAE is partially funded by the CERCA program of the Generalitat de Catalunya.

%%%%%%%%%%%%%%%%%%%%%%%%%%%%Appendix
%%%%%%%%%%%%%%%%%%%%%%%%%%%%%%%%%%%%%%%%%
\appendix

%%%%%%%%%%%%%%%%%%%%%%%%%%%%%%%%%%%%%
\section{Gauge boson spectra}\label{app:gaugegoldspectra}
%%%%%%%%%%%%%%%%%%%%%%%%%%%%%%%%%%%%%%%%%%%%%%%%%%%%%%%%%%%%%%
We provide the spectra for the $Z$, $W$ and photon for illustration in comparison to corresponding BD-spectra. This is utilized to evaluate
the corresponding energy densities for the respective fields. See main text for details.

%%%%%%%%%%%%%%%%%%%%%%%%%%%%%%%%%%%%%
\begin{figure}[htbp]
\centering
\includegraphics[height= 3.2cm]{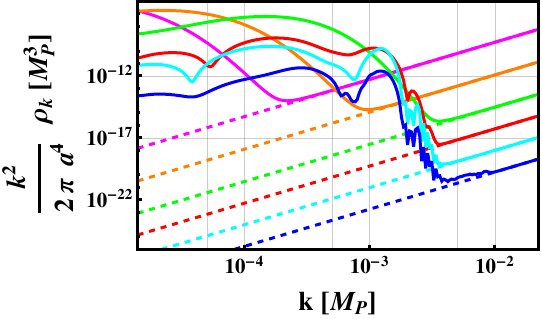}
\includegraphics[height= 3.2cm]{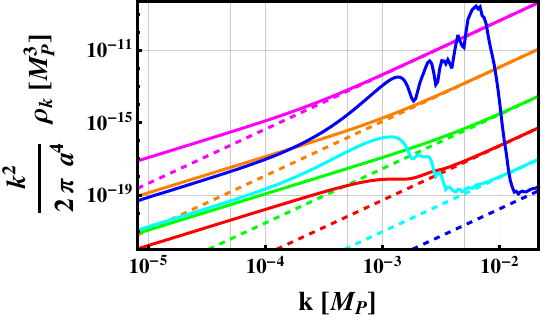}
\includegraphics[height= 3.2cm]{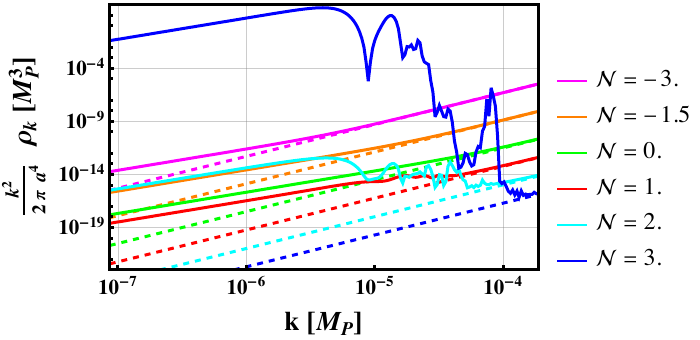}
\caption{Spectra of the transverse mode of the $Z$ boson and the corresponding BD spectra for different values of $\mathcal{N}$.
\label{plot:gaugeZ-spectra}}
\end{figure}
%%%%%%%%%%%%%%%%%%%%%%%%%%%%%%%%%%%%%

%%%%%%%%%%%%%%%%%%%%%%%%%%%%%%%%%%%%%
\begin{figure}[htbp]
\centering
\includegraphics[height= 3.2cm]{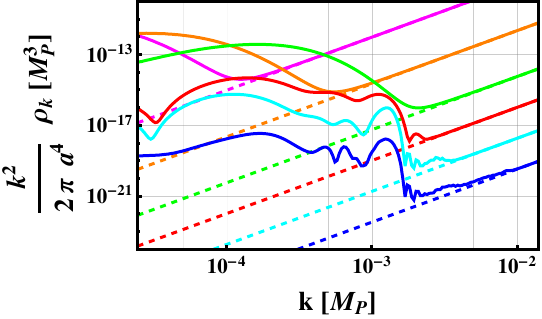}
\includegraphics[height= 3.2cm]{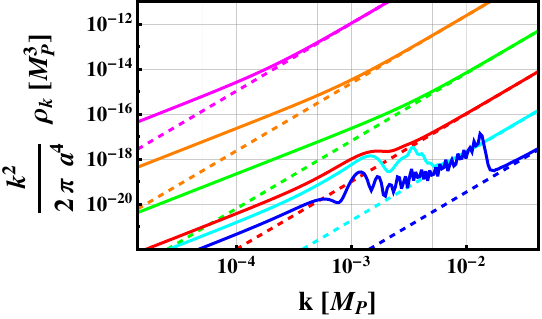}
\includegraphics[height= 3.2cm]{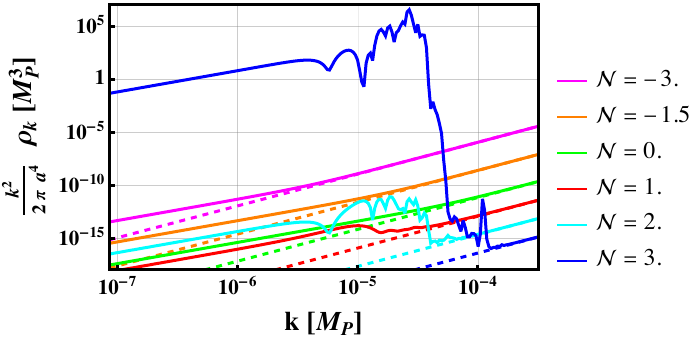}
\caption{Same figure as Fig.~\ref{plot:gaugeZ-spectra} but for the transverse $W$ boson.
\label{plot:gaugeW-spectra}}
\end{figure}
%%%%%%%%%%%%%%%%%%%%%%%%%%%%%%%%%%%%%

%%%%%%%%%%%%%%%%%%%
\begin{figure}[htbp]
\centering
\includegraphics[height= 3.2cm]{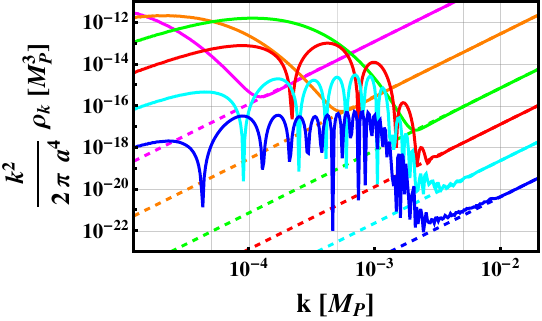}
\includegraphics[height= 3.2cm]{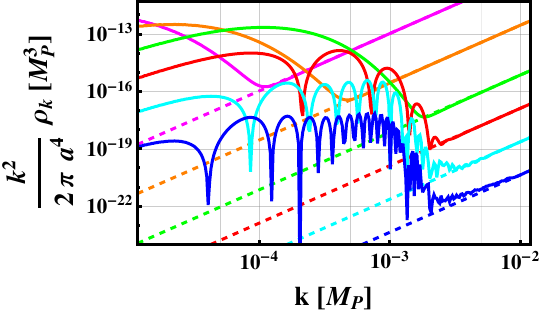}
\includegraphics[height= 3.2cm]{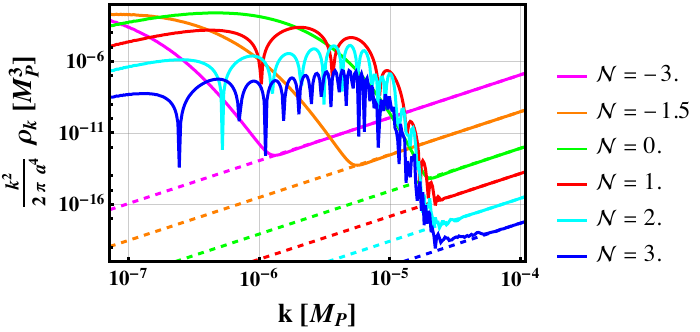}
\caption{Same figure as Fig.~\ref{plot:gaugeZ-spectra} but for the photon.
\label{plot:gaugeA-spectra}}
\end{figure}
%%%%%%%%%%%%%%%%%%%

%%%%%%%%%%%%%%%%%%%%%%%%%%%%%%%%%%%%%  
\renewcommand{\emph}{}
\bibliography{references}

%merlin.mbs apsrev4-1.bst 2010-07-25 4.21a (PWD, AO, DPC) hacked
%Control: key (0)
%Control: author (8) initials jnrlst
%Control: editor formatted (1) identically to author
%Control: production of article title (-1) disabled
%Control: page (0) single
%Control: year (1) truncated
%Control: production of eprint (0) enabled
\begin{thebibliography}{83}%
\makeatletter
\providecommand \@ifxundefined [1]{%
 \@ifx{#1\undefined}
}%
\providecommand \@ifnum [1]{%
 \ifnum #1\expandafter \@firstoftwo
 \else \expandafter \@secondoftwo
 \fi
}%
\providecommand \@ifx [1]{%
 \ifx #1\expandafter \@firstoftwo
 \else \expandafter \@secondoftwo
 \fi
}%
\providecommand \natexlab [1]{#1}%
\providecommand \enquote  [1]{``#1''}%
\providecommand \bibnamefont  [1]{#1}%
\providecommand \bibfnamefont [1]{#1}%
\providecommand \citenamefont [1]{#1}%
\providecommand \href@noop [0]{\@secondoftwo}%
\providecommand \href [0]{\begingroup \@sanitize@url \@href}%
\providecommand \@href[1]{\@@startlink{#1}\@@href}%
\providecommand \@@href[1]{\endgroup#1\@@endlink}%
\providecommand \@sanitize@url [0]{\catcode `\\12\catcode `\$12\catcode
  `\&12\catcode `\#12\catcode `\^12\catcode `\_12\catcode `\%12\relax}%
\providecommand \@@startlink[1]{}%
\providecommand \@@endlink[0]{}%
\providecommand \url  [0]{\begingroup\@sanitize@url \@url }%
\providecommand \@url [1]{\endgroup\@href {#1}{\urlprefix }}%
\providecommand \urlprefix  [0]{URL }%
\providecommand \Eprint [0]{\href }%
\providecommand \doibase [0]{http://dx.doi.org/}%
\providecommand \selectlanguage [0]{\@gobble}%
\providecommand \bibinfo  [0]{\@secondoftwo}%
\providecommand \bibfield  [0]{\@secondoftwo}%
\providecommand \translation [1]{[#1]}%
\providecommand \BibitemOpen [0]{}%
\providecommand \bibitemStop [0]{}%
\providecommand \bibitemNoStop [0]{.\EOS\space}%
\providecommand \EOS [0]{\spacefactor3000\relax}%
\providecommand \BibitemShut  [1]{\csname bibitem#1\endcsname}%
\let\auto@bib@innerbib\@empty
%</preamble>
\bibitem [{\citenamefont {Cado}\ \emph {et~al.}(2024)\citenamefont {Cado},
  \citenamefont {Englert}, \citenamefont {Modak},\ and\ \citenamefont
  {Quir\'os}}]{Cado:2023zbm}%
  \BibitemOpen
  \bibfield  {author} {\bibinfo {author} {\bibfnamefont {Y.}~\bibnamefont
  {Cado}}, \bibinfo {author} {\bibfnamefont {C.}~\bibnamefont {Englert}},
  \bibinfo {author} {\bibfnamefont {T.}~\bibnamefont {Modak}}, \ and\ \bibinfo
  {author} {\bibfnamefont {M.}~\bibnamefont {Quir\'os}},\ }\href {\doibase
  10.1103/PhysRevD.109.043026} {\bibfield  {journal} {\bibinfo  {journal}
  {Phys. Rev. D}\ }\textbf {\bibinfo {volume} {109}},\ \bibinfo {pages}
  {043026} (\bibinfo {year} {2024})},\ \Eprint
  {http://arxiv.org/abs/2312.10414} {arXiv:2312.10414 [astro-ph.CO]}
  \BibitemShut {NoStop}%
\bibitem [{\citenamefont {Shaposhnikov}(1987)}]{Shaposhnikov:1987tw}%
  \BibitemOpen
  \bibfield  {author} {\bibinfo {author} {\bibfnamefont {M.~E.}\ \bibnamefont
  {Shaposhnikov}},\ }\href {\doibase 10.1016/0550-3213(87)90127-1} {\bibfield
  {journal} {\bibinfo  {journal} {Nucl. Phys. B}\ }\textbf {\bibinfo {volume}
  {287}},\ \bibinfo {pages} {757} (\bibinfo {year} {1987})}\BibitemShut
  {NoStop}%
\bibitem [{\citenamefont {Kamada}\ and\ \citenamefont
  {Long}(2016{\natexlab{a}})}]{Kamada:2016eeb}%
  \BibitemOpen
  \bibfield  {author} {\bibinfo {author} {\bibfnamefont {K.}~\bibnamefont
  {Kamada}}\ and\ \bibinfo {author} {\bibfnamefont {A.~J.}\ \bibnamefont
  {Long}},\ }\href {\doibase 10.1103/PhysRevD.94.063501} {\bibfield  {journal}
  {\bibinfo  {journal} {Phys. Rev. D}\ }\textbf {\bibinfo {volume} {94}},\
  \bibinfo {pages} {063501} (\bibinfo {year} {2016}{\natexlab{a}})},\ \Eprint
  {http://arxiv.org/abs/1606.08891} {arXiv:1606.08891 [astro-ph.CO]}
  \BibitemShut {NoStop}%
\bibitem [{\citenamefont {Durrer}\ \emph {et~al.}(2022)\citenamefont {Durrer},
  \citenamefont {Sobol},\ and\ \citenamefont {Vilchinskii}}]{Durrer:2022emo}%
  \BibitemOpen
  \bibfield  {author} {\bibinfo {author} {\bibfnamefont {R.}~\bibnamefont
  {Durrer}}, \bibinfo {author} {\bibfnamefont {O.}~\bibnamefont {Sobol}}, \
  and\ \bibinfo {author} {\bibfnamefont {S.}~\bibnamefont {Vilchinskii}},\
  }\href {\doibase 10.1103/PhysRevD.106.123520} {\bibfield  {journal} {\bibinfo
   {journal} {Phys. Rev. D}\ }\textbf {\bibinfo {volume} {106}},\ \bibinfo
  {pages} {123520} (\bibinfo {year} {2022})},\ \Eprint
  {http://arxiv.org/abs/2207.05030} {arXiv:2207.05030 [gr-qc]} \BibitemShut
  {NoStop}%
\bibitem [{\citenamefont {Durrer}\ \emph {et~al.}(2023)\citenamefont {Durrer},
  \citenamefont {Sobol},\ and\ \citenamefont {Vilchinskii}}]{Durrer:2023rhc}%
  \BibitemOpen
  \bibfield  {author} {\bibinfo {author} {\bibfnamefont {R.}~\bibnamefont
  {Durrer}}, \bibinfo {author} {\bibfnamefont {O.}~\bibnamefont {Sobol}}, \
  and\ \bibinfo {author} {\bibfnamefont {S.}~\bibnamefont {Vilchinskii}},\
  }\href {\doibase 10.1103/PhysRevD.108.043540} {\bibfield  {journal} {\bibinfo
   {journal} {Phys. Rev. D}\ }\textbf {\bibinfo {volume} {108}},\ \bibinfo
  {pages} {043540} (\bibinfo {year} {2023})},\ \Eprint
  {http://arxiv.org/abs/2303.04583} {arXiv:2303.04583 [gr-qc]} \BibitemShut
  {NoStop}%
\bibitem [{\citenamefont {Savchenko}\ and\ \citenamefont
  {Shtanov}(2018)}]{Savchenko:2018pdr}%
  \BibitemOpen
  \bibfield  {author} {\bibinfo {author} {\bibfnamefont {O.}~\bibnamefont
  {Savchenko}}\ and\ \bibinfo {author} {\bibfnamefont {Y.}~\bibnamefont
  {Shtanov}},\ }\href {\doibase 10.1088/1475-7516/2018/10/040} {\bibfield
  {journal} {\bibinfo  {journal} {JCAP}\ }\textbf {\bibinfo {volume} {10}},\
  \bibinfo {pages} {040} (\bibinfo {year} {2018})},\ \Eprint
  {http://arxiv.org/abs/1808.06193} {arXiv:1808.06193 [astro-ph.CO]}
  \BibitemShut {NoStop}%
\bibitem [{\citenamefont {Subramanian}(2016)}]{Subramanian:2015lua}%
  \BibitemOpen
  \bibfield  {author} {\bibinfo {author} {\bibfnamefont {K.}~\bibnamefont
  {Subramanian}},\ }\href {\doibase 10.1088/0034-4885/79/7/076901} {\bibfield
  {journal} {\bibinfo  {journal} {Rept. Prog. Phys.}\ }\textbf {\bibinfo
  {volume} {79}},\ \bibinfo {pages} {076901} (\bibinfo {year} {2016})},\
  \Eprint {http://arxiv.org/abs/1504.02311} {arXiv:1504.02311 [astro-ph.CO]}
  \BibitemShut {NoStop}%
\bibitem [{\citenamefont {Durrer}\ and\ \citenamefont
  {Neronov}(2013)}]{Durrer:2013pga}%
  \BibitemOpen
  \bibfield  {author} {\bibinfo {author} {\bibfnamefont {R.}~\bibnamefont
  {Durrer}}\ and\ \bibinfo {author} {\bibfnamefont {A.}~\bibnamefont
  {Neronov}},\ }\href {\doibase 10.1007/s00159-013-0062-7} {\bibfield
  {journal} {\bibinfo  {journal} {Astron. Astrophys. Rev.}\ }\textbf {\bibinfo
  {volume} {21}},\ \bibinfo {pages} {62} (\bibinfo {year} {2013})},\ \Eprint
  {http://arxiv.org/abs/1303.7121} {arXiv:1303.7121 [astro-ph.CO]} \BibitemShut
  {NoStop}%
\bibitem [{\citenamefont {Anber}\ and\ \citenamefont
  {Sorbo}(2006)}]{Anber:2006xt}%
  \BibitemOpen
  \bibfield  {author} {\bibinfo {author} {\bibfnamefont {M.~M.}\ \bibnamefont
  {Anber}}\ and\ \bibinfo {author} {\bibfnamefont {L.}~\bibnamefont {Sorbo}},\
  }\href {\doibase 10.1088/1475-7516/2006/10/018} {\bibfield  {journal}
  {\bibinfo  {journal} {JCAP}\ }\textbf {\bibinfo {volume} {10}},\ \bibinfo
  {pages} {018} (\bibinfo {year} {2006})},\ \Eprint
  {http://arxiv.org/abs/astro-ph/0606534} {arXiv:astro-ph/0606534} \BibitemShut
  {NoStop}%
\bibitem [{\citenamefont {Bamba}(2006)}]{Bamba:2006km}%
  \BibitemOpen
  \bibfield  {author} {\bibinfo {author} {\bibfnamefont {K.}~\bibnamefont
  {Bamba}},\ }\href {\doibase 10.1103/PhysRevD.74.123504} {\bibfield  {journal}
  {\bibinfo  {journal} {Phys. Rev. D}\ }\textbf {\bibinfo {volume} {74}},\
  \bibinfo {pages} {123504} (\bibinfo {year} {2006})},\ \Eprint
  {http://arxiv.org/abs/hep-ph/0611152} {arXiv:hep-ph/0611152} \BibitemShut
  {NoStop}%
\bibitem [{\citenamefont {Bamba}\ \emph {et~al.}(2008)\citenamefont {Bamba},
  \citenamefont {Geng},\ and\ \citenamefont {Ho}}]{Bamba:2007hf}%
  \BibitemOpen
  \bibfield  {author} {\bibinfo {author} {\bibfnamefont {K.}~\bibnamefont
  {Bamba}}, \bibinfo {author} {\bibfnamefont {C.~Q.}\ \bibnamefont {Geng}}, \
  and\ \bibinfo {author} {\bibfnamefont {S.~H.}\ \bibnamefont {Ho}},\ }\href
  {\doibase 10.1016/j.physletb.2008.05.027} {\bibfield  {journal} {\bibinfo
  {journal} {Phys. Lett. B}\ }\textbf {\bibinfo {volume} {664}},\ \bibinfo
  {pages} {154} (\bibinfo {year} {2008})},\ \Eprint
  {http://arxiv.org/abs/0712.1523} {arXiv:0712.1523 [hep-ph]} \BibitemShut
  {NoStop}%
\bibitem [{\citenamefont {Anber}\ and\ \citenamefont
  {Sorbo}(2010)}]{Anber:2009ua}%
  \BibitemOpen
  \bibfield  {author} {\bibinfo {author} {\bibfnamefont {M.~M.}\ \bibnamefont
  {Anber}}\ and\ \bibinfo {author} {\bibfnamefont {L.}~\bibnamefont {Sorbo}},\
  }\href {\doibase 10.1103/PhysRevD.81.043534} {\bibfield  {journal} {\bibinfo
  {journal} {Phys. Rev. D}\ }\textbf {\bibinfo {volume} {81}},\ \bibinfo
  {pages} {043534} (\bibinfo {year} {2010})},\ \Eprint
  {http://arxiv.org/abs/0908.4089} {arXiv:0908.4089 [hep-th]} \BibitemShut
  {NoStop}%
\bibitem [{\citenamefont {Anber}\ and\ \citenamefont
  {Sabancilar}(2015)}]{Anber:2015yca}%
  \BibitemOpen
  \bibfield  {author} {\bibinfo {author} {\bibfnamefont {M.~M.}\ \bibnamefont
  {Anber}}\ and\ \bibinfo {author} {\bibfnamefont {E.}~\bibnamefont
  {Sabancilar}},\ }\href {\doibase 10.1103/PhysRevD.92.101501} {\bibfield
  {journal} {\bibinfo  {journal} {Phys. Rev. D}\ }\textbf {\bibinfo {volume}
  {92}},\ \bibinfo {pages} {101501} (\bibinfo {year} {2015})},\ \Eprint
  {http://arxiv.org/abs/1507.00744} {arXiv:1507.00744 [hep-th]} \BibitemShut
  {NoStop}%
\bibitem [{\citenamefont {Cado}\ and\ \citenamefont
  {Sabancilar}(2017)}]{Cado:2016kdp}%
  \BibitemOpen
  \bibfield  {author} {\bibinfo {author} {\bibfnamefont {Y.}~\bibnamefont
  {Cado}}\ and\ \bibinfo {author} {\bibfnamefont {E.}~\bibnamefont
  {Sabancilar}},\ }\href {\doibase 10.1088/1475-7516/2017/04/047} {\bibfield
  {journal} {\bibinfo  {journal} {JCAP}\ }\textbf {\bibinfo {volume} {04}},\
  \bibinfo {pages} {047} (\bibinfo {year} {2017})},\ \Eprint
  {http://arxiv.org/abs/1611.02293} {arXiv:1611.02293 [hep-ph]} \BibitemShut
  {NoStop}%
\bibitem [{\citenamefont {Kamada}\ and\ \citenamefont
  {Long}(2016{\natexlab{b}})}]{Kamada:2016cnb}%
  \BibitemOpen
  \bibfield  {author} {\bibinfo {author} {\bibfnamefont {K.}~\bibnamefont
  {Kamada}}\ and\ \bibinfo {author} {\bibfnamefont {A.~J.}\ \bibnamefont
  {Long}},\ }\href {\doibase 10.1103/PhysRevD.94.123509} {\bibfield  {journal}
  {\bibinfo  {journal} {Phys. Rev. D}\ }\textbf {\bibinfo {volume} {94}},\
  \bibinfo {pages} {123509} (\bibinfo {year} {2016}{\natexlab{b}})},\ \Eprint
  {http://arxiv.org/abs/1610.03074} {arXiv:1610.03074 [hep-ph]} \BibitemShut
  {NoStop}%
\bibitem [{\citenamefont {Jim\'enez}\ \emph {et~al.}(2017)\citenamefont
  {Jim\'enez}, \citenamefont {Kamada}, \citenamefont {Schmitz},\ and\
  \citenamefont {Xu}}]{Jimenez:2017cdr}%
  \BibitemOpen
  \bibfield  {author} {\bibinfo {author} {\bibfnamefont {D.}~\bibnamefont
  {Jim\'enez}}, \bibinfo {author} {\bibfnamefont {K.}~\bibnamefont {Kamada}},
  \bibinfo {author} {\bibfnamefont {K.}~\bibnamefont {Schmitz}}, \ and\
  \bibinfo {author} {\bibfnamefont {X.-J.}\ \bibnamefont {Xu}},\ }\href
  {\doibase 10.1088/1475-7516/2017/12/011} {\bibfield  {journal} {\bibinfo
  {journal} {JCAP}\ }\textbf {\bibinfo {volume} {12}},\ \bibinfo {pages} {011}
  (\bibinfo {year} {2017})},\ \Eprint {http://arxiv.org/abs/1707.07943}
  {arXiv:1707.07943 [hep-ph]} \BibitemShut {NoStop}%
\bibitem [{\citenamefont {Domcke}\ \emph {et~al.}(2019)\citenamefont {Domcke},
  \citenamefont {von Harling}, \citenamefont {Morgante},\ and\ \citenamefont
  {Mukaida}}]{Domcke:2019mnd}%
  \BibitemOpen
  \bibfield  {author} {\bibinfo {author} {\bibfnamefont {V.}~\bibnamefont
  {Domcke}}, \bibinfo {author} {\bibfnamefont {B.}~\bibnamefont {von Harling}},
  \bibinfo {author} {\bibfnamefont {E.}~\bibnamefont {Morgante}}, \ and\
  \bibinfo {author} {\bibfnamefont {K.}~\bibnamefont {Mukaida}},\ }\href
  {\doibase 10.1088/1475-7516/2019/10/032} {\bibfield  {journal} {\bibinfo
  {journal} {JCAP}\ }\textbf {\bibinfo {volume} {10}},\ \bibinfo {pages} {032}
  (\bibinfo {year} {2019})},\ \Eprint {http://arxiv.org/abs/1905.13318}
  {arXiv:1905.13318 [hep-ph]} \BibitemShut {NoStop}%
\bibitem [{\citenamefont {Cado}\ \emph {et~al.}(2021)\citenamefont {Cado},
  \citenamefont {von Harling}, \citenamefont {Mass\'o},\ and\ \citenamefont
  {Quir\'os}}]{Cado:2021bia}%
  \BibitemOpen
  \bibfield  {author} {\bibinfo {author} {\bibfnamefont {Y.}~\bibnamefont
  {Cado}}, \bibinfo {author} {\bibfnamefont {B.}~\bibnamefont {von Harling}},
  \bibinfo {author} {\bibfnamefont {E.}~\bibnamefont {Mass\'o}}, \ and\
  \bibinfo {author} {\bibfnamefont {M.}~\bibnamefont {Quir\'os}},\ }\href
  {\doibase 10.1088/1475-7516/2021/07/049} {\bibfield  {journal} {\bibinfo
  {journal} {JCAP}\ }\textbf {\bibinfo {volume} {07}},\ \bibinfo {pages} {049}
  (\bibinfo {year} {2021})},\ \Eprint {http://arxiv.org/abs/2102.13650}
  {arXiv:2102.13650 [hep-ph]} \BibitemShut {NoStop}%
\bibitem [{\citenamefont {Cado}\ and\ \citenamefont
  {Quir\'os}(2022{\natexlab{a}})}]{Cado:2022evn}%
  \BibitemOpen
  \bibfield  {author} {\bibinfo {author} {\bibfnamefont {Y.}~\bibnamefont
  {Cado}}\ and\ \bibinfo {author} {\bibfnamefont {M.}~\bibnamefont
  {Quir\'os}},\ }\href {\doibase 10.1103/PhysRevD.106.055018} {\bibfield
  {journal} {\bibinfo  {journal} {Phys. Rev. D}\ }\textbf {\bibinfo {volume}
  {106}},\ \bibinfo {pages} {055018} (\bibinfo {year} {2022}{\natexlab{a}})},\
  \Eprint {http://arxiv.org/abs/2201.06422} {arXiv:2201.06422 [hep-ph]}
  \BibitemShut {NoStop}%
\bibitem [{\citenamefont {Cado}\ and\ \citenamefont
  {Quir\'os}(2022{\natexlab{b}})}]{Cado:2022pxk}%
  \BibitemOpen
  \bibfield  {author} {\bibinfo {author} {\bibfnamefont {Y.}~\bibnamefont
  {Cado}}\ and\ \bibinfo {author} {\bibfnamefont {M.}~\bibnamefont
  {Quir\'os}},\ }\href {\doibase 10.1103/PhysRevD.106.123527} {\bibfield
  {journal} {\bibinfo  {journal} {Phys. Rev. D}\ }\textbf {\bibinfo {volume}
  {106}},\ \bibinfo {pages} {123527} (\bibinfo {year} {2022}{\natexlab{b}})},\
  \Eprint {http://arxiv.org/abs/2208.10977} {arXiv:2208.10977 [hep-ph]}
  \BibitemShut {NoStop}%
\bibitem [{\citenamefont {Salvio}\ and\ \citenamefont
  {Mazumdar}(2015)}]{Salvio:2015kka}%
  \BibitemOpen
  \bibfield  {author} {\bibinfo {author} {\bibfnamefont {A.}~\bibnamefont
  {Salvio}}\ and\ \bibinfo {author} {\bibfnamefont {A.}~\bibnamefont
  {Mazumdar}},\ }\href {\doibase 10.1016/j.physletb.2015.09.020} {\bibfield
  {journal} {\bibinfo  {journal} {Phys. Lett. B}\ }\textbf {\bibinfo {volume}
  {750}},\ \bibinfo {pages} {194} (\bibinfo {year} {2015})},\ \Eprint
  {http://arxiv.org/abs/1506.07520} {arXiv:1506.07520 [hep-ph]} \BibitemShut
  {NoStop}%
\bibitem [{\citenamefont {Ema}(2017)}]{Ema:2017rqn}%
  \BibitemOpen
  \bibfield  {author} {\bibinfo {author} {\bibfnamefont {Y.}~\bibnamefont
  {Ema}},\ }\href {\doibase 10.1016/j.physletb.2017.04.060} {\bibfield
  {journal} {\bibinfo  {journal} {Phys. Lett. B}\ }\textbf {\bibinfo {volume}
  {770}},\ \bibinfo {pages} {403} (\bibinfo {year} {2017})},\ \Eprint
  {http://arxiv.org/abs/1701.07665} {arXiv:1701.07665 [hep-ph]} \BibitemShut
  {NoStop}%
\bibitem [{\citenamefont {Pi}\ \emph {et~al.}(2018)\citenamefont {Pi},
  \citenamefont {Zhang}, \citenamefont {Huang},\ and\ \citenamefont
  {Sasaki}}]{Pi:2017gih}%
  \BibitemOpen
  \bibfield  {author} {\bibinfo {author} {\bibfnamefont {S.}~\bibnamefont
  {Pi}}, \bibinfo {author} {\bibfnamefont {Y.-l.}\ \bibnamefont {Zhang}},
  \bibinfo {author} {\bibfnamefont {Q.-G.}\ \bibnamefont {Huang}}, \ and\
  \bibinfo {author} {\bibfnamefont {M.}~\bibnamefont {Sasaki}},\ }\href
  {\doibase 10.1088/1475-7516/2018/05/042} {\bibfield  {journal} {\bibinfo
  {journal} {JCAP}\ }\textbf {\bibinfo {volume} {05}},\ \bibinfo {pages} {042}
  (\bibinfo {year} {2018})},\ \Eprint {http://arxiv.org/abs/1712.09896}
  {arXiv:1712.09896 [astro-ph.CO]} \BibitemShut {NoStop}%
\bibitem [{\citenamefont {Gorbunov}\ and\ \citenamefont
  {Tokareva}(2019)}]{Gorbunov:2018llf}%
  \BibitemOpen
  \bibfield  {author} {\bibinfo {author} {\bibfnamefont {D.}~\bibnamefont
  {Gorbunov}}\ and\ \bibinfo {author} {\bibfnamefont {A.}~\bibnamefont
  {Tokareva}},\ }\href {\doibase 10.1016/j.physletb.2018.11.015} {\bibfield
  {journal} {\bibinfo  {journal} {Phys. Lett. B}\ }\textbf {\bibinfo {volume}
  {788}},\ \bibinfo {pages} {37} (\bibinfo {year} {2019})},\ \Eprint
  {http://arxiv.org/abs/1807.02392} {arXiv:1807.02392 [hep-ph]} \BibitemShut
  {NoStop}%
\bibitem [{\citenamefont {Gundhi}\ and\ \citenamefont
  {Steinwachs}(2020)}]{Gundhi:2018wyz}%
  \BibitemOpen
  \bibfield  {author} {\bibinfo {author} {\bibfnamefont {A.}~\bibnamefont
  {Gundhi}}\ and\ \bibinfo {author} {\bibfnamefont {C.~F.}\ \bibnamefont
  {Steinwachs}},\ }\href {\doibase 10.1016/j.nuclphysb.2020.114989} {\bibfield
  {journal} {\bibinfo  {journal} {Nucl. Phys. B}\ }\textbf {\bibinfo {volume}
  {954}},\ \bibinfo {pages} {114989} (\bibinfo {year} {2020})},\ \Eprint
  {http://arxiv.org/abs/1810.10546} {arXiv:1810.10546 [hep-th]} \BibitemShut
  {NoStop}%
\bibitem [{\citenamefont {He}\ \emph {et~al.}(2019)\citenamefont {He},
  \citenamefont {Jinno}, \citenamefont {Kamada}, \citenamefont {Park},
  \citenamefont {Starobinsky},\ and\ \citenamefont {Yokoyama}}]{He:2018mgb}%
  \BibitemOpen
  \bibfield  {author} {\bibinfo {author} {\bibfnamefont {M.}~\bibnamefont
  {He}}, \bibinfo {author} {\bibfnamefont {R.}~\bibnamefont {Jinno}}, \bibinfo
  {author} {\bibfnamefont {K.}~\bibnamefont {Kamada}}, \bibinfo {author}
  {\bibfnamefont {S.~C.}\ \bibnamefont {Park}}, \bibinfo {author}
  {\bibfnamefont {A.~A.}\ \bibnamefont {Starobinsky}}, \ and\ \bibinfo {author}
  {\bibfnamefont {J.}~\bibnamefont {Yokoyama}},\ }\href {\doibase
  10.1016/j.physletb.2019.02.008} {\bibfield  {journal} {\bibinfo  {journal}
  {Phys. Lett. B}\ }\textbf {\bibinfo {volume} {791}},\ \bibinfo {pages} {36}
  (\bibinfo {year} {2019})},\ \Eprint {http://arxiv.org/abs/1812.10099}
  {arXiv:1812.10099 [hep-ph]} \BibitemShut {NoStop}%
\bibitem [{\citenamefont {Cheong}\ \emph {et~al.}(2021)\citenamefont {Cheong},
  \citenamefont {Lee},\ and\ \citenamefont {Park}}]{Cheong:2019vzl}%
  \BibitemOpen
  \bibfield  {author} {\bibinfo {author} {\bibfnamefont {D.~Y.}\ \bibnamefont
  {Cheong}}, \bibinfo {author} {\bibfnamefont {S.~M.}\ \bibnamefont {Lee}}, \
  and\ \bibinfo {author} {\bibfnamefont {S.~C.}\ \bibnamefont {Park}},\ }\href
  {\doibase 10.1088/1475-7516/2021/01/032} {\bibfield  {journal} {\bibinfo
  {journal} {JCAP}\ }\textbf {\bibinfo {volume} {01}},\ \bibinfo {pages} {032}
  (\bibinfo {year} {2021})},\ \Eprint {http://arxiv.org/abs/1912.12032}
  {arXiv:1912.12032 [hep-ph]} \BibitemShut {NoStop}%
\bibitem [{\citenamefont {He}\ \emph {et~al.}(2021)\citenamefont {He},
  \citenamefont {Jinno}, \citenamefont {Kamada}, \citenamefont {Starobinsky},\
  and\ \citenamefont {Yokoyama}}]{He:2020ivk}%
  \BibitemOpen
  \bibfield  {author} {\bibinfo {author} {\bibfnamefont {M.}~\bibnamefont
  {He}}, \bibinfo {author} {\bibfnamefont {R.}~\bibnamefont {Jinno}}, \bibinfo
  {author} {\bibfnamefont {K.}~\bibnamefont {Kamada}}, \bibinfo {author}
  {\bibfnamefont {A.~A.}\ \bibnamefont {Starobinsky}}, \ and\ \bibinfo {author}
  {\bibfnamefont {J.}~\bibnamefont {Yokoyama}},\ }\href {\doibase
  10.1088/1475-7516/2021/01/066} {\bibfield  {journal} {\bibinfo  {journal}
  {JCAP}\ }\textbf {\bibinfo {volume} {01}},\ \bibinfo {pages} {066} (\bibinfo
  {year} {2021})},\ \Eprint {http://arxiv.org/abs/2007.10369} {arXiv:2007.10369
  [hep-ph]} \BibitemShut {NoStop}%
\bibitem [{\citenamefont {He}(2021)}]{He:2020qcb}%
  \BibitemOpen
  \bibfield  {author} {\bibinfo {author} {\bibfnamefont {M.}~\bibnamefont
  {He}},\ }\href {\doibase 10.1088/1475-7516/2021/05/021} {\bibfield  {journal}
  {\bibinfo  {journal} {JCAP}\ }\textbf {\bibinfo {volume} {05}},\ \bibinfo
  {pages} {021} (\bibinfo {year} {2021})},\ \Eprint
  {http://arxiv.org/abs/2010.11717} {arXiv:2010.11717 [hep-ph]} \BibitemShut
  {NoStop}%
\bibitem [{\citenamefont {Starobinsky}(1980)}]{Starobinsky:1980te}%
  \BibitemOpen
  \bibfield  {author} {\bibinfo {author} {\bibfnamefont {A.~A.}\ \bibnamefont
  {Starobinsky}},\ }\href {\doibase 10.1016/0370-2693(80)90670-X} {\bibfield
  {journal} {\bibinfo  {journal} {Phys. Lett. B}\ }\textbf {\bibinfo {volume}
  {91}},\ \bibinfo {pages} {99} (\bibinfo {year} {1980})}\BibitemShut {NoStop}%
\bibitem [{\citenamefont {Starobinsky}(1983)}]{Starobinsky:1983zz}%
  \BibitemOpen
  \bibfield  {author} {\bibinfo {author} {\bibfnamefont {A.~A.}\ \bibnamefont
  {Starobinsky}},\ }\href@noop {} {\bibfield  {journal} {\bibinfo  {journal}
  {Sov. Astron. Lett.}\ }\textbf {\bibinfo {volume} {9}},\ \bibinfo {pages}
  {302} (\bibinfo {year} {1983})}\BibitemShut {NoStop}%
\bibitem [{\citenamefont {Vilenkin}(1985)}]{Vilenkin:1985md}%
  \BibitemOpen
  \bibfield  {author} {\bibinfo {author} {\bibfnamefont {A.}~\bibnamefont
  {Vilenkin}},\ }\href {\doibase 10.1103/PhysRevD.32.2511} {\bibfield
  {journal} {\bibinfo  {journal} {Phys. Rev. D}\ }\textbf {\bibinfo {volume}
  {32}},\ \bibinfo {pages} {2511} (\bibinfo {year} {1985})}\BibitemShut
  {NoStop}%
\bibitem [{\citenamefont {Mijic}\ \emph {et~al.}(1986)\citenamefont {Mijic},
  \citenamefont {Morris},\ and\ \citenamefont {Suen}}]{Mijic:1986iv}%
  \BibitemOpen
  \bibfield  {author} {\bibinfo {author} {\bibfnamefont {M.~B.}\ \bibnamefont
  {Mijic}}, \bibinfo {author} {\bibfnamefont {M.~S.}\ \bibnamefont {Morris}}, \
  and\ \bibinfo {author} {\bibfnamefont {W.-M.}\ \bibnamefont {Suen}},\ }\href
  {\doibase 10.1103/PhysRevD.34.2934} {\bibfield  {journal} {\bibinfo
  {journal} {Phys. Rev. D}\ }\textbf {\bibinfo {volume} {34}},\ \bibinfo
  {pages} {2934} (\bibinfo {year} {1986})}\BibitemShut {NoStop}%
\bibitem [{\citenamefont {Maeda}(1988)}]{Maeda:1987xf}%
  \BibitemOpen
  \bibfield  {author} {\bibinfo {author} {\bibfnamefont {K.-i.}\ \bibnamefont
  {Maeda}},\ }\href {\doibase 10.1103/PhysRevD.37.858} {\bibfield  {journal}
  {\bibinfo  {journal} {Phys. Rev. D}\ }\textbf {\bibinfo {volume} {37}},\
  \bibinfo {pages} {858} (\bibinfo {year} {1988})}\BibitemShut {NoStop}%
\bibitem [{\citenamefont {Bezrukov}\ and\ \citenamefont
  {Shaposhnikov}(2008)}]{Bezrukov:2007ep}%
  \BibitemOpen
  \bibfield  {author} {\bibinfo {author} {\bibfnamefont {F.~L.}\ \bibnamefont
  {Bezrukov}}\ and\ \bibinfo {author} {\bibfnamefont {M.}~\bibnamefont
  {Shaposhnikov}},\ }\href {\doibase 10.1016/j.physletb.2007.11.072} {\bibfield
   {journal} {\bibinfo  {journal} {Phys. Lett. B}\ }\textbf {\bibinfo {volume}
  {659}},\ \bibinfo {pages} {703} (\bibinfo {year} {2008})},\ \Eprint
  {http://arxiv.org/abs/0710.3755} {arXiv:0710.3755 [hep-th]} \BibitemShut
  {NoStop}%
\bibitem [{\citenamefont {Barvinsky}\ \emph {et~al.}(2008)\citenamefont
  {Barvinsky}, \citenamefont {Kamenshchik},\ and\ \citenamefont
  {Starobinsky}}]{Barvinsky:2008ia}%
  \BibitemOpen
  \bibfield  {author} {\bibinfo {author} {\bibfnamefont {A.~O.}\ \bibnamefont
  {Barvinsky}}, \bibinfo {author} {\bibfnamefont {A.~Y.}\ \bibnamefont
  {Kamenshchik}}, \ and\ \bibinfo {author} {\bibfnamefont {A.~A.}\ \bibnamefont
  {Starobinsky}},\ }\href {\doibase 10.1088/1475-7516/2008/11/021} {\bibfield
  {journal} {\bibinfo  {journal} {JCAP}\ }\textbf {\bibinfo {volume} {11}},\
  \bibinfo {pages} {021} (\bibinfo {year} {2008})},\ \Eprint
  {http://arxiv.org/abs/0809.2104} {arXiv:0809.2104 [hep-ph]} \BibitemShut
  {NoStop}%
\bibitem [{\citenamefont {Bezrukov}\ \emph {et~al.}(2011)\citenamefont
  {Bezrukov}, \citenamefont {Magnin}, \citenamefont {Shaposhnikov},\ and\
  \citenamefont {Sibiryakov}}]{Bezrukov:2010jz}%
  \BibitemOpen
  \bibfield  {author} {\bibinfo {author} {\bibfnamefont {F.}~\bibnamefont
  {Bezrukov}}, \bibinfo {author} {\bibfnamefont {A.}~\bibnamefont {Magnin}},
  \bibinfo {author} {\bibfnamefont {M.}~\bibnamefont {Shaposhnikov}}, \ and\
  \bibinfo {author} {\bibfnamefont {S.}~\bibnamefont {Sibiryakov}},\ }\href
  {\doibase 10.1007/JHEP01(2011)016} {\bibfield  {journal} {\bibinfo  {journal}
  {JHEP}\ }\textbf {\bibinfo {volume} {01}},\ \bibinfo {pages} {016} (\bibinfo
  {year} {2011})},\ \Eprint {http://arxiv.org/abs/1008.5157} {arXiv:1008.5157
  [hep-ph]} \BibitemShut {NoStop}%
\bibitem [{\citenamefont {Bezrukov}(2013)}]{Bezrukov:2013fka}%
  \BibitemOpen
  \bibfield  {author} {\bibinfo {author} {\bibfnamefont {F.}~\bibnamefont
  {Bezrukov}},\ }\href {\doibase 10.1088/0264-9381/30/21/214001} {\bibfield
  {journal} {\bibinfo  {journal} {Class. Quant. Grav.}\ }\textbf {\bibinfo
  {volume} {30}},\ \bibinfo {pages} {214001} (\bibinfo {year} {2013})},\
  \Eprint {http://arxiv.org/abs/1307.0708} {arXiv:1307.0708 [hep-ph]}
  \BibitemShut {NoStop}%
\bibitem [{\citenamefont {De~Simone}\ \emph {et~al.}(2009)\citenamefont
  {De~Simone}, \citenamefont {Hertzberg},\ and\ \citenamefont
  {Wilczek}}]{DeSimone:2008ei}%
  \BibitemOpen
  \bibfield  {author} {\bibinfo {author} {\bibfnamefont {A.}~\bibnamefont
  {De~Simone}}, \bibinfo {author} {\bibfnamefont {M.~P.}\ \bibnamefont
  {Hertzberg}}, \ and\ \bibinfo {author} {\bibfnamefont {F.}~\bibnamefont
  {Wilczek}},\ }\href {\doibase 10.1016/j.physletb.2009.05.054} {\bibfield
  {journal} {\bibinfo  {journal} {Phys. Lett. B}\ }\textbf {\bibinfo {volume}
  {678}},\ \bibinfo {pages} {1} (\bibinfo {year} {2009})},\ \Eprint
  {http://arxiv.org/abs/0812.4946} {arXiv:0812.4946 [hep-ph]} \BibitemShut
  {NoStop}%
\bibitem [{\citenamefont {Bezrukov}\ \emph
  {et~al.}(2009{\natexlab{a}})\citenamefont {Bezrukov}, \citenamefont
  {Magnin},\ and\ \citenamefont {Shaposhnikov}}]{Bezrukov:2008ej}%
  \BibitemOpen
  \bibfield  {author} {\bibinfo {author} {\bibfnamefont {F.~L.}\ \bibnamefont
  {Bezrukov}}, \bibinfo {author} {\bibfnamefont {A.}~\bibnamefont {Magnin}}, \
  and\ \bibinfo {author} {\bibfnamefont {M.}~\bibnamefont {Shaposhnikov}},\
  }\href {\doibase 10.1016/j.physletb.2009.03.035} {\bibfield  {journal}
  {\bibinfo  {journal} {Phys. Lett. B}\ }\textbf {\bibinfo {volume} {675}},\
  \bibinfo {pages} {88} (\bibinfo {year} {2009}{\natexlab{a}})},\ \Eprint
  {http://arxiv.org/abs/0812.4950} {arXiv:0812.4950 [hep-ph]} \BibitemShut
  {NoStop}%
\bibitem [{\citenamefont {Barvinsky}\ \emph {et~al.}(2012)\citenamefont
  {Barvinsky}, \citenamefont {Kamenshchik}, \citenamefont {Kiefer},
  \citenamefont {Starobinsky},\ and\ \citenamefont
  {Steinwachs}}]{Barvinsky:2009ii}%
  \BibitemOpen
  \bibfield  {author} {\bibinfo {author} {\bibfnamefont {A.~O.}\ \bibnamefont
  {Barvinsky}}, \bibinfo {author} {\bibfnamefont {A.~Y.}\ \bibnamefont
  {Kamenshchik}}, \bibinfo {author} {\bibfnamefont {C.}~\bibnamefont {Kiefer}},
  \bibinfo {author} {\bibfnamefont {A.~A.}\ \bibnamefont {Starobinsky}}, \ and\
  \bibinfo {author} {\bibfnamefont {C.~F.}\ \bibnamefont {Steinwachs}},\ }\href
  {\doibase 10.1140/epjc/s10052-012-2219-3} {\bibfield  {journal} {\bibinfo
  {journal} {Eur. Phys. J. C}\ }\textbf {\bibinfo {volume} {72}},\ \bibinfo
  {pages} {2219} (\bibinfo {year} {2012})},\ \Eprint
  {http://arxiv.org/abs/0910.1041} {arXiv:0910.1041 [hep-ph]} \BibitemShut
  {NoStop}%
\bibitem [{\citenamefont {Spokoiny}(1984)}]{Spokoiny:1984bd}%
  \BibitemOpen
  \bibfield  {author} {\bibinfo {author} {\bibfnamefont {B.~L.}\ \bibnamefont
  {Spokoiny}},\ }\href {\doibase 10.1016/0370-2693(84)90587-2} {\bibfield
  {journal} {\bibinfo  {journal} {Phys. Lett. B}\ }\textbf {\bibinfo {volume}
  {147}},\ \bibinfo {pages} {39} (\bibinfo {year} {1984})}\BibitemShut
  {NoStop}%
\bibitem [{\citenamefont {Futamase}\ and\ \citenamefont
  {Maeda}(1989)}]{Futamase:1987ua}%
  \BibitemOpen
  \bibfield  {author} {\bibinfo {author} {\bibfnamefont {T.}~\bibnamefont
  {Futamase}}\ and\ \bibinfo {author} {\bibfnamefont {K.-i.}\ \bibnamefont
  {Maeda}},\ }\href {\doibase 10.1103/PhysRevD.39.399} {\bibfield  {journal}
  {\bibinfo  {journal} {Phys. Rev. D}\ }\textbf {\bibinfo {volume} {39}},\
  \bibinfo {pages} {399} (\bibinfo {year} {1989})}\BibitemShut {NoStop}%
\bibitem [{\citenamefont {Salopek}\ \emph {et~al.}(1989)\citenamefont
  {Salopek}, \citenamefont {Bond},\ and\ \citenamefont
  {Bardeen}}]{Salopek:1988qh}%
  \BibitemOpen
  \bibfield  {author} {\bibinfo {author} {\bibfnamefont {D.~S.}\ \bibnamefont
  {Salopek}}, \bibinfo {author} {\bibfnamefont {J.~R.}\ \bibnamefont {Bond}}, \
  and\ \bibinfo {author} {\bibfnamefont {J.~M.}\ \bibnamefont {Bardeen}},\
  }\href {\doibase 10.1103/PhysRevD.40.1753} {\bibfield  {journal} {\bibinfo
  {journal} {Phys. Rev. D}\ }\textbf {\bibinfo {volume} {40}},\ \bibinfo
  {pages} {1753} (\bibinfo {year} {1989})}\BibitemShut {NoStop}%
\bibitem [{\citenamefont {Fakir}\ and\ \citenamefont
  {Unruh}(1990)}]{Fakir:1990eg}%
  \BibitemOpen
  \bibfield  {author} {\bibinfo {author} {\bibfnamefont {R.}~\bibnamefont
  {Fakir}}\ and\ \bibinfo {author} {\bibfnamefont {W.~G.}\ \bibnamefont
  {Unruh}},\ }\href {\doibase 10.1103/PhysRevD.41.1783} {\bibfield  {journal}
  {\bibinfo  {journal} {Phys. Rev. D}\ }\textbf {\bibinfo {volume} {41}},\
  \bibinfo {pages} {1783} (\bibinfo {year} {1990})}\BibitemShut {NoStop}%
\bibitem [{\citenamefont {Amendola}\ \emph {et~al.}(1990)\citenamefont
  {Amendola}, \citenamefont {Litterio},\ and\ \citenamefont
  {Occhionero}}]{Amendola:1990nn}%
  \BibitemOpen
  \bibfield  {author} {\bibinfo {author} {\bibfnamefont {L.}~\bibnamefont
  {Amendola}}, \bibinfo {author} {\bibfnamefont {M.}~\bibnamefont {Litterio}},
  \ and\ \bibinfo {author} {\bibfnamefont {F.}~\bibnamefont {Occhionero}},\
  }\href {\doibase 10.1142/S0217751X90001653} {\bibfield  {journal} {\bibinfo
  {journal} {Int. J. Mod. Phys. A}\ }\textbf {\bibinfo {volume} {5}},\ \bibinfo
  {pages} {3861} (\bibinfo {year} {1990})}\BibitemShut {NoStop}%
\bibitem [{\citenamefont {Kaiser}(1995)}]{Kaiser:1994vs}%
  \BibitemOpen
  \bibfield  {author} {\bibinfo {author} {\bibfnamefont {D.~I.}\ \bibnamefont
  {Kaiser}},\ }\href {\doibase 10.1103/PhysRevD.52.4295} {\bibfield  {journal}
  {\bibinfo  {journal} {Phys. Rev. D}\ }\textbf {\bibinfo {volume} {52}},\
  \bibinfo {pages} {4295} (\bibinfo {year} {1995})},\ \Eprint
  {http://arxiv.org/abs/astro-ph/9408044} {arXiv:astro-ph/9408044} \BibitemShut
  {NoStop}%
\bibitem [{\citenamefont {Cervantes-Cota}\ and\ \citenamefont
  {Dehnen}(1995)}]{Cervantes-Cota:1995ehs}%
  \BibitemOpen
  \bibfield  {author} {\bibinfo {author} {\bibfnamefont {J.~L.}\ \bibnamefont
  {Cervantes-Cota}}\ and\ \bibinfo {author} {\bibfnamefont {H.}~\bibnamefont
  {Dehnen}},\ }\href {\doibase 10.1016/0550-3213(95)00128-X} {\bibfield
  {journal} {\bibinfo  {journal} {Nucl. Phys. B}\ }\textbf {\bibinfo {volume}
  {442}},\ \bibinfo {pages} {391} (\bibinfo {year} {1995})},\ \Eprint
  {http://arxiv.org/abs/astro-ph/9505069} {arXiv:astro-ph/9505069} \BibitemShut
  {NoStop}%
\bibitem [{\citenamefont {Komatsu}\ and\ \citenamefont
  {Futamase}(1999)}]{Komatsu:1999mt}%
  \BibitemOpen
  \bibfield  {author} {\bibinfo {author} {\bibfnamefont {E.}~\bibnamefont
  {Komatsu}}\ and\ \bibinfo {author} {\bibfnamefont {T.}~\bibnamefont
  {Futamase}},\ }\href {\doibase 10.1103/PhysRevD.59.064029} {\bibfield
  {journal} {\bibinfo  {journal} {Phys. Rev. D}\ }\textbf {\bibinfo {volume}
  {59}},\ \bibinfo {pages} {064029} (\bibinfo {year} {1999})},\ \Eprint
  {http://arxiv.org/abs/astro-ph/9901127} {arXiv:astro-ph/9901127} \BibitemShut
  {NoStop}%
\bibitem [{\citenamefont {Akrami}\ \emph {et~al.}(2020)\citenamefont {Akrami}
  \emph {et~al.}}]{Planck:2018jri}%
  \BibitemOpen
  \bibfield  {author} {\bibinfo {author} {\bibfnamefont {Y.}~\bibnamefont
  {Akrami}} \emph {et~al.} (\bibinfo {collaboration} {Planck}),\ }\href
  {\doibase 10.1051/0004-6361/201833887} {\bibfield  {journal} {\bibinfo
  {journal} {Astron. Astrophys.}\ }\textbf {\bibinfo {volume} {641}},\ \bibinfo
  {pages} {A10} (\bibinfo {year} {2020})},\ \Eprint
  {http://arxiv.org/abs/1807.06211} {arXiv:1807.06211 [astro-ph.CO]}
  \BibitemShut {NoStop}%
\bibitem [{\citenamefont {DeCross}\ \emph
  {et~al.}(2018{\natexlab{a}})\citenamefont {DeCross}, \citenamefont {Kaiser},
  \citenamefont {Prabhu}, \citenamefont {Prescod-Weinstein},\ and\
  \citenamefont {Sfakianakis}}]{DeCross:2015uza}%
  \BibitemOpen
  \bibfield  {author} {\bibinfo {author} {\bibfnamefont {M.~P.}\ \bibnamefont
  {DeCross}}, \bibinfo {author} {\bibfnamefont {D.~I.}\ \bibnamefont {Kaiser}},
  \bibinfo {author} {\bibfnamefont {A.}~\bibnamefont {Prabhu}}, \bibinfo
  {author} {\bibfnamefont {C.}~\bibnamefont {Prescod-Weinstein}}, \ and\
  \bibinfo {author} {\bibfnamefont {E.~I.}\ \bibnamefont {Sfakianakis}},\
  }\href {\doibase 10.1103/PhysRevD.97.023526} {\bibfield  {journal} {\bibinfo
  {journal} {Phys. Rev. D}\ }\textbf {\bibinfo {volume} {97}},\ \bibinfo
  {pages} {023526} (\bibinfo {year} {2018}{\natexlab{a}})},\ \Eprint
  {http://arxiv.org/abs/1510.08553} {arXiv:1510.08553 [astro-ph.CO]}
  \BibitemShut {NoStop}%
\bibitem [{\citenamefont {Ema}\ \emph {et~al.}(2017)\citenamefont {Ema},
  \citenamefont {Jinno}, \citenamefont {Mukaida},\ and\ \citenamefont
  {Nakayama}}]{Ema:2016dny}%
  \BibitemOpen
  \bibfield  {author} {\bibinfo {author} {\bibfnamefont {Y.}~\bibnamefont
  {Ema}}, \bibinfo {author} {\bibfnamefont {R.}~\bibnamefont {Jinno}}, \bibinfo
  {author} {\bibfnamefont {K.}~\bibnamefont {Mukaida}}, \ and\ \bibinfo
  {author} {\bibfnamefont {K.}~\bibnamefont {Nakayama}},\ }\href {\doibase
  10.1088/1475-7516/2017/02/045} {\bibfield  {journal} {\bibinfo  {journal}
  {JCAP}\ }\textbf {\bibinfo {volume} {02}},\ \bibinfo {pages} {045} (\bibinfo
  {year} {2017})},\ \Eprint {http://arxiv.org/abs/1609.05209} {arXiv:1609.05209
  [hep-ph]} \BibitemShut {NoStop}%
\bibitem [{\citenamefont {Sfakianakis}\ and\ \citenamefont {van~de
  Vis}(2019)}]{Sfakianakis:2018lzf}%
  \BibitemOpen
  \bibfield  {author} {\bibinfo {author} {\bibfnamefont {E.~I.}\ \bibnamefont
  {Sfakianakis}}\ and\ \bibinfo {author} {\bibfnamefont {J.}~\bibnamefont
  {van~de Vis}},\ }\href {\doibase 10.1103/PhysRevD.99.083519} {\bibfield
  {journal} {\bibinfo  {journal} {Phys. Rev. D}\ }\textbf {\bibinfo {volume}
  {99}},\ \bibinfo {pages} {083519} (\bibinfo {year} {2019})},\ \Eprint
  {http://arxiv.org/abs/1810.01304} {arXiv:1810.01304 [hep-ph]} \BibitemShut
  {NoStop}%
\bibitem [{\citenamefont {Bezrukov}\ \emph {et~al.}(2019)\citenamefont
  {Bezrukov}, \citenamefont {Gorbunov}, \citenamefont {Shepherd},\ and\
  \citenamefont {Tokareva}}]{Bezrukov:2019ylq}%
  \BibitemOpen
  \bibfield  {author} {\bibinfo {author} {\bibfnamefont {F.}~\bibnamefont
  {Bezrukov}}, \bibinfo {author} {\bibfnamefont {D.}~\bibnamefont {Gorbunov}},
  \bibinfo {author} {\bibfnamefont {C.}~\bibnamefont {Shepherd}}, \ and\
  \bibinfo {author} {\bibfnamefont {A.}~\bibnamefont {Tokareva}},\ }\href
  {\doibase 10.1016/j.physletb.2019.06.064} {\bibfield  {journal} {\bibinfo
  {journal} {Phys. Lett. B}\ }\textbf {\bibinfo {volume} {795}},\ \bibinfo
  {pages} {657} (\bibinfo {year} {2019})},\ \Eprint
  {http://arxiv.org/abs/1904.04737} {arXiv:1904.04737 [hep-ph]} \BibitemShut
  {NoStop}%
\bibitem [{\citenamefont {Garcia-Bellido}\ \emph {et~al.}(2009)\citenamefont
  {Garcia-Bellido}, \citenamefont {Figueroa},\ and\ \citenamefont
  {Rubio}}]{Garcia-Bellido:2008ycs}%
  \BibitemOpen
  \bibfield  {author} {\bibinfo {author} {\bibfnamefont {J.}~\bibnamefont
  {Garcia-Bellido}}, \bibinfo {author} {\bibfnamefont {D.~G.}\ \bibnamefont
  {Figueroa}}, \ and\ \bibinfo {author} {\bibfnamefont {J.}~\bibnamefont
  {Rubio}},\ }\href {\doibase 10.1103/PhysRevD.79.063531} {\bibfield  {journal}
  {\bibinfo  {journal} {Phys. Rev. D}\ }\textbf {\bibinfo {volume} {79}},\
  \bibinfo {pages} {063531} (\bibinfo {year} {2009})},\ \Eprint
  {http://arxiv.org/abs/0812.4624} {arXiv:0812.4624 [hep-ph]} \BibitemShut
  {NoStop}%
\bibitem [{\citenamefont {Bezrukov}\ and\ \citenamefont
  {Shepherd}(2020)}]{Bezrukov:2020txg}%
  \BibitemOpen
  \bibfield  {author} {\bibinfo {author} {\bibfnamefont {F.}~\bibnamefont
  {Bezrukov}}\ and\ \bibinfo {author} {\bibfnamefont {C.}~\bibnamefont
  {Shepherd}},\ }\href {\doibase 10.1088/1475-7516/2020/12/028} {\bibfield
  {journal} {\bibinfo  {journal} {JCAP}\ }\textbf {\bibinfo {volume} {12}},\
  \bibinfo {pages} {028} (\bibinfo {year} {2020})},\ \Eprint
  {http://arxiv.org/abs/2007.10978} {arXiv:2007.10978 [hep-ph]} \BibitemShut
  {NoStop}%
\bibitem [{\citenamefont {Figueroa}\ \emph {et~al.}(2023)\citenamefont
  {Figueroa}, \citenamefont {Florio}, \citenamefont {Opferkuch},\ and\
  \citenamefont {Stefanek}}]{Figueroa:2021iwm}%
  \BibitemOpen
  \bibfield  {author} {\bibinfo {author} {\bibfnamefont {D.~G.}\ \bibnamefont
  {Figueroa}}, \bibinfo {author} {\bibfnamefont {A.}~\bibnamefont {Florio}},
  \bibinfo {author} {\bibfnamefont {T.}~\bibnamefont {Opferkuch}}, \ and\
  \bibinfo {author} {\bibfnamefont {B.~A.}\ \bibnamefont {Stefanek}},\ }\href
  {\doibase 10.21468/SciPostPhys.15.3.077} {\bibfield  {journal} {\bibinfo
  {journal} {SciPost Phys.}\ }\textbf {\bibinfo {volume} {15}},\ \bibinfo
  {pages} {077} (\bibinfo {year} {2023})},\ \Eprint
  {http://arxiv.org/abs/2112.08388} {arXiv:2112.08388 [astro-ph.CO]}
  \BibitemShut {NoStop}%
\bibitem [{\citenamefont {Jeong}\ \emph {et~al.}(2023)\citenamefont {Jeong},
  \citenamefont {Kamada}, \citenamefont {Starobinsky},\ and\ \citenamefont
  {Yokoyama}}]{Jeong:2023zrv}%
  \BibitemOpen
  \bibfield  {author} {\bibinfo {author} {\bibfnamefont {H.}~\bibnamefont
  {Jeong}}, \bibinfo {author} {\bibfnamefont {K.}~\bibnamefont {Kamada}},
  \bibinfo {author} {\bibfnamefont {A.~A.}\ \bibnamefont {Starobinsky}}, \ and\
  \bibinfo {author} {\bibfnamefont {J.}~\bibnamefont {Yokoyama}},\ }\href
  {\doibase 10.1088/1475-7516/2023/11/023} {\bibfield  {journal} {\bibinfo
  {journal} {JCAP}\ }\textbf {\bibinfo {volume} {11}},\ \bibinfo {pages} {023}
  (\bibinfo {year} {2023})},\ \Eprint {http://arxiv.org/abs/2305.14273}
  {arXiv:2305.14273 [hep-ph]} \BibitemShut {NoStop}%
\bibitem [{\citenamefont {Gong}\ and\ \citenamefont
  {Tanaka}(2011)}]{Gong:2011uw}%
  \BibitemOpen
  \bibfield  {author} {\bibinfo {author} {\bibfnamefont {J.-O.}\ \bibnamefont
  {Gong}}\ and\ \bibinfo {author} {\bibfnamefont {T.}~\bibnamefont {Tanaka}},\
  }\href {\doibase 10.1088/1475-7516/2012/02/E01} {\bibfield  {journal}
  {\bibinfo  {journal} {JCAP}\ }\textbf {\bibinfo {volume} {03}},\ \bibinfo
  {pages} {015} (\bibinfo {year} {2011})},\ \bibinfo {note} {[Erratum: JCAP 02,
  E01 (2012)]},\ \Eprint {http://arxiv.org/abs/1101.4809} {arXiv:1101.4809
  [astro-ph.CO]} \BibitemShut {NoStop}%
\bibitem [{\citenamefont {Kaiser}\ \emph {et~al.}(2013)\citenamefont {Kaiser},
  \citenamefont {Mazenc},\ and\ \citenamefont {Sfakianakis}}]{Kaiser:2012ak}%
  \BibitemOpen
  \bibfield  {author} {\bibinfo {author} {\bibfnamefont {D.~I.}\ \bibnamefont
  {Kaiser}}, \bibinfo {author} {\bibfnamefont {E.~A.}\ \bibnamefont {Mazenc}},
  \ and\ \bibinfo {author} {\bibfnamefont {E.~I.}\ \bibnamefont
  {Sfakianakis}},\ }\href {\doibase 10.1103/PhysRevD.87.064004} {\bibfield
  {journal} {\bibinfo  {journal} {Phys. Rev. D}\ }\textbf {\bibinfo {volume}
  {87}},\ \bibinfo {pages} {064004} (\bibinfo {year} {2013})},\ \Eprint
  {http://arxiv.org/abs/1210.7487} {arXiv:1210.7487 [astro-ph.CO]} \BibitemShut
  {NoStop}%
\bibitem [{\citenamefont {Kodama}\ and\ \citenamefont
  {Sasaki}(1984)}]{Kodama:1984ziu}%
  \BibitemOpen
  \bibfield  {author} {\bibinfo {author} {\bibfnamefont {H.}~\bibnamefont
  {Kodama}}\ and\ \bibinfo {author} {\bibfnamefont {M.}~\bibnamefont
  {Sasaki}},\ }\href {\doibase 10.1143/PTPS.78.1} {\bibfield  {journal}
  {\bibinfo  {journal} {Prog. Theor. Phys. Suppl.}\ }\textbf {\bibinfo {volume}
  {78}},\ \bibinfo {pages} {1} (\bibinfo {year} {1984})}\BibitemShut {NoStop}%
\bibitem [{\citenamefont {Mukhanov}\ \emph {et~al.}(1992)\citenamefont
  {Mukhanov}, \citenamefont {Feldman},\ and\ \citenamefont
  {Brandenberger}}]{Mukhanov:1990me}%
  \BibitemOpen
  \bibfield  {author} {\bibinfo {author} {\bibfnamefont {V.~F.}\ \bibnamefont
  {Mukhanov}}, \bibinfo {author} {\bibfnamefont {H.~A.}\ \bibnamefont
  {Feldman}}, \ and\ \bibinfo {author} {\bibfnamefont {R.~H.}\ \bibnamefont
  {Brandenberger}},\ }\href {\doibase 10.1016/0370-1573(92)90044-Z} {\bibfield
  {journal} {\bibinfo  {journal} {Phys. Rept.}\ }\textbf {\bibinfo {volume}
  {215}},\ \bibinfo {pages} {203} (\bibinfo {year} {1992})}\BibitemShut
  {NoStop}%
\bibitem [{\citenamefont {Malik}\ and\ \citenamefont
  {Wands}(2009)}]{Malik:2008im}%
  \BibitemOpen
  \bibfield  {author} {\bibinfo {author} {\bibfnamefont {K.~A.}\ \bibnamefont
  {Malik}}\ and\ \bibinfo {author} {\bibfnamefont {D.}~\bibnamefont {Wands}},\
  }\href {\doibase 10.1016/j.physrep.2009.03.001} {\bibfield  {journal}
  {\bibinfo  {journal} {Phys. Rept.}\ }\textbf {\bibinfo {volume} {475}},\
  \bibinfo {pages} {1} (\bibinfo {year} {2009})},\ \Eprint
  {http://arxiv.org/abs/0809.4944} {arXiv:0809.4944 [astro-ph]} \BibitemShut
  {NoStop}%
\bibitem [{\citenamefont {Elliston}\ \emph {et~al.}(2012)\citenamefont
  {Elliston}, \citenamefont {Seery},\ and\ \citenamefont
  {Tavakol}}]{Elliston:2012ab}%
  \BibitemOpen
  \bibfield  {author} {\bibinfo {author} {\bibfnamefont {J.}~\bibnamefont
  {Elliston}}, \bibinfo {author} {\bibfnamefont {D.}~\bibnamefont {Seery}}, \
  and\ \bibinfo {author} {\bibfnamefont {R.}~\bibnamefont {Tavakol}},\ }\href
  {\doibase 10.1088/1475-7516/2012/11/060} {\bibfield  {journal} {\bibinfo
  {journal} {JCAP}\ }\textbf {\bibinfo {volume} {11}},\ \bibinfo {pages} {060}
  (\bibinfo {year} {2012})},\ \Eprint {http://arxiv.org/abs/1208.6011}
  {arXiv:1208.6011 [astro-ph.CO]} \BibitemShut {NoStop}%
\bibitem [{\citenamefont {Sasaki}(1986)}]{Sasaki:1986hm}%
  \BibitemOpen
  \bibfield  {author} {\bibinfo {author} {\bibfnamefont {M.}~\bibnamefont
  {Sasaki}},\ }\href {\doibase 10.1143/PTP.76.1036} {\bibfield  {journal}
  {\bibinfo  {journal} {Prog. Theor. Phys.}\ }\textbf {\bibinfo {volume}
  {76}},\ \bibinfo {pages} {1036} (\bibinfo {year} {1986})}\BibitemShut
  {NoStop}%
\bibitem [{\citenamefont {Mukhanov}(1988)}]{Mukhanov:1988jd}%
  \BibitemOpen
  \bibfield  {author} {\bibinfo {author} {\bibfnamefont {V.~F.}\ \bibnamefont
  {Mukhanov}},\ }\href@noop {} {\bibfield  {journal} {\bibinfo  {journal} {Sov.
  Phys. JETP}\ }\textbf {\bibinfo {volume} {67}},\ \bibinfo {pages} {1297}
  (\bibinfo {year} {1988})}\BibitemShut {NoStop}%
\bibitem [{\citenamefont {Amin}\ \emph {et~al.}(2014)\citenamefont {Amin},
  \citenamefont {Hertzberg}, \citenamefont {Kaiser},\ and\ \citenamefont
  {Karouby}}]{Amin:2014eta}%
  \BibitemOpen
  \bibfield  {author} {\bibinfo {author} {\bibfnamefont {M.~A.}\ \bibnamefont
  {Amin}}, \bibinfo {author} {\bibfnamefont {M.~P.}\ \bibnamefont {Hertzberg}},
  \bibinfo {author} {\bibfnamefont {D.~I.}\ \bibnamefont {Kaiser}}, \ and\
  \bibinfo {author} {\bibfnamefont {J.}~\bibnamefont {Karouby}},\ }\href
  {\doibase 10.1142/S0218271815300037} {\bibfield  {journal} {\bibinfo
  {journal} {Int. J. Mod. Phys. D}\ }\textbf {\bibinfo {volume} {24}},\
  \bibinfo {pages} {1530003} (\bibinfo {year} {2014})},\ \Eprint
  {http://arxiv.org/abs/1410.3808} {arXiv:1410.3808 [hep-ph]} \BibitemShut
  {NoStop}%
\bibitem [{\citenamefont {DeCross}\ \emph
  {et~al.}(2018{\natexlab{b}})\citenamefont {DeCross}, \citenamefont {Kaiser},
  \citenamefont {Prabhu}, \citenamefont {Prescod-Weinstein},\ and\
  \citenamefont {Sfakianakis}}]{DeCross:2016cbs}%
  \BibitemOpen
  \bibfield  {author} {\bibinfo {author} {\bibfnamefont {M.~P.}\ \bibnamefont
  {DeCross}}, \bibinfo {author} {\bibfnamefont {D.~I.}\ \bibnamefont {Kaiser}},
  \bibinfo {author} {\bibfnamefont {A.}~\bibnamefont {Prabhu}}, \bibinfo
  {author} {\bibfnamefont {C.}~\bibnamefont {Prescod-Weinstein}}, \ and\
  \bibinfo {author} {\bibfnamefont {E.~I.}\ \bibnamefont {Sfakianakis}},\
  }\href {\doibase 10.1103/PhysRevD.97.023528} {\bibfield  {journal} {\bibinfo
  {journal} {Phys. Rev. D}\ }\textbf {\bibinfo {volume} {97}},\ \bibinfo
  {pages} {023528} (\bibinfo {year} {2018}{\natexlab{b}})},\ \Eprint
  {http://arxiv.org/abs/1610.08916} {arXiv:1610.08916 [astro-ph.CO]}
  \BibitemShut {NoStop}%
\bibitem [{\citenamefont {Felder}\ \emph {et~al.}(2001)\citenamefont {Felder},
  \citenamefont {Kofman},\ and\ \citenamefont {Linde}}]{Felder:2001kt}%
  \BibitemOpen
  \bibfield  {author} {\bibinfo {author} {\bibfnamefont {G.~N.}\ \bibnamefont
  {Felder}}, \bibinfo {author} {\bibfnamefont {L.}~\bibnamefont {Kofman}}, \
  and\ \bibinfo {author} {\bibfnamefont {A.~D.}\ \bibnamefont {Linde}},\ }\href
  {\doibase 10.1103/PhysRevD.64.123517} {\bibfield  {journal} {\bibinfo
  {journal} {Phys. Rev. D}\ }\textbf {\bibinfo {volume} {64}},\ \bibinfo
  {pages} {123517} (\bibinfo {year} {2001})},\ \Eprint
  {http://arxiv.org/abs/hep-th/0106179} {arXiv:hep-th/0106179} \BibitemShut
  {NoStop}%
\bibitem [{\citenamefont {Lozanov}\ and\ \citenamefont
  {Amin}(2016)}]{Lozanov:2016pac}%
  \BibitemOpen
  \bibfield  {author} {\bibinfo {author} {\bibfnamefont {K.~D.}\ \bibnamefont
  {Lozanov}}\ and\ \bibinfo {author} {\bibfnamefont {M.~A.}\ \bibnamefont
  {Amin}},\ }\href {\doibase 10.1088/1475-7516/2016/06/032} {\bibfield
  {journal} {\bibinfo  {journal} {JCAP}\ }\textbf {\bibinfo {volume} {06}},\
  \bibinfo {pages} {032} (\bibinfo {year} {2016})},\ \Eprint
  {http://arxiv.org/abs/1603.05663} {arXiv:1603.05663 [hep-ph]} \BibitemShut
  {NoStop}%
\bibitem [{\citenamefont {Bezrukov}\ \emph
  {et~al.}(2009{\natexlab{b}})\citenamefont {Bezrukov}, \citenamefont
  {Gorbunov},\ and\ \citenamefont {Shaposhnikov}}]{Bezrukov:2008ut}%
  \BibitemOpen
  \bibfield  {author} {\bibinfo {author} {\bibfnamefont {F.}~\bibnamefont
  {Bezrukov}}, \bibinfo {author} {\bibfnamefont {D.}~\bibnamefont {Gorbunov}},
  \ and\ \bibinfo {author} {\bibfnamefont {M.}~\bibnamefont {Shaposhnikov}},\
  }\href {\doibase 10.1088/1475-7516/2009/06/029} {\bibfield  {journal}
  {\bibinfo  {journal} {JCAP}\ }\textbf {\bibinfo {volume} {06}},\ \bibinfo
  {pages} {029} (\bibinfo {year} {2009}{\natexlab{b}})},\ \Eprint
  {http://arxiv.org/abs/0812.3622} {arXiv:0812.3622 [hep-ph]} \BibitemShut
  {NoStop}%
\bibitem [{\citenamefont {Domcke}\ and\ \citenamefont
  {Mukaida}(2018)}]{Domcke:2018eki}%
  \BibitemOpen
  \bibfield  {author} {\bibinfo {author} {\bibfnamefont {V.}~\bibnamefont
  {Domcke}}\ and\ \bibinfo {author} {\bibfnamefont {K.}~\bibnamefont
  {Mukaida}},\ }\href {\doibase 10.1088/1475-7516/2018/11/020} {\bibfield
  {journal} {\bibinfo  {journal} {JCAP}\ }\textbf {\bibinfo {volume} {11}},\
  \bibinfo {pages} {020} (\bibinfo {year} {2018})},\ \Eprint
  {http://arxiv.org/abs/1806.08769} {arXiv:1806.08769 [hep-ph]} \BibitemShut
  {NoStop}%
\bibitem [{\citenamefont {Aghanim}\ \emph {et~al.}(2020)\citenamefont {Aghanim}
  \emph {et~al.}}]{Planck:2018vyg}%
  \BibitemOpen
  \bibfield  {author} {\bibinfo {author} {\bibfnamefont {N.}~\bibnamefont
  {Aghanim}} \emph {et~al.} (\bibinfo {collaboration} {Planck}),\ }\href
  {\doibase 10.1051/0004-6361/201833910} {\bibfield  {journal} {\bibinfo
  {journal} {Astron. Astrophys.}\ }\textbf {\bibinfo {volume} {641}},\ \bibinfo
  {pages} {A6} (\bibinfo {year} {2020})},\ \bibinfo {note} {[Erratum:
  Astron.Astrophys. 652, C4 (2021)]},\ \Eprint
  {http://arxiv.org/abs/1807.06209} {arXiv:1807.06209 [astro-ph.CO]}
  \BibitemShut {NoStop}%
\bibitem [{\citenamefont {Workman}\ \emph {et~al.}(2022)\citenamefont {Workman}
  \emph {et~al.}}]{ParticleDataGroup:2022pth}%
  \BibitemOpen
  \bibfield  {author} {\bibinfo {author} {\bibfnamefont {R.~L.}\ \bibnamefont
  {Workman}} \emph {et~al.} (\bibinfo {collaboration} {Particle Data Group}),\
  }\href {\doibase 10.1093/ptep/ptac097} {\bibfield  {journal} {\bibinfo
  {journal} {PTEP}\ }\textbf {\bibinfo {volume} {2022}},\ \bibinfo {pages}
  {083C01} (\bibinfo {year} {2022})}\BibitemShut {NoStop}%
\bibitem [{\citenamefont {Sakharov}(1967)}]{Sakharov:1967dj}%
  \BibitemOpen
  \bibfield  {author} {\bibinfo {author} {\bibfnamefont {A.~D.}\ \bibnamefont
  {Sakharov}},\ }\href {\doibase 10.1070/PU1991v034n05ABEH002497} {\bibfield
  {journal} {\bibinfo  {journal} {Pisma Zh. Eksp. Teor. Fiz.}\ }\textbf
  {\bibinfo {volume} {5}},\ \bibinfo {pages} {32} (\bibinfo {year}
  {1967})}\BibitemShut {NoStop}%
\bibitem [{\citenamefont {Kajantie}\ \emph {et~al.}(1997)\citenamefont
  {Kajantie}, \citenamefont {Laine}, \citenamefont {Rummukainen},\ and\
  \citenamefont {Shaposhnikov}}]{Kajantie:1996qd}%
  \BibitemOpen
  \bibfield  {author} {\bibinfo {author} {\bibfnamefont {K.}~\bibnamefont
  {Kajantie}}, \bibinfo {author} {\bibfnamefont {M.}~\bibnamefont {Laine}},
  \bibinfo {author} {\bibfnamefont {K.}~\bibnamefont {Rummukainen}}, \ and\
  \bibinfo {author} {\bibfnamefont {M.~E.}\ \bibnamefont {Shaposhnikov}},\
  }\href {\doibase 10.1016/S0550-3213(97)00164-8} {\bibfield  {journal}
  {\bibinfo  {journal} {Nucl. Phys. B}\ }\textbf {\bibinfo {volume} {493}},\
  \bibinfo {pages} {413} (\bibinfo {year} {1997})},\ \Eprint
  {http://arxiv.org/abs/hep-lat/9612006} {arXiv:hep-lat/9612006} \BibitemShut
  {NoStop}%
\bibitem [{\citenamefont {D'Onofrio}\ and\ \citenamefont
  {Rummukainen}(2016)}]{DOnofrio:2015gop}%
  \BibitemOpen
  \bibfield  {author} {\bibinfo {author} {\bibfnamefont {M.}~\bibnamefont
  {D'Onofrio}}\ and\ \bibinfo {author} {\bibfnamefont {K.}~\bibnamefont
  {Rummukainen}},\ }\href {\doibase 10.1103/PhysRevD.93.025003} {\bibfield
  {journal} {\bibinfo  {journal} {Phys. Rev. D}\ }\textbf {\bibinfo {volume}
  {93}},\ \bibinfo {pages} {025003} (\bibinfo {year} {2016})},\ \Eprint
  {http://arxiv.org/abs/1508.07161} {arXiv:1508.07161 [hep-ph]} \BibitemShut
  {NoStop}%
\bibitem [{\citenamefont {Cado}(2023)}]{Cado:2023jty}%
  \BibitemOpen
  \bibfield  {author} {\bibinfo {author} {\bibfnamefont {Y.~D.}\ \bibnamefont
  {Cado}},\ }\emph {\bibinfo {title} {{Baryogenesis and Inflation from the
  Higgs sector}}},\ \href@noop {} {Ph.D. thesis},\ \bibinfo  {school}
  {Barcelona, Autonoma U.} (\bibinfo {year} {2023})\BibitemShut {NoStop}%
\bibitem [{\citenamefont {D'Onofrio}\ \emph {et~al.}(2014)\citenamefont
  {D'Onofrio}, \citenamefont {Rummukainen},\ and\ \citenamefont
  {Tranberg}}]{DOnofrio:2014rug}%
  \BibitemOpen
  \bibfield  {author} {\bibinfo {author} {\bibfnamefont {M.}~\bibnamefont
  {D'Onofrio}}, \bibinfo {author} {\bibfnamefont {K.}~\bibnamefont
  {Rummukainen}}, \ and\ \bibinfo {author} {\bibfnamefont {A.}~\bibnamefont
  {Tranberg}},\ }\href {\doibase 10.1103/PhysRevLett.113.141602} {\bibfield
  {journal} {\bibinfo  {journal} {Phys. Rev. Lett.}\ }\textbf {\bibinfo
  {volume} {113}},\ \bibinfo {pages} {141602} (\bibinfo {year} {2014})},\
  \Eprint {http://arxiv.org/abs/1404.3565} {arXiv:1404.3565 [hep-ph]}
  \BibitemShut {NoStop}%
\bibitem [{\citenamefont {Baym}\ and\ \citenamefont
  {Heiselberg}(1997)}]{Baym:1997gq}%
  \BibitemOpen
  \bibfield  {author} {\bibinfo {author} {\bibfnamefont {G.}~\bibnamefont
  {Baym}}\ and\ \bibinfo {author} {\bibfnamefont {H.}~\bibnamefont
  {Heiselberg}},\ }\href {\doibase 10.1103/PhysRevD.56.5254} {\bibfield
  {journal} {\bibinfo  {journal} {Phys. Rev. D}\ }\textbf {\bibinfo {volume}
  {56}},\ \bibinfo {pages} {5254} (\bibinfo {year} {1997})},\ \Eprint
  {http://arxiv.org/abs/astro-ph/9704214} {arXiv:astro-ph/9704214} \BibitemShut
  {NoStop}%
\bibitem [{\citenamefont {Arnold}\ \emph {et~al.}(2000)\citenamefont {Arnold},
  \citenamefont {Moore},\ and\ \citenamefont {Yaffe}}]{Arnold:2000dr}%
  \BibitemOpen
  \bibfield  {author} {\bibinfo {author} {\bibfnamefont {P.~B.}\ \bibnamefont
  {Arnold}}, \bibinfo {author} {\bibfnamefont {G.~D.}\ \bibnamefont {Moore}}, \
  and\ \bibinfo {author} {\bibfnamefont {L.~G.}\ \bibnamefont {Yaffe}},\ }\href
  {\doibase 10.1088/1126-6708/2000/11/001} {\bibfield  {journal} {\bibinfo
  {journal} {JHEP}\ }\textbf {\bibinfo {volume} {11}},\ \bibinfo {pages} {001}
  (\bibinfo {year} {2000})},\ \Eprint {http://arxiv.org/abs/hep-ph/0010177}
  {arXiv:hep-ph/0010177} \BibitemShut {NoStop}%
\bibitem [{\citenamefont {Joyce}\ and\ \citenamefont
  {Shaposhnikov}(1997)}]{Joyce:1997uy}%
  \BibitemOpen
  \bibfield  {author} {\bibinfo {author} {\bibfnamefont {M.}~\bibnamefont
  {Joyce}}\ and\ \bibinfo {author} {\bibfnamefont {M.~E.}\ \bibnamefont
  {Shaposhnikov}},\ }\href {\doibase 10.1103/PhysRevLett.79.1193} {\bibfield
  {journal} {\bibinfo  {journal} {Phys. Rev. Lett.}\ }\textbf {\bibinfo
  {volume} {79}},\ \bibinfo {pages} {1193} (\bibinfo {year} {1997})},\ \Eprint
  {http://arxiv.org/abs/astro-ph/9703005} {arXiv:astro-ph/9703005} \BibitemShut
  {NoStop}%
\bibitem [{\citenamefont {Cado}\ and\ \citenamefont
  {Quir\'os}(2023)}]{Cado:2023gan}%
  \BibitemOpen
  \bibfield  {author} {\bibinfo {author} {\bibfnamefont {Y.}~\bibnamefont
  {Cado}}\ and\ \bibinfo {author} {\bibfnamefont {M.}~\bibnamefont
  {Quir\'os}},\ }\href {\doibase 10.1103/PhysRevD.108.023508} {\bibfield
  {journal} {\bibinfo  {journal} {Phys. Rev. D}\ }\textbf {\bibinfo {volume}
  {108}},\ \bibinfo {pages} {023508} (\bibinfo {year} {2023})},\ \Eprint
  {http://arxiv.org/abs/2303.12932} {arXiv:2303.12932 [hep-ph]} \BibitemShut
  {NoStop}%
\end{thebibliography}%
%%%%%%%%%%%%%%%%%%%%%%%%%%%%%%%%%%%%%  

\end{document}